\documentclass[12pt]{article}
\usepackage[utf8]{inputenc}

\usepackage{graphicx}
\usepackage{physics}
\usepackage{amsmath}
\usepackage{mathtools}
\usepackage{cases}
\usepackage{amssymb}
\usepackage{amsfonts}
\usepackage{multicol}
\usepackage{enumitem}
\usepackage{caption}
\usepackage{subcaption}
\usepackage{mwe}
\usepackage{amsthm}
\usepackage{blindtext}
\usepackage{yfonts}
\usepackage{multirow}
\usepackage[tablename=Tabela]{caption}
\usepackage[usenames, dvipsnames]{color}

\usepackage[a4paper, total={5.5in, 9in},left=4cm]{geometry}

\usepackage{pgf,tikz}
\usepackage{mathrsfs}

\usepackage{tikz}
\usetikzlibrary{positioning}
\usetikzlibrary{arrows}
\usetikzlibrary{backgrounds}
\usetikzlibrary{decorations.markings}

\graphicspath{ {Img/} }

\usepackage[nottoc]{tocbibind}

\newcommand*{\BeginNoToc}{%
  \addtocontents{toc}{%
    \edef\protect\SavedTocDepth{\protect\the\protect\value{tocdepth}}%
  }%
  \addtocontents{toc}{%
    \protect\setcounter{tocdepth}{-10}%
  }%
}
\newcommand*{\EndNoToc}{%
  \addtocontents{toc}{%
    \protect\setcounter{tocdepth}{\protect\SavedTocDepth}%
  }%
}

\newcommand{\bld}{\textbf}
\newcommand{\sumRinG}{\sum_{\bld{R}\in G}}
\newcommand{\epikdr}{e^{i \bld{k} \cdot \bld{R}}}
\newcommand{\psik}{\psi_\bld{k}}

\newcommand{\bR}{\bld{R}}

\newcommand{\red}{\textcolor{red}}

\newcommand{\ZZ}{\mathbb{Z}}
\newcommand{\NN}{\mathbb{N}}
\newcommand{\RR}{\mathbb{R}}

\title{MasterT}
\author{Manuel Santos}
\date{June 2017}

\begin{document}

\begin{titlepage}
    
    \newcommand{\HRule}{\rule{\linewidth}{0.5mm}} 


\includegraphics[width = 4cm]{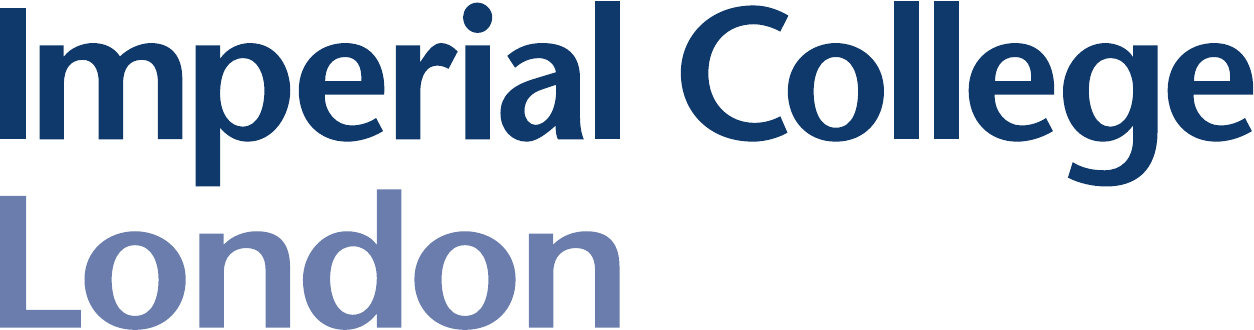}\\[0.5cm] 

\center 


\textsc{\Large Imperial College London}\\[0.5cm] 
\textsc{\large Department of Mathematics}\\[0.5cm] 


\HRule \\[0.4cm]
{ \huge \bfseries Topological Band Systems and Finite Size Effects}\\ 
\HRule \\[1.5cm]
 

\begin{minipage}{0.4\textwidth}
\begin{flushleft} \large
\emph{Author:}\\
Manuel B. Santos
\end{flushleft}
\end{minipage}
~
\begin{minipage}{0.4\textwidth}
\begin{flushright} \large
\emph{Supervisor:} \\
Dr Ryan Barnett
\end{flushright}
\end{minipage}\\[4cm]

\vfill 
Submitted in partial fulfillment of the requirements for the MSc degree in Applied Mathematics of Imperial College London\\[0.5cm]

\makeatletter
June 2017 
\makeatother

    


\end{titlepage}

\newpage

\pagenumbering{roman}
\renewcommand{\abstractname}{}
\begin{abstract}
\normalsize
The recent discoveries about topological insulator have been promoting theoretical and experimental research. In this dissertation, the basic concepts of topological insulators and the Quantum Hall Effect are reviewed focusing the discussion on edge states and its band structure. Lattice models with pierced magnetism are described and the Hofstadter model is presented for bounded systems with and without an in-site disorder. An overview of the experimental procedure based on cold atoms in optical lattices with synthetic dimensions is given. In order to understand to what extent these small systems of cold atoms mimic the behaviour of a topological insulator, an analysis of some finite size effects is provided and a deduction of the gap opening in the band structure is presented using perturbation theory.
\end{abstract}

\newpage

\section*{}

\vfill

The work contained in this thesis is my own work unless otherwise stated.

\

Signed: Manuel Santos

\newpage

\renewcommand{\abstractname}{}
\begin{abstract}
\normalsize
I would like to express my sincere gratitude to my supervisor, Dr Ryan Barnett, for his valuable guidance; to my parents for giving me this opportunity; and to Teresinha, Zé Reis and all who lived with me throughout this year for their support.
\end{abstract}

\newpage

\BeginNoToc

\listoffigures

\EndNoToc

\newpage

\tableofcontents

\newpage
\pagenumbering{arabic}
\section{Introduction}

In the past few years, condensed-matter physics has seen from theoretical predictions the rising of a new and exciting phase of matter: topological insulators. The close relationship between these and some quantum effects has been boosting theoretical and experimental research \cite{Hasan_2010_TopoInsul}. One of these effects is called Quantum Hall Effect (QHE) and it concerns the quantized values of the so called Hall conductivity. It is observed when a magnetic field pierces finite two dimensional systems with a longitudinal current applied. In fact, QHE allows the realization of a topological insulator, where the bulk behaves like a classical insulator but the edges of the system allow current to flow. Some of the features of this new phase of matter, such as robustness to impurities, make it attractive and useful for applications to quantum computation or spintronics.

In addition to great physical interest, this effect is strongly influenced by several branches of Mathematics such as Topology and Differential Geometry. They play a decisive role in the values of the transverse conductance of the system, $\sigma_{xy}$. Based on a mathematical result that relates the number of holes in an object with its geometry (Gauss-Bonnet Theorem), it was found that $\sigma_{xy}$ is given in integer multiples of $\frac{e^2}{2\pi \hbar}$. We will see that this quantization is directly related to the filling of a specific type of energy levels: Landau levels. Interestingly, there is no relationship between the conductance values and the amount of disorder in the system. However, the accuracy of the QHE increases with the level of impurities, making them the cause of the existence of resistivity plateaux for certain intervals of magnetic fields.

Since the QHE is only visible for large magnetic fields and very low temperatures, the experimental effort and investment are big. Therefore, it is important to develop less demanding systems that mimic its behaviour. The use of synthetic magnetic fields with cold atoms has been crucial to study and explore these systems \cite{Price_2017,Lacki_2016,DM_Celi_2014,DM_Hugel_2014,DM_Stuhl_2015,Mancini_2015,Lin_2009,Aidelsburger_2013,Miyake_2013,Jim_2012}. However, it is important to understand in which conditions the experimental apparatus and the results obtained actually depicts a topological insulator. In the case of large systems, the gapless energy dispersion of the edge states supports the existence of edge currents. However, for small systems, the band structure is deformed leading to a gap opening which could influence some topological properties of the topological insulators. In fact, one of the features of these cold-atomic gases experiments is that the system size is small, typically with just three sites in one of the directions for discretized systems. Regarding this issue, one of the main themes of this dissertation is to investigate the types of topological effects that will survive for these small sized systems. 

This dissertation is organized as follows:
\begin{itemize}
    \item In chapter \ref{theory} an overview of the background material is given allowing the understanding of the Quantum Hall Effect and some of its most important properties: edge states and Chern number. Then a description of lattice models in the presence of magnetism is given. This chapter is largely based on several sources from the literature \cite{Hall_1879,Lect_QHE,Klitzing_1980_IQHE,Suddards_2012_QHE,Landau_1977_Book,Fradkin_2013_Book,Halperin_1982,Hasan_2010_TopoInsul,Moore_2010,Laughlin_1981,Berry_1984,Thouless_1985,Thouless_1982,Simon_2013_Book,Aharonov_1959,Peierls_1933}. 
    \item The well known Hofstadter model is described in chapter \ref{hofsModel}. Then, we perform some numerical simulations of the band structure of a system using this model. This helps to gain some insight into the effects of the size of the system and other parameters on the dispersion relation. Also, to test the robustness of edge states in this Quantum Hall regime, we simulate the band structure in the presence of a random potential representing in-site impurities. All calculations and numerical simulations presented are my own though similar results can be found in some sources \cite{DM_Stuhl_2015,DM_Mugel_2017,DM_Hatsugai_2016,DM_Celi_2014,DM_Hugel_2014,YH_1993}. Here we look for a different parameter regime than those usually discussed.
    \item In chapter \ref{expProc} we intend to give a general overview of the experimental procedure used to represent the Hofstadter model based on cold atoms \cite{DM_Stuhl_2015,DM_Celi_2014,Mancini_2015,Lin_2009,Jim_2012,Aidelsburger_2013,Oliver_2006}.
    \item Finally, in chapter \ref{problemSection} we deduce expressions for the dispersion relation of the top and bottom edge states in a semi-infinite system. To my knowledge, these relations are not present in the literature. After that, we discover that finite size systems open a gap in the dispersion relation of the edge states and we find an asymptotic expression of this energy gap which depends exponentially on the system size. This result would be of immediate interest to those working experimentally on topological insulators that are trying to understand topological phenomena in finite size systems.
\end{itemize}

\newpage

\section{Theoretical background} \label{theory}

In this thesis, we are interested in looking at two dimensional systems under a perpendicular magnetic field. In order to understand the relationship between topological insulators and the Quantum Hall Effect (QHE), we start this section with an explanation of the Classical Hall Effect. After that, we describe the quantum system in the continuum case and some of its topological properties. Finally, we see how to define these systems in a lattice. These lattice systems will, in fact, reduce to the continuum case in the limit when the lattice constant is very small. This provides with the background for the next chapter where quantum phenomena and lattice systems under a magnetic field come together in the Hofstadter model.

\subsection{Quantum Hall Effect}

\subsubsection{Classical Hall Effect}

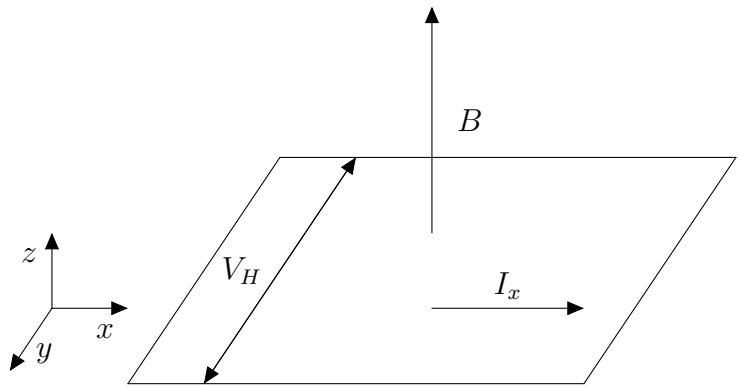
\begin{figure}[h]
\begin{center}
\begin{tikzpicture}[line cap=round,line join=round,>=triangle 45,x=1.0cm,y=1.0cm]

\draw[color=black] (0,0) -- (2,3);
\draw[color=black] (0,0) -- (6,0);
\draw[color=black] (2,3) -- (8,3);
\draw[color=black] (6,0) -- (8,3);

\node at (4.5,3.5) {$B$};
\draw[->,color=black] (4,2) -- (4,5);
\node at (5,1.3) {$I_x$};
\draw[->,color=black] (4,1) -- (6,1);
\node at (1.5,1.5) {$V_H$};
\draw[->,color=black] (1,0) -- (3,3);
\draw[->,color=black] (3,3) -- (1,0);

\node at (-0.3,0.7) {$x$};
\draw[->] (-1,1) -- (0,1);
\node at (-1.3,1.7) {$z$};
\draw[->] (-1,1) -- (-1,2);
\node at (-1.1,0.4) {$y$};
\draw[->] (-1,1) -- (-1.5547, 0.16795);

\end{tikzpicture}
\end{center}
\caption{Classical Hall Effect.}
\label{fig:classicalHallEffect}
\end{figure}

We start briefly to explain the Classical Hall Effect which was discovered in 1879 by Edwin Hall \cite{Hall_1879}. Following closely the analysis carried in \cite{Lect_QHE}, we consider a finite two-dimensional rectangle with an applied current flowing in the $x$-direction from left to right and a perpendicular magnetic field \bld{B} in the $z$-direction as shown in Fig.~\ref{fig:classicalHallEffect}.
The path of the electrons is bent due to the magnetic field and a charge is built up along the edge of the rectangle. In a scenario described by Eq.~(\ref{eq:EoM}) without electric field, we know electrons start moving in circles with frequency $\omega_\bld{B} = \frac{e \bld{B}}{m}$ (\textit{cyclotron frequency}).

\begin{equation}
\label{eq:EoM}
    m \dv{\bld{v}}{t} = - e \bld{v} \times \bld{B},
\end{equation}
where $m$, $\bld{v}$ and $e$ are the mass, velocity and charge of the electron, respectively.

The \textit{Hall Effect} is the voltage created in the $y$-direction, $V_H$. An equilibrium is attained when the electric field in the $y$-direction, $\bld{E}_y$, cancels the action of the magnetic field $\bld{B}$ in $\bld{E}_x$. So, considering the Drude model, where electrons are considered to act like pinballs, the system is described by the following expression:

\begin{equation}
\label{eq:DrudeModel}
    m \dv{\bld{v}}{t} = -e \bld{E} - e \bld{v} \times \bld{B} - \frac{m\bld{v}}{\tau},
\end{equation}
where $\bld{E}$ is the electric field and $\tau$ is a term representing friction. Assuming we are in equilibrium, $\dv{\bld{v}}{t}=0$, and solving Eq.~(\ref{eq:DrudeModel}), we get the Ohm's law:

\begin{equation}
\label{eq:OhmLaw}
\bld{J} = \sigma \bld{E},
\end{equation}
where $\bld{J}$ represents the current ($\bld{J}=-n e \bld{v}$),
\begin{eqnarray*}
\sigma = \frac{\sigma_{DC}}{1+ \sigma^2 \tau^2} \mqty(1 & -\omega_\bld{B}\tau \\
\omega_\bld{B}\tau & 1)  \quad \textrm{and} \quad \sigma_{DC} = \frac{n e^2 \tau}{m}.
\end{eqnarray*}
The anti-diagonal terms of $\sigma$ are responsible for the Hall Effect. Due to historical reasons, let us consider the resistivity: 

\begin{equation*}
\rho = \sigma^{-1}  =  \frac{1}{\sigma_{DC}} \mqty(1 & \omega_\bld{B}\tau \\
-\omega_\bld{B}\tau & 1) = \mqty(\rho_{xx} & \rho_{xy} \\ -\rho_{xy} & \rho_{xx}),
\end{equation*}
where 

\begin{equation}
\label{eq:classicalHR}
\rho_{xx} = \frac{m}{n e^2 \tau} \textrm{,} \quad \rho_{xy} = \frac{B}{n e} \quad \textrm{and $n$ is the electron density.}  
\end{equation}
Thus, the Classical Hall Effect claims $\rho_{xx}$ does not depend on the magnetic field while $\rho_{xy}$ is linearly dependent on $B$. On the other hand, $\rho_{xx}$ depends on the material ($\tau$) while $\rho_{xy}$ is not affected by it.

\begin{figure}[h]
\begin{center}
\begin{tikzpicture}[line cap=round,line join=round,>=triangle 45,x=1.0cm,y=1.0cm]

\node at (2,-0.4) {$B$};
\draw[->,color=black] (0,0) -- (0,4);

\node at (4.4,2) {\red{$\rho_{xy}$}};
\draw[color=red] (0,0) -- (3.7,3.7);
\draw[->,color=red] (4,0) -- (4,4);

\node at (-0.4,2) {$\rho_{xx}$};
\draw[color=black] (0,1) -- (3.7,1);
\draw[->,color=black] (0,0) -- (4,0);

\end{tikzpicture}
\end{center}
\caption{Prediction for the resistivities $\rho_{xx}$ and $\rho_{xy}$ in the classical picture.}
\label{fig:predictRest}
\end{figure}
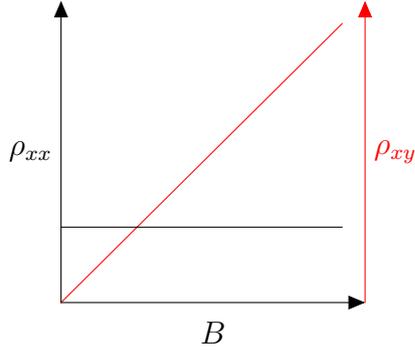

\subsubsection{Quantum Hall Effect}

\begin{figure}[h]
    \centering
    \includegraphics[width=8cm, height=7cm]{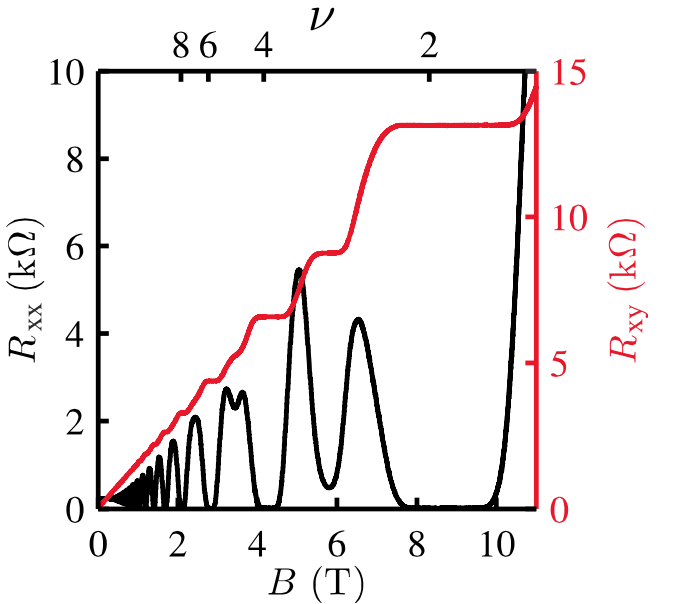}
    \caption{Experimental evidence of the QHE taken from \cite{Suddards_2012_QHE}.}
    \label{fig:IQHE}
\end{figure}

In 1980, Klaus von Klitzing observed for the first time the \textit{Integer Quantum Hall Effect} (IQHE) under extremely low temperatures and strong magnetic fields \cite{Klitzing_1980_IQHE}. Along with Dorda and Pepper, they recorded the resistivities to be similar to those shown in Fig.~\ref{fig:IQHE}. 
Unlike the prediction made by Hall (Fig.~\ref{fig:predictRest}), the Hall resistivity is a step function with plateaux set on
\begin{equation}
\label{eq:hallResistivity}
\rho_{xy} = \frac{2\pi \hbar}{e^2}\frac{1}{\nu},
\end{equation}
where $\nu \in \ZZ$. Interestingly, the longitudinal resistivity, $\rho_{xx}$ ($R_{xx}$ in Fig.~\ref{fig:IQHE}), vanishes when $\rho_{xy}$ ($R_{xy}$ in Fig.~\ref{fig:IQHE}) is in a plateau and has sharp spikes when $\nu$ change number.

In order to better understand why $\nu$ is quantized, let us take a quick overview of the behaviour of a single electron in a magnetic field under quantum mechanic rules. Here we assume electrons are spinless and there is no interaction between them. In this case, the Hamiltonian of the system is given by 

\begin{equation}
\label{eq:landauHam}
H = \frac{1}{2 m} (\bld{p} + e\bld{A})^2,
\end{equation}
where $\bld{B} = (0,0,B) = \nabla \times \bld{A}$. We want to find the eigenvalues and eigenfunctions of the Schr\"odinger Equation (SE):

\begin{equation}
\label{eq:SE}
    H\psi = E\psi.
\end{equation}
It is possible to find the eigenenergies of (\ref{eq:SE}) in a purely algebraic method similar to the quantum harmonic oscillator. As it is mentioned in \cite{Landau_1977_Book} the eigenenergies are 
\begin{equation}
\label{eq:landauLevels}
    E_n = \hbar \omega_\bld{B} (n + \frac{1}{2}),
\end{equation}
where $n \in \NN$. These energy levels are known as \textit{Landau levels}.

To find the eigenfunctions it is necessary to choose a particular gauge potential $\bld{A}$. Most commonly it is chosen the \textit{Landau Gauge} given by $\bld{A} = (0,Bx)$. Then the Hamiltonian (\ref{eq:landauHam}) can be written as
\begin{equation}
\label{eq:landauGaugeHam}
    H = \frac{1}{2 m}\big(p_x^2 + (p_y+eBx)^2\big).
\end{equation}
As it is proposed in \cite{Lect_QHE}, due to translational invariance in the $y$-direction we can take as an ansatz the following function:
\begin{equation}
\label{eq:landauAnsatz}
    \psi_k(x,y) = e^{i ky} f_k(x).
\end{equation}
Applying (\ref{eq:landauAnsatz}) into (\ref{eq:landauGaugeHam}) it can be concluded that the wavefunctions are given by
\begin{equation}
\label{eq:landauFunction}
    \psi_k(x,y) = C e^{i ky} H_n(x+k l_B^2) e^{-(x+kl_B^2)/2l_B^2},
\end{equation}
where $C$ is a normalization constant, $l_B = \sqrt{\frac{\hbar}{eB}}$ is called the magnetic length, $H_n$ is an Hermite polynomial, $n \in \NN$ and $k \in \RR$ are quantum numbers.

Note that the eigenfunctions (\ref{eq:landauFunction}) depend on both $n$ and $k$ while the eigenenergies (\ref{eq:landauLevels}) only depend on $n$. This is a clue that there is a degeneracy. As analyzed in \cite{Lect_QHE}, in a finite system the number of states for each Landau level that fit inside a rectangle with area $A$ is given by
\begin{equation}
\label{eq:numbStates}
N = \frac{e B A}{2\pi\hbar} = \frac{BA}{\Phi_0} \quad \textrm{and} \quad \Phi_0 = \frac{2 \pi \hbar}{e},
\end{equation}
where $\Phi_0$ is known as the \textit{quantum of flux}. So, an electron travelling in a cyclotron orbit occupies an area of $A = \frac{2\pi\hbar}{B e} = 2\pi l_B^2$. This can be easily deduced by setting $N=1$ in (\ref{eq:numbStates}). 
By knowing the number of filled Landau levels ($\nu$) we get the total number of electrons, $N_{tot}=N\nu$, and then we can conclude that the density of electrons is given by

\begin{equation}
\label{eq:densityStates}
n = \frac{N_{tot}}{A} = \frac{B}{\Phi_0}\nu.
\end{equation}
From (\ref{eq:densityStates}) and the classical expression (\ref{eq:classicalHR}) we can deduce that the Hall resistivity agrees with the experimental predictions (\ref{eq:hallResistivity}),
\begin{equation*}
\rho_{xy} = \frac{B}{n e} = \frac{\Phi_0}{\nu e} = \frac{2\pi \hbar}{e^2}\frac{1}{\nu}.
\end{equation*}

\

In the next two sections, we explore some topological properties. We start by describing the reason for edge states to appear in finite systems and how the disorder can explain the existence and accuracy of the plateaux. After this, we give a brief overview of an important topological invariant: Chern number.

\subsubsection{Edge states} \label{Edge_States}

\begin{figure}[h]
\begin{center}
\begin{tikzpicture}[line cap=round,line join=round,>=triangle 45,x=1.0cm,y=1.0cm]
\draw [gray,very thin, xstep=1.0cm,ystep=1.0cm] (-4.2,3) grid (5.2,-2);

\clip(-5.9,3.9) rectangle (6.9,-2.9);

    \draw[
        decoration={markings, mark=at position 0.90 with {\arrow{>}}},
        postaction={decorate}
        ]
        (-2.5,1.5) circle (0.45cm);
        \clip(-5.9,3.9) rectangle (6.9,-2.9);
    \draw[
        decoration={markings, mark=at position 0.90 with {\arrow{>}}},
        postaction={decorate}
        ]
        (1.5,1.5) circle (0.45cm);
    \draw[
        decoration={markings, mark=at position 0.90 with {\arrow{>}}},
        postaction={decorate}
        ]
        (-0.5,1.5) circle (0.45cm);
    \draw[
        decoration={markings, mark=at position 0.90 with {\arrow{>}}},
        postaction={decorate}
        ]
        (3.5,1.5) circle (0.45cm);
    \draw[
        decoration={markings, mark=at position 0.90 with {\arrow{>}}},
        postaction={decorate}
        ]
        (-2.5,-0.5) circle (0.45cm);
    \draw[
        decoration={markings, mark=at position 0.90 with {\arrow{>}}},
        postaction={decorate}
        ]
        (-0.5,-0.5) circle (0.45cm);
    \draw[
        decoration={markings, mark=at position 0.90 with {\arrow{>}}},
        postaction={decorate}
        ]
        (1.5,-0.5) circle (0.45cm);
    \draw[
        decoration={markings, mark=at position 0.90 with {\arrow{>}}},
        postaction={decorate}
        ]
        (3.5,-0.5) circle (0.45cm);
\node at (0.5,0.5) {Cyclotron orbits};
        
\draw [line width=1.2pt] (-4.2,3.)-- (5.2,3.);
\draw [line width=1.2pt] (-4.2,-2.)-- (5.2,-2.);

\node [above] at (0.5,3) {Edge states};
\draw [->] (-4,3) arc (-150:-30:15.5pt);
\draw [->] (-3,3) arc (-150:-30:15.5pt);
\draw [->] (-2,3) arc (-150:-30:15.5pt);
\draw [->] (-1,3) arc (-150:-30:15.5pt);
\draw [->] (0,3) arc (-150:-30:15.5pt);
\draw [->] (1,3) arc (-150:-30:15.5pt);
\draw [->] (2,3) arc (-150:-30:15.5pt);
\draw [->] (3,3) arc (-150:-30:15.5pt);
\draw [->] (4,3) arc (-150:-30:15.5pt);
\draw [<-] (-4,-2) arc (150:30:15.5pt);
\draw [<-] (-3,-2) arc (150:30:15.5pt);
\draw [<-] (-2,-2) arc (150:30:15.5pt);
\draw [<-] (-1,-2) arc (150:30:15.5pt);
\draw [<-] (0,-2) arc (150:30:15.5pt);
\draw [<-] (1,-2) arc (150:30:15.5pt);
\draw [<-] (2,-2) arc (150:30:15.5pt);
\draw [<-] (3,-2) arc (150:30:15.5pt);
\draw [<-] (4,-2) arc (150:30:15.5pt);
\node [below] at (0.5,-2) {Edge states};
\end{tikzpicture}
\end{center}
\caption{Cyclotron orbits.}
\label{fig:cyclotronOrbit}
\end{figure}
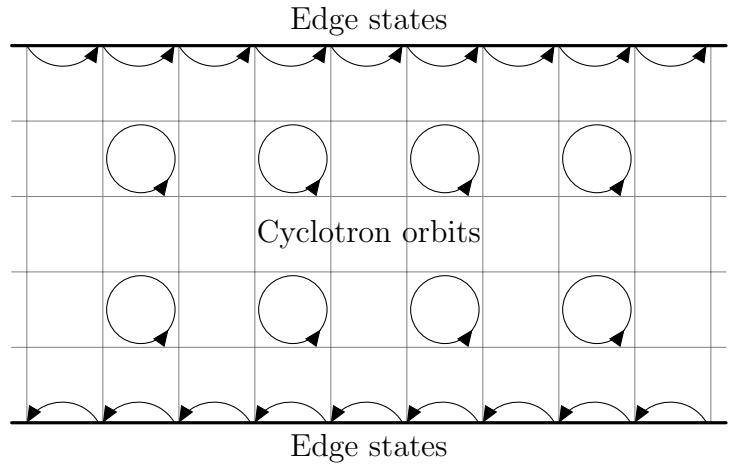
Let us consider a finite system with hard boundaries. Classically, the edge states can be thought as electrons bouncing back in the hard wall. As the magnetic field is constant, the motion of the electrons has only one direction, say anti-clockwise. This results in opposite skipping orbits in each side of the system. This is shown in Fig.~\ref{fig:cyclotronOrbit}.

The direction in which these edge states propagate depends exclusively on the orientation of the magnetic field. As pointed by E. Fradkin in \cite{Fradkin_2013_Book} based on the original article by B. I. Halperin \cite{Halperin_1982}, that is the reason for the edge currents not to be affected by impurities in the system. Since electrons cannot go back the only possibility is to overcome the impurities. This also implies that there are no \textit{localized states} on the edges but only \textit{extended states}.

So, we have a material which is transformed into an insulator in the bulk with conducting edges. As defined in \cite{Hasan_2010_TopoInsul}, this type of electronic materials are called \textit{topological insulators}. In \cite{Moore_2010}, Joel E. Moore provides with a more intuitive overview of topological insulators and describes its possible applications to spintronic and quantum computing.

\begin{figure}[h]
    \begin{subfigure}[t]{\textwidth}
        \centering
        \includegraphics[width=3cm, height=3cm]{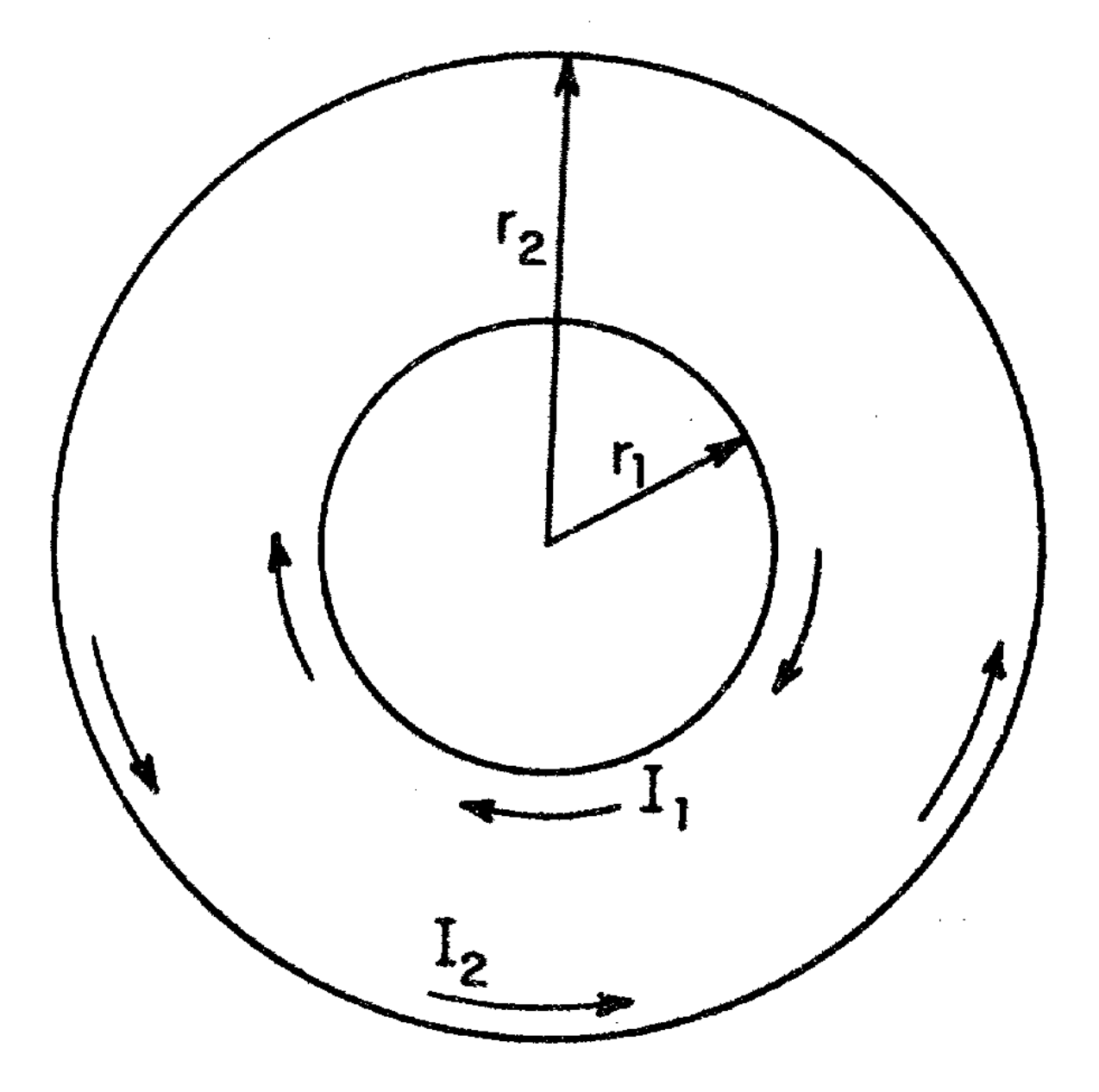}
        \caption{Annulus geometry.}
        \label{fig:annulusGeometry}
    \end{subfigure}
    
    \begin{subfigure}[t]{0.5\textwidth}
        \centering
        \includegraphics[width=4cm,height=3.5cm]{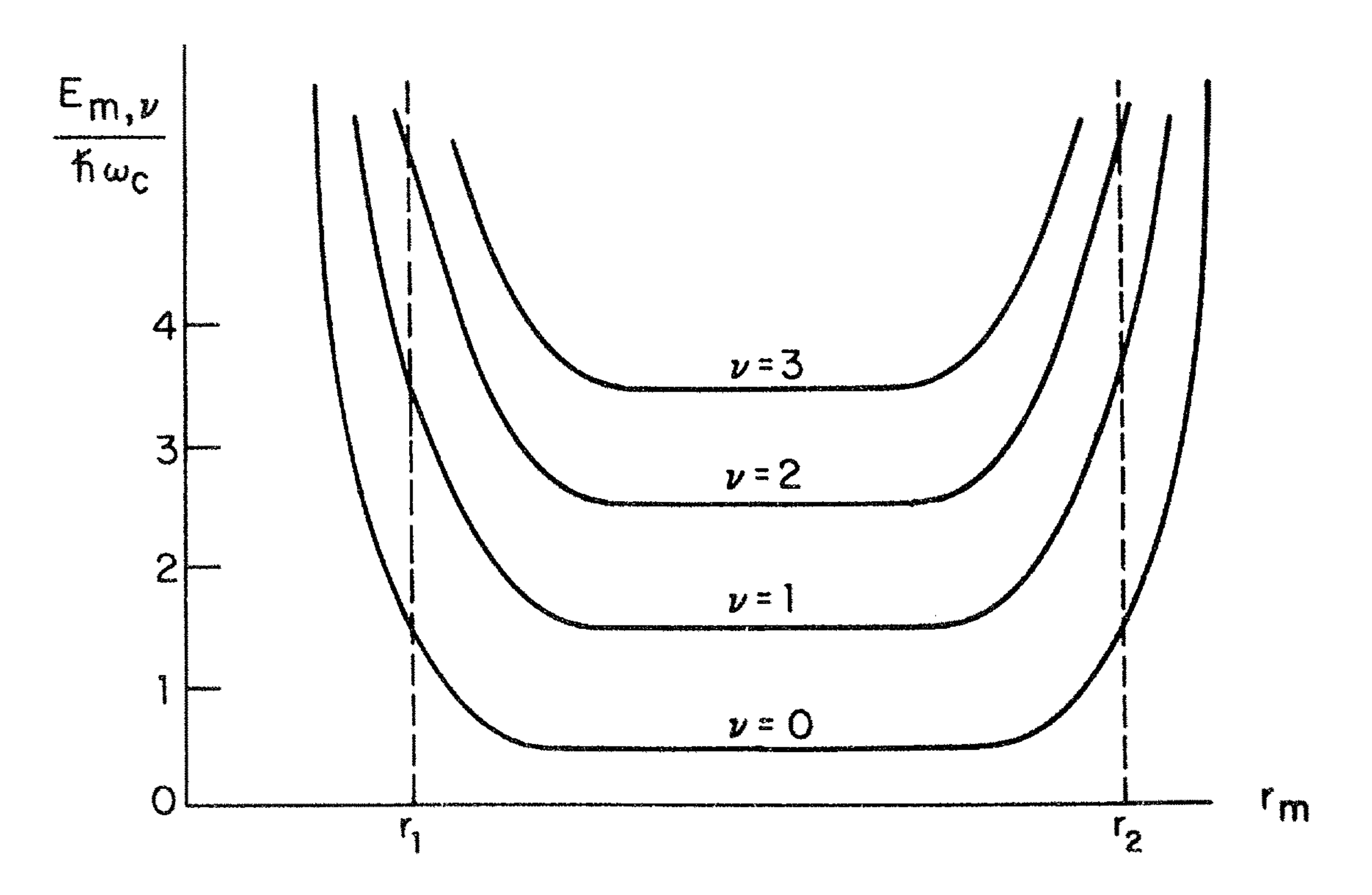}
        \caption{Dispersion relation for different energy levels.}
        \label{fig:energyPotentialCorbino}
    \end{subfigure}
    ~
    \begin{subfigure}[t]{0.5\textwidth}
        \centering
        \includegraphics[width=4cm,height=3.5cm]{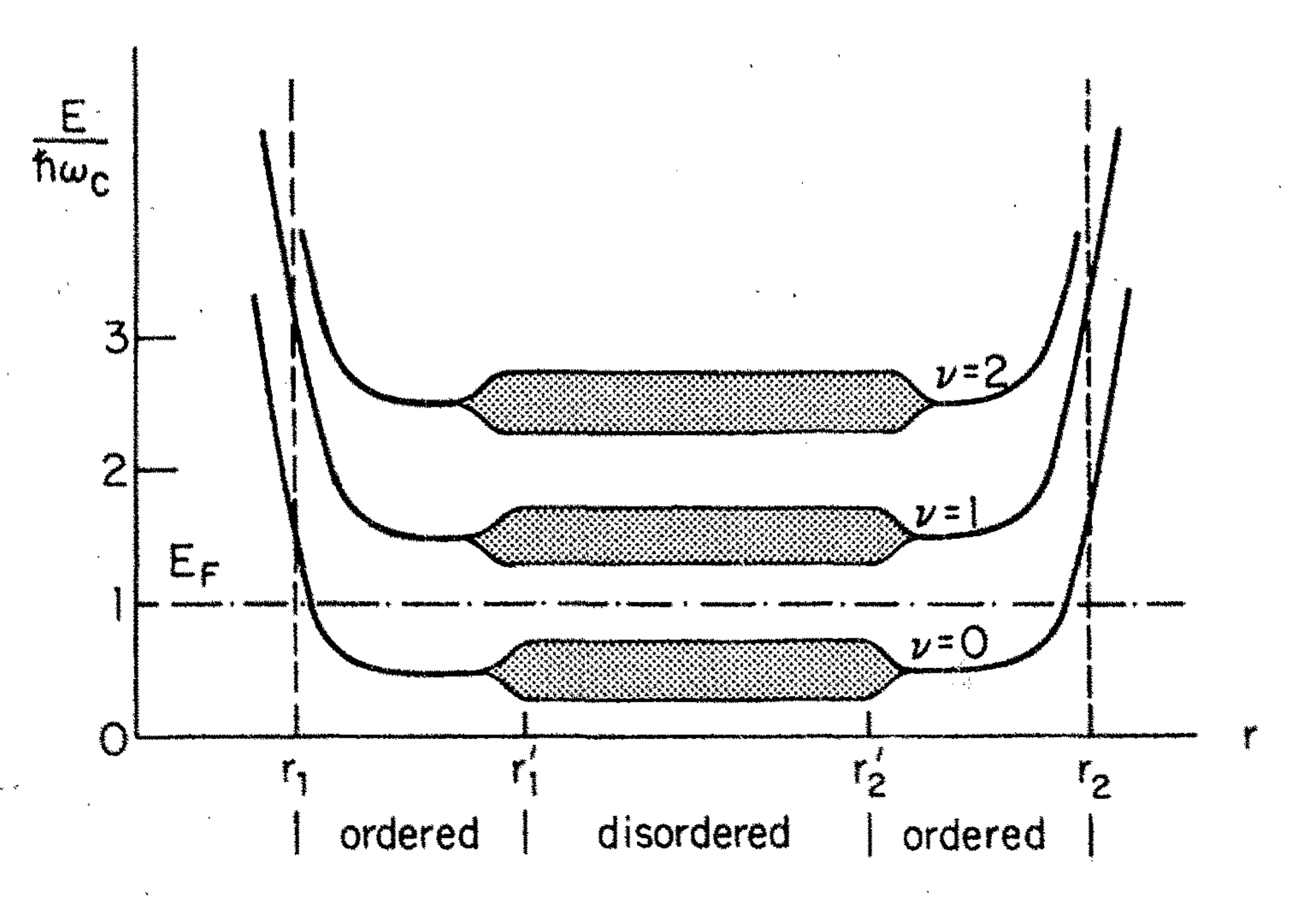}
        \caption{Dispersion relation with disorder.}
        \label{fig:energyPotentialCorbinoDisorder}
    \end{subfigure}
    \caption{Images taken from \cite{Halperin_1982}.}
\end{figure}
This effect can also be seen from a quantum point of view. Based on \cite{Halperin_1982}, let us take an annulus geometry as shown in Fig.~\ref{fig:annulusGeometry}. The potential, $V(x)$, that describes the boundary conditions can be thought as constant in the bulk and rising near $r_1$ and $r_2$. The resulting energy levels are similar to the Landau levels in the bulk but start rising when they get near the edges, $r_1$ and $r_2$. This is shown in Fig.~\ref{fig:energyPotentialCorbino}. So, if the Fermi Energy is set between $\nu=0$ and $\nu=1$, it will intersect the energy levels near the frontier. This means that near the edges the material behaves like a metal and in the bulk, it behaves like an insulator.

\

Until now, we have just explained why the Hall resistivity has quantized values given by (\ref{eq:hallResistivity}). But it is also important to understand why the resistivity stays fixed in plateaux for certain intervals of the magnetic field. Here is where the impurities in the system come into play. 

In fact, with the introduction of disorder, the energy bands change from those in Fig.~\ref{fig:energyPotentialCorbino} to Fig.~\ref{fig:energyPotentialCorbinoDisorder}. In \cite{Halperin_1982}, Halperin reaches a contradiction assuming there are only localized states for energy bands between $r_1'$ and $r_2'$. From that, he concludes that there must be at least one extended state. As pointed in \cite{Lect_QHE}, it turns out that the localized states are located in the edge of the energy band and the extended states populate its centre. As localized states keep the electrons in small regions, they do not allow electrons to go from one edge to another. This means they cannot transport charge. So, the only states that interfere with conductivity are the extended states, which allow electrons to move throughout the system. 

Following the analysis that is done in \cite{Lect_QHE}, suppose that the electron density is fixed to be $n$ and that all the extended states are filled. Looking at expression (\ref{eq:numbStates}), we easily conclude that if $B$ decreases the number of electrons per Landau level, $N$, decreases. Now, recalling that the area $A$ is fixed, we have that the number filled Landau levels $\nu$ increases (Eq.~\ref{eq:densityStates}). Since the Fermi Energy gives the energy value of the most energetic electron in the system we deduce that it must increase. As there are still some localized states to be filled, the conductivity is not affected and stays constant in a plateau. So, if the disorder is increased until a certain amount, the number of localized states available increases and the plateaux would be clearer. 

\subsubsection{Chern number}

In this section, let us see the role of Topology and Differential Geometry in the QHE and deduce a formula for the Hall conductance. We will see that it is directly related with a \textit{topological invariant} called \textit{Chern number}. This property is a strong theoretical supporter of the robustness of the edge states under the inclusion of impurities.

In 1981 \cite{Laughlin_1981}, R. B. Laughlin argued that by imposing a system a change of flux equivalent to a flux quantum, there would be a transfer of $n$ electrons from one edge to another. A few years later, M. V. Berry \cite{Berry_1984} discovered that a quantum system would acquire a \textit{geometrical phase} factor $e^{i\gamma}$ by slowly varying a parameter, $\Phi$, on which the Hamiltonian, $H(\Phi)$, depends. This phase difference is called \textit{Berry phase}. Berry also found an explicit formula for this phase difference based on the following expression:
\begin{equation}
k_i = -i\ev{\pdv{\Phi_i}}{\psi_n},
\end{equation}
where $\Phi$ is taken to have two components, $\Phi_x$ and $\Phi_y$. This is called the \textit{Berry connection}. As it is commonly done in Differential Geometry, a connection defines a curvature by the following expression:
\begin{equation}
\label{eq:curvature}
K_{xy} = \pdv{k_x}{\Phi_y} - \pdv{k_y}{\Phi_x} =  -i\Bigg[ \pdv{\Phi_y}\ip{\psi_n}{\pdv{\psi_n}{\Phi_x}} - \pdv{\Phi_x}\ip{\psi_n}{\pdv{\psi_n}{\Phi_y}}  \Bigg].
\end{equation}
In 1985, \cite{Thouless_1985} Qian Niu, D. J. Thouless, and Yong-Shi Wu generalized an earlier formula of the Hall conductivity which relies on the Kubo formula \cite{Thouless_1982}. They concluded that it can be written as:
\begin{equation}
\label{eq:curvatureConductivity}
\sigma_{xy} =  -i\hbar\Bigg[ \pdv{\Phi_y}\ip{\psi_n}{\pdv{\psi_n}{\Phi_x}} - \pdv{\Phi_x}\ip{\psi_n}{\pdv{\psi_n}{\Phi_y}}  \Bigg],
\end{equation}
which is closely related with the curvature of the space of parameters in (\ref{eq:curvature}). $\hbar$ denotes the reduced Planck's constant.
Define $\theta_i = \frac{2\pi\Phi_i}{\Phi_0}$ where $\Phi_i \in [0,2\pi)$ and $\Phi_0$ the flux quantum. Using $\theta_i$ in (\ref{eq:curvature}) instead of $\Phi_i$, we get
\begin{equation}
\label{eq:cConductivity}
\sigma_{xy} =  -\frac{e^2}{\hbar}K_{xy},
\end{equation}
where $e$ is the electron charge. As it is done in \cite{Fradkin_2013_Book}, averaging the Hall conductance
\begin{equation}
\label{eq:cConductivityGB}
\langle\sigma_{xy}\rangle =  -\frac{e^2}{2\pi\hbar}\iint_{0}^{2\pi}\frac{K_{xy}}{2\pi}\ d^2\theta = -\frac{e^2}{2\pi\hbar}C.
\end{equation}
By the Gauss-Bonnet Theorem it is known that $C\in\ZZ$. Thus we arrive to the Integer Quantum Hall Effect again. $C$ is called the \textit{first Chern number}, or shortly, \textit{Chern number}. This formula points out the deep relation between the Hall conductance and the topology of the space of parameters. We can see that there is no relation between its average and the amount of disorder in the system or the type of material used in experiments.

\

Until now we have explained the theory behind the QHE in the continuum case and the role of edge states and disorder in the robustness of this effect. In the next section, we see how to discretize the system by using lattice models and how to consider magnetic fields in those cases. This sets the playground to mimic topological insulators and to explore some of its properties.

\subsection{Lattice model}

Our goal is to understand some finite size effects in the energy band of the system. We will make use of lattice models under magnetic fields to perform some numerical simulations and, later, to deduce analytic expressions describing some of the effects. Now, let us start to describe a lattice system in the absence of magnetism. As we will see in section \ref{expProc}, these systems are experimentally described by fixed cold atoms in a lattice. Due to this feature, it is sensible to describe these systems by the Linear Combination of Atomic Orbitals (LCAO) procedure \cite{Simon_2013_Book}.

\subsubsection{Hamiltonian}

Let us see how to obtain an Hamiltonian that describes systems in lattices. Using Bracket notation, the SE is given by:

\begin{equation}
\label{eq:SEBKNotation}
    H\ket{\psi} = E\ket{\psi}.
\end{equation}
In the continuous case the Hamiltonian of a system with a potential $V$ can be written as follows:
\begin{equation}
\label{eq:HamCont}
    H = \frac{1}{2}p^2 + V(\bld{x}).
\end{equation}
As we are interested in the Tight Binding approximation \cite{Simon_2013_Book} the Hamiltonian must be written in its discrete formula. Here we consider that electrons are just allowed to hop to its nearest-neighbourhood sites. Following the LCAO procedure, we only consider one orbit per unit site. Other derivations exist in the literature but the present one was found to be the clearest through discussions with the supervisor. Considering a unit cell with more than one site, the orbits are labelled according to the position of the unit cell ($\bld{R}$) and its position inside the unit cell ($m$). So, let $\ket{\bld{R},m}$ denotes localized orbitals in unit cell $\bld{R}$ centred on $m$. Furthermore, let us consider the projection operator:
\begin{equation}
\label{eq:DiscOp}
    Q = \sum_{\bld{R},m} 
    \dyad{\bld{R},m}{\bld{R},m}.
\end{equation}
Note that the more elements we consider in (\ref{eq:DiscOp}), the more it resembles the identity operator and the more complete description of the full system we get. In order to solve the SE (\ref{eq:SEBKNotation}) in the space generated by $\{ \ket{\bld{R},m} \}_{\bld{R},m}$, we have to restrict the Hamiltonian in (\ref{eq:HamCont}) as follows:

\begin{eqnarray*}
    H_{TB} &=& Q H Q\\
    &=& \sum_{\bld{R},\bld{R}',m,m'} \ket{\bld{R},m} \mel{\bR,m}{H}{\bR,m'} \bra{\bR,m'},
\end{eqnarray*} 
where $t = \mel{\bR,m}{H}{\bR,m'}$ are called the \textit{hopping terms}. They describe the likeliness of a particle to jump from a particular site to another. Next, we consider a give a simple example of a square lattice describing a Torus.

\subsubsection{Energy dispersion on a Torus}

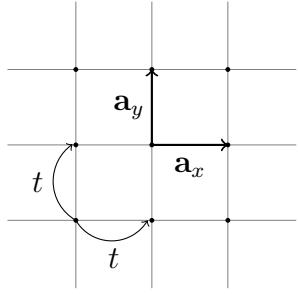
\begin{figure}[h]
\begin{center}
\begin{tikzpicture}

\draw[step=1cm,gray,very thin] (-0.9,-0.9) grid (2.9,2.9);

\draw[fill] (0,0) circle [radius=0.025];
\draw[fill] (1,0) circle [radius=0.025];
\draw[fill] (2,0) circle [radius=0.025];
\draw[fill] (0,1) circle [radius=0.025];
\draw[fill] (1,1) circle [radius=0.025];
\draw[fill] (2,1) circle [radius=0.025];
\draw[fill] (0,2) circle [radius=0.025];
\draw[fill] (1,2) circle [radius=0.025];
\draw[fill] (2,2) circle [radius=0.025];

\node at (1.5,0.7) {$\bld{a}_x$};
\node at (0.7,1.5) {$\bld{a}_y$};
\node at (0.5,-0.5) {$t$};
\node at (-0.5,0.5) {$t$};
\draw [->, thick] (1,1) -- (2,1);
\draw [->, thick] (1,1) -- (1,2);
\draw [->] (0,0) arc (-150:-30:15.5pt);
\draw [->] (0,0) arc (240:125:17pt);

\end{tikzpicture}
\end{center}
\caption{Particle in a lattice.}
\label{fig:partLattice}
\end{figure}

Let us consider the case where a particle moves in a square lattice (Fig.~\ref{fig:partLattice}). Define the distance between the sites to be $a$ in both directions. As there is no magnetic field the unit cell is composed by just one site. The Hamiltonian is:
\begin{equation}
\label{eq:Ham1}
    H_{TB} = - t \sumRinG \big(\dyad{\bld{R} + \bld{a}_y}{\bld{R}} + \dyad{\bld{R} + \bld{a}_x}{\bld{R}} + h.c\big).
\end{equation}

As we want to consider a lattice describing a Torus, in this section, we take a finite system with periodic boundary conditions and with $N$ elements. We have that the system is translation invariant meaning we can use the Bloch's Theorem. Using the normal procedure \cite{Simon_2013_Book} and applying the Fourier transform to go to momentum space the wavefunction can be described by:
\begin{equation}
\label{eq:ansatz1D}
    \ket{\psi_\bld{k}} = \frac{1}{\sqrt{N}}\sumRinG \epikdr \ket{\bld{R}}.
\end{equation}
Note that in this case, the momentum space parameters, $k_x$ and $k_y$, take values in the interval $[-\frac{\pi}{a},\frac{\pi}{a}]$, called \textit{Brillouin Zone}. Since this space is periodic in both directions we can, in fact, identify it with a two dimensional Torus. The states (\ref{eq:ansatz1D}) are eigenstates of $H_{TB}$,
\begin{equation*}
    H_{TB}\ket{\psi_\bld{k}} = E(\bld{k})\ket{\psi_\bld{k}},
\end{equation*}
as it can be verified:
\begin{align*}
    H_{TB} \ket{\bld{R}'} &= - t(\ket{\bld{R}'+\bld{a}_y} + \ket{\bld{R}'-\bld{a}_y} + \ket{\bld{R}'+\bld{a}_x} + \ket{\bld{R}'-\bld{a}_x}) \\
    \begin{split}\mel{\bld{R}}{H_{TB}}{\psik}  &= \bra{\bld{R}} \frac{1}{\sqrt{N}}\sum_{\bld{R}'} e^{i \bld{k}\cdot\bld{R}'} (-t) \big( \ket{\bld{R}'+\bld{a}_y} + \\ & \qquad \qquad \qquad \qquad \ket{\bld{R}'-\bld{a}_y} + \ket{\bld{R}'+\bld{a}_x} + \ket{\bld{R}'-\bld{a}_x}   \big)
    \end{split}\\
    &= - 2 \frac{t e^{i \bld{k}\cdot\bld{R}}}{\sqrt{N}}\big( \cos{(\bld{k}\cdot\bld{a}_x)} + \cos{(\bld{k}\cdot\bld{a}_y)} \big) \\
    \ip{\bld{R}}{\psik} &= \frac{\epikdr}{\sqrt{N}}.
\end{align*}

\begin{figure}[!h]
\centering
\includegraphics[width=9.5cm, height=6.2cm]{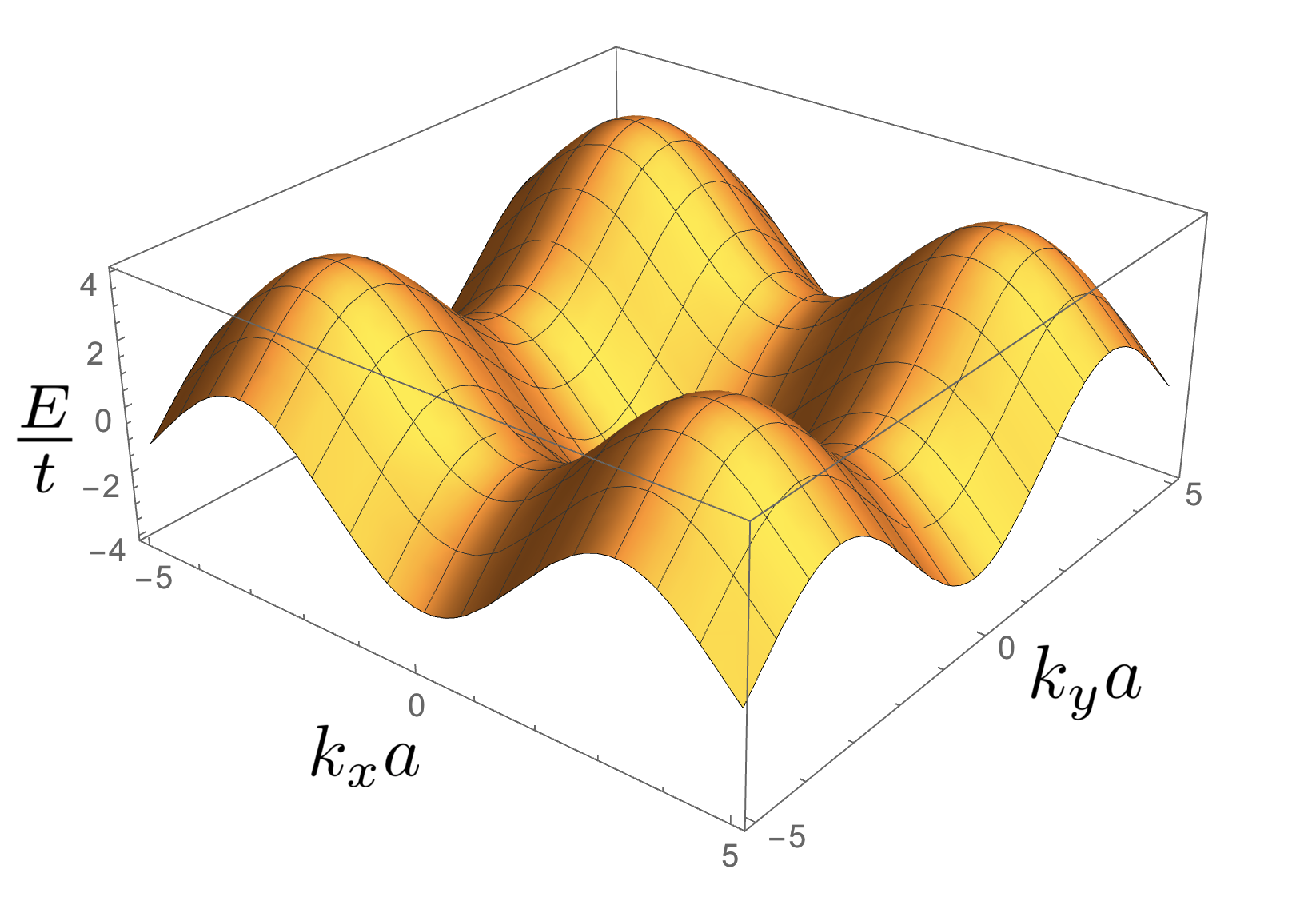}
\caption{Energy dispersion on a Torus.}
\label{fig:DispTorus}
\end{figure}
Thus we have that,
\begin{equation}
\label{eq:energy1}
    E(\bld{k}) = -2 t (\cos{(\bld{k}\cdot\bld{a}_x)} + \cos{(\bld{k}\cdot\bld{a}_y)}),
\end{equation}
which is shown in Fig.~\ref{fig:DispTorus}.

\subsubsection{Magnetism in a lattice}

In order to mimic the behaviour of topological insulators, we need to integrate a magnetic field in our lattice. Since we have to change the Hamiltonian (\ref{eq:Ham1}) let us gain some intuition about what happens when an electron goes around a square in the lattice.

\begin{figure}[h]
\begin{center}
\begin{tikzpicture}

\draw[step=2cm,gray,very thin] (-0.9,-0.9) grid (2.9,2.9);

\draw[fill] (0,0) circle [radius=0.05];
\draw[fill] (2,0) circle [radius=0.05];
\draw[fill] (0,2) circle [radius=0.05];
\draw[fill] (2,2) circle [radius=0.05];

\draw [<-] (1.4,1.5) arc (50:-230:17pt);

\node at (1,1) {$\Phi$};

\node at (1,-0.3) {$A_{m,n}^x$};
\node at (2.7,1) {$A_{m+1,n}^y$};
\node at (1,2.3) {$A_{m,n+1}^x$};
\node at (-0.5,1) {$A_{m,n}^y$};

\end{tikzpicture}
\end{center}
\caption{Gauge field.}
\label{fig:flux}
\end{figure}
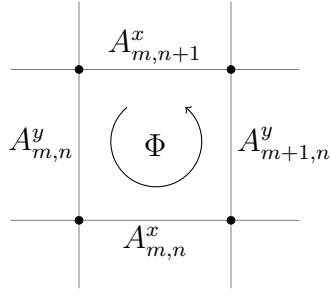

As it was first found by Y. Aharonov and D. Bohm \cite{Aharonov_1959}, when an electron with charge $e$ performs a circular motion around a magnetic flux, $\Phi$, it picks up an Aharonov-Bohm phase $e^{ie\Phi/\hbar}$. This is a particular case of a Berry phase where the connection, $k_i$, is proportional to the electromagnetic potential $\bld{A}$ \cite{Lect_QHE},
\begin{equation}
\label{eq:relationAharonovPotential}
k_i = \frac{e}{\hbar}A_i.
\end{equation}
This comes from the fact that
\begin{equation}
\label{eq:exp1}
e^{-ie\Phi/\hbar} = \exp (-i\frac{e}{\hbar}\oint_C \bld{A}\cdot d\bld{x}).
\end{equation}
So, in order to introduce the magnetic field we have to choose a particular gauge field $\bld{A}$ around the lattice satisfying the following:
\begin{equation}
\label{eq:fluxRelLattice}
\Phi = a(A_{m,n}^x + A_{m+1,n}^y - A_{m,n+1}^x - A_{m,n}^y),
\end{equation}
where $a$ is the space between sites in the lattice. This can be visualized in Fig.~\ref{fig:flux}.

Let us take $\ket{\psi_0}$ to be a solution of the SE (\ref{eq:SEBKNotation}) with the Hamiltonian (\ref{eq:HamCont}) describing a system without magnetism. As we pointed in (\ref{eq:landauHam}), the Hamiltonian in the presence of a magnetic field, $\bld{B}$, and potential $\bld{V}$ is given by
\begin{equation}
\label{eq:hamMagPot}
    H = \frac{(\bld{p}+e\bld{A})^2}{2m} + \bld{V}.
\end{equation}
After the work of R. Peierls \cite{Peierls_1933}, in order to obtain a solution of the SE with the Hamiltonian given by (\ref{eq:hamMagPot}) we just need to adjust $\ket{\psi_0}$ to 
\begin{equation}
\label{eq:changePeierls}
\ket{\psi} = \exp (-i\frac{e}{\hbar}\int \bld{A}\cdot d\bld{x}) \ket{\psi_0}.
\end{equation}
Then, in the tight-binding approximation, the hopping terms $t$ in the Hamiltonian (\ref{eq:Ham1}) are substituted by $t\exp(-i A_{\bld{R}}^j)$, where $j=x,y$. Writing the Hamiltonian in the second quantized notation we have
\begin{equation}
\label{eq:HamArt2ndQuant2}
   H = -t \sum_\bld{R} \big( e^{-i A_{\bld{R}}^x} \bld{c}^\dagger_{\bld{R}+\bld{a}_x} \bld{c}_{\bld{R}} +  e^{-i A_{\bld{R}}^y} \bld{c}^\dagger_{\bld{R}+\bld{a}_y} \bld{c}_{\bld{R}} + h.c. \big),
\end{equation}
where $\bld{c}^\dagger_{\bld{R}}$ and $\bld{c}_{\bld{R}}$ are the creation and annihilation operators in the site $\bld{R}$ and with the following anti-commutation relations:
\begin{eqnarray*}
&\{\bld{c}_{\bld{R}}, \bld{c}^\dagger_{\bld{R'}}\} = \bld{c}_{\bld{R}} \bld{c}^\dagger_{\bld{R'}} + \bld{c}^\dagger_{\bld{R'}} \bld{c}_{\bld{R}} = \delta_{\bld{R} \bld{R'}}\\
&\{\bld{c}_{\bld{R}}, \bld{c}_{\bld{R+}}\} = \{\bld{c}^\dagger_{\bld{R}}, \bld{c}^\dagger_{\bld{R'}}\} = 0.\\
\end{eqnarray*}

\newpage

\section{Hofstadter model} \label{hofsModel}

\begin{figure}[h]

    \begin{subfigure}[t]{0.5\textwidth}
\begin{center}
\begin{tikzpicture}

\draw[step=1cm,gray,very thin] (-0.9,-0.9) grid (4.9,3.9);

\draw[fill] (0,0) circle [radius=0.025];
\draw[fill] (1,0) circle [radius=0.025];
\draw[fill] (2,0) circle [radius=0.025];
\draw[fill] (3,0) circle [radius=0.025];
\draw[fill] (4,0) circle [radius=0.025];
\draw[fill] (0,1) circle [radius=0.025];
\draw[fill] (1,1) circle [radius=0.025];
\draw[fill] (2,1) circle [radius=0.025];
\draw[fill] (3,1) circle [radius=0.025];
\draw[fill] (4,1) circle [radius=0.025];
\draw[fill] (0,2) circle [radius=0.025];
\draw[fill] (1,2) circle [radius=0.025];
\draw[fill] (2,2) circle [radius=0.025];
\draw[fill] (3,2) circle [radius=0.025];
\draw[fill] (4,2) circle [radius=0.025];
\draw[fill] (0,3) circle [radius=0.025];
\draw[fill] (1,3) circle [radius=0.025];
\draw[fill] (2,3) circle [radius=0.025];
\draw[fill] (3,3) circle [radius=0.025];
\draw[fill] (4,3) circle [radius=0.025];

\node [above right] at (0.3,-0.8) {$t_x$};
\node [above right] at (0.3,0.2) {$t_x e^{i\phi}$};
\node [above right] at (0.3, 1.2) {$t_x e^{i 2\phi}$};
\node at (-0.6, 0.5) {$t_y$};
\node at (2.5,1.5) {$\bld{a}_y$};
\node at (2.5,-0.3) {$\bld{a}_x$};
\draw [->, thick] (2,0) -- (3,0);
\draw [->, thick] (2,0) -- (2,3);
\draw [->] (0,0) arc (-150:-30:15.5pt);
\draw [->] (0,1) arc (-150:-30:15.5pt);
\draw [->] (0,2) arc (-150:-30:15.5pt);
\draw [->] (0,0) arc (240:125:17pt);
\draw [->] (0,1) arc (240:125:17pt);

\draw [->, thick] (-1.2,-1) -- (-1.2,0);
\node at (-1.4,-0.5) {$y$};
\draw [->, thick] (-1.2,-1) -- (-0.2,-1);
\node at (-0.65,-1.25) {$x$};

\node at (-0.7, 2.5) {Unit cell};
\draw (-0.35,-0.35) rectangle (0.25,2.25);

\end{tikzpicture}
\end{center}
\caption{Hofstadter model for $M=3$.}
\label{fig:discreteModelM3}
    \end{subfigure}%
~ 
    \begin{subfigure}[t]{0.5\textwidth}
        \centering
        \includegraphics[height=1.8in]{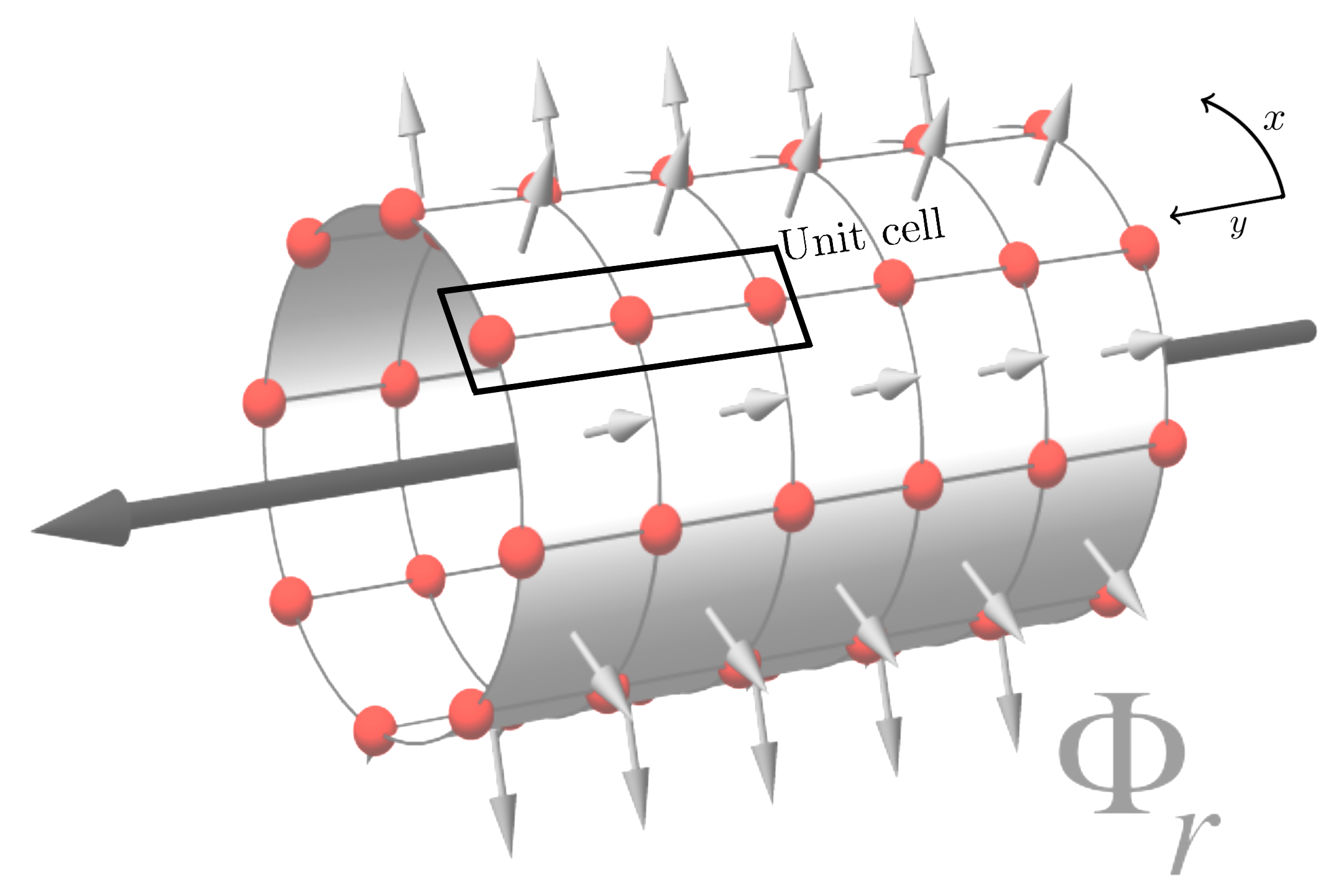}
        \caption{Cylinder geometry, image taken from \cite{Lacki_2016} and edited.}
        \label{fig:cylinder}
    \end{subfigure}%
\caption{Hofstadter model.}
\label{fig:HofCylinder}
\end{figure}
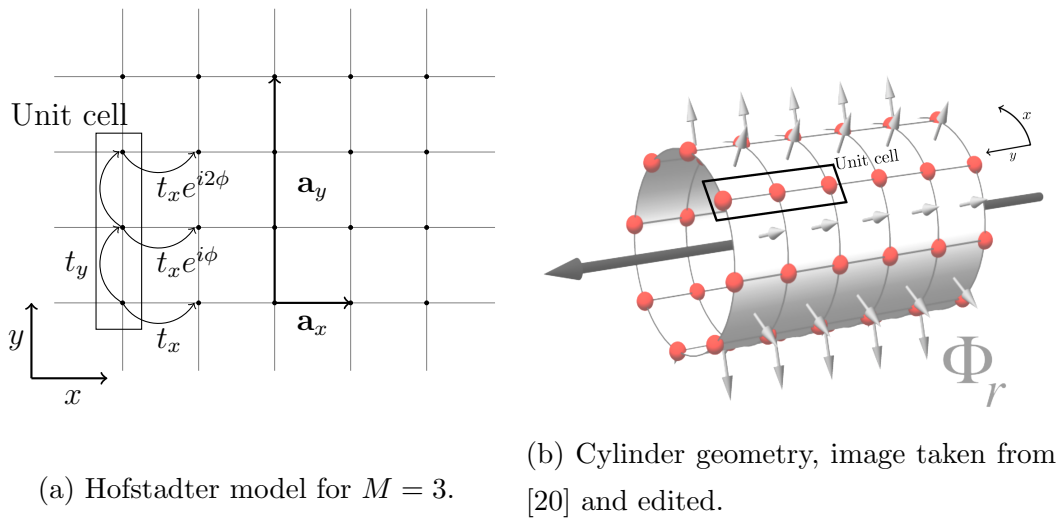

We have previously seen that a magnetic field can be defined in a lattice by the phase acquired by an electron travelling through it. Let us consider the system described in Fig.~\ref{fig:discreteModelM3} with $M$ sites per unit cell and periodic conditions along the $x$-direction. We will first deduce the Hamiltonian describing a periodic system in the $y$-direction and with closed boundary conditions (Torus). After that, we consider a bounded system with open boundary conditions. This can be thought as a cylinder, Fig.~\ref{fig:cylinder}. Based on some experimental work \cite{DM_Stuhl_2015}, we will be interested in analyzing the band structure of a model with just $M=3$.

\

In this case $\phi = \frac{2\pi}{M}$ for $M\in\NN$ and the distance between each node is given by $a$, $\bld{a}_y=(0,Ma)$ and $\bld{a}_x=(a,0)$. Thus, the Hamiltonian of the system with open boundary conditions may be written as:
\begin{multline}
\label{eq:HamArt2ndQuant}
   H = -t_y \sum_\bld{R} \Bigg[ \Big( \sum_{j=1}^{M-1} \bld{c}^\dagger_{\bld{R},j+1} \bld{c}_{\bld{R},j} + h.c. \Big) + \bld{c}^\dagger_{\bld{R}+\bld{a}_y,1} \bld{c}_{\bld{R},M} + h.c. \Bigg]\\
   -t_x \sum_{\bld{R}} \sum_{j=1}^M \big( \bld{c}^\dagger_{\bld{R}+\bld{a}_x,j} \bld{c}_{\bld{R},j}\ \omega^{j-1} + h.c. \big),
\end{multline}
where we sum over all unit cells $\bld{R}\in \bld{G}$ and $\omega=e^{i\phi}$. To go to momentum space we need to perform a Fourier transform:
\begin{eqnarray*}
\bld{c}_{\bld{R},j} &=& \frac{1}{\sqrt{n}}\sum_{\bld{k}} \bld{c}_{\bld{k},j}\ e^{i \bld{k}\cdot\bld{R}},  \\
\bld{c}_{\bld{k},j} &=& \frac{1}{\sqrt{n}}\sum_{\bld{R}} \bld{c}_{\bld{R},j}\ e^{-i \bld{k}\cdot\bld{R}}, 
\end{eqnarray*}
where $n$ is the total number of unit cells. Substituting these two expressions into the Hamiltonian (\ref{eq:HamArt2ndQuant}) and taking into account that $\sum_{\bld{R}} e^{i (\bld{k}-\bld{k}')\cdot \bld{R}} = n\ \delta_{\bld{k}\bld{k}'}$, we have: 
\begin{multline}
\label{eq:HamArt2ndQuantFT}
   H = \sum_{\bld{k}} \Bigg[  -t_y \Big( \sum_{j=1}^{M-1} \bld{c}^\dagger_{\bld{k},j} \bld{c}_{\bld{k},j+1} + h.c. \Big) +  \bld{c}^\dagger_{\bld{k},M} \bld{c}_{\bld{k},1}\ e^{-i \bld{k}\cdot\bld{a}_y} + h.c. \\
   -t_x  \sum_{j=1}^M\big( \bld{c}^\dagger_{\bld{k},j} \bld{c}_{\bld{k},j}\ e^{-i\bld{k}\cdot\bld{a}_x} \omega^{j-1} + h.c. \big)  \Bigg],
\end{multline}
which is equivalent to,

\begin{eqnarray*}
H = \sum_{\bld{k}} \mqty(\bld{c}_{\bld{k},1}^\dagger & ... & \bld{c}_{\bld{k},M}^\dagger)\ \mathcal{H}\  \mqty(\bld{c}_{\bld{k},1} \\ ... \\ \bld{c}_{\bld{k},M}).\\
\end{eqnarray*}
So, we can write $\mathcal{H}$ as,

\begin{eqnarray*}
\mathcal{H} = \left(\begin{smallmatrix}
-2 t_x \cos\big(\bld{k}\cdot\bld{a}_x\big) & -t_y & 0 & ... & 0 & -t_y e^{i\bld{k}\cdot\bld{a}_y} \\
-t_y & -2 t_x \cos\big(\bld{k}\cdot\bld{a}_x + \phi\big) & -t_y & \ddots &  & 0\\
0 & -t_y &  &  &  & \vdots\\
\vdots & & & \ddots & & 0 \\
0 & & & &  & -t_y\\
-t_y e^{-i \bld{k}\cdot\bld{a}_y} & 0 & \dots & 0 & -t_y & -2 t_x \cos\big(\bld{k}\cdot\bld{a}_x + (M-1)\phi\big) 
\end{smallmatrix}\right)
\end{eqnarray*}

\subsection{Bounded systems} \label{bounded_systems}

\begin{figure}[h]
\begin{center}
\begin{tikzpicture}

\draw[step=0.5cm,gray,very thin] (-0.2,0) grid (3.7,2.5);

\draw[fill] (0,0) circle [radius=0.025];
\draw[fill] (0.5,0) circle [radius=0.025];
\draw[fill] (1,0) circle [radius=0.025];
\draw[fill] (1.5,0) circle [radius=0.025];
\draw[fill] (2,0) circle [radius=0.025];
\draw[fill] (2.5,0) circle [radius=0.025];
\draw[fill] (3,0) circle [radius=0.025];
\draw[fill] (3.5,0) circle [radius=0.025];
\draw[fill] (0,0.5) circle [radius=0.025];
\draw[fill] (0.5,0.5) circle [radius=0.025];
\draw[fill] (1,0.5) circle [radius=0.025];
\draw[fill] (1.5,0.5) circle [radius=0.025];
\draw[fill] (2,0.5) circle [radius=0.025];
\draw[fill] (2.5,0.5) circle [radius=0.025];
\draw[fill] (3,0.5) circle [radius=0.025];
\draw[fill] (3.5,0.5) circle [radius=0.025];
\draw[fill] (0,1) circle [radius=0.025];
\draw[fill] (0.5,1) circle [radius=0.025];
\draw[fill] (1,1) circle [radius=0.025];
\draw[fill] (1.5,1) circle [radius=0.025];
\draw[fill] (2,1) circle [radius=0.025];
\draw[fill] (2.5,1) circle [radius=0.025];
\draw[fill] (3,1) circle [radius=0.025];
\draw[fill] (3.5,1) circle [radius=0.025];
\draw[fill] (0,1.5) circle [radius=0.025];
\draw[fill] (0.5,1.5) circle [radius=0.025];
\draw[fill] (1,1.5) circle [radius=0.025];
\draw[fill] (1.5,1.5) circle [radius=0.025];
\draw[fill] (2,1.5) circle [radius=0.025];
\draw[fill] (2.5,1.5) circle [radius=0.025];
\draw[fill] (3,1.5) circle [radius=0.025];
\draw[fill] (3.5,1.5) circle [radius=0.025];
\draw[fill] (0,2) circle [radius=0.025];
\draw[fill] (0.5,2) circle [radius=0.025];
\draw[fill] (1,2) circle [radius=0.025];
\draw[fill] (1.5,2) circle [radius=0.025];
\draw[fill] (2,2) circle [radius=0.025];
\draw[fill] (2.5,2) circle [radius=0.025];
\draw[fill] (3,2) circle [radius=0.025];
\draw[fill] (3.5,2) circle [radius=0.025];
\draw[fill] (0,2.5) circle [radius=0.025];
\draw[fill] (0.5,2.5) circle [radius=0.025];
\draw[fill] (1,2.5) circle [radius=0.025];
\draw[fill] (1.5,2.5) circle [radius=0.025];
\draw[fill] (2,2.5) circle [radius=0.025];
\draw[fill] (2.5,2.5) circle [radius=0.025];
\draw[fill] (3,2.5) circle [radius=0.025];
\draw[fill] (3.5,2.5) circle [radius=0.025];

\node at (-0.5, 3) {Site ($m$)};
\node at (-0.5, 2.5) {$1$};
\node at (-0.5, 2) {$2$};
\node at (-0.5, 1.5) {$3$};
\node at (-0.5, 1) {$4$};
\node at (-0.5, 0.5) {$5$};
\node at (-0.5, 0) {$6$};


\draw [->] (0.5,0) arc (150:30:7.8pt);
\draw [->] (0.5,0.5) arc (150:30:7.8pt);
\draw [->] (0.5,1) arc (150:30:7.8pt);
\draw [->] (0.5,1.5) arc (-150:-30:7.8pt);
\draw [->] (0.5,2) arc (-150:-30:7.8pt);
\draw [->] (0.5,2.5) arc (-150:-30:7.8pt);

\draw [->] (0.5,0) arc (230:125:8.1pt);
\draw [->] (0.5,0.5) arc (230:125:8.1pt);
\draw [->] (0.5,1) arc (230:125:8.1pt);
\draw [->] (0.5,1.5) arc (230:125:8.1pt);
\draw [->] (0.5,2) arc (230:125:8.1pt);

\draw (0.35,-0.1) rectangle (0.65,1.1);
\draw (0.35,1.4) rectangle (0.65,2.6);

\draw (0.9,-0.1) rectangle (1.1,1.1);
\draw (0.9,1.4) rectangle (1.1,2.6);

\draw (1.4,-0.1) rectangle (1.6,1.1);
\draw (1.4,1.4) rectangle (1.6,2.6);

\draw (1.9,-0.1) rectangle (2.1,1.1);
\draw (1.9,1.4) rectangle (2.1,2.6);

\draw (2.4,-0.1) rectangle (2.6,1.1);
\draw (2.4,1.4) rectangle (2.6,2.6);

\draw (2.9,-0.1) rectangle (3.1,1.1);
\draw (2.9,1.4) rectangle (3.1,2.6);

\draw [line width=1pt] (-0.2,0)-- (3.7,0);
\draw [line width=1pt] (-0.2,2.5)-- (3.7,2.5);

\end{tikzpicture}
\end{center}
\caption{Bounded system for $M=3$ and $N=2$.}
\label{fig:boundedSystem}
\end{figure}
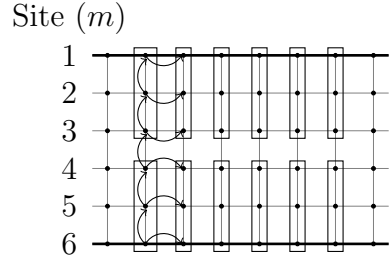

We are particularly interested in a system with two boundaries. This means that a particle cannot hop through the boundary and thus the terms $-t_y e^{\pm i\bld{k}\cdot\bld{a}_y}$ in $\mathcal{H}$ are set to zero. Furthermore, as we aim to vary the length of the system in the $y$-direction without changing the length of the unit cell, we have to take $\mathcal{H}$ to be:

\begin{eqnarray}
\label{eq:hamBoundary}
\mathcal{H} = \mqty(A & B & 0 & ...  & 0 \\
B^\dagger & A & B &  & \vdots\\
0 & B^\dagger & \ddots &  & 0\\
\vdots & & & & B\\
0 & \dots & 0 & B^\dagger & A ),
\end{eqnarray}
where $A$ and $B$ are $M$-by-$M$ matrices given by

\begin{eqnarray*}
A = \mqty(-2 t_x \cos(\bld{k}\cdot\bld{a}_x) & -t_y & 0 & ... & 0 \\
-t_y & \ddots &  &  & \vdots\\
0 & &  & & 0\\
\vdots & & & \ddots & -t_y\\
0 &  \dots & 0 & -t_y & -2 t_x \cos(\bld{k}\cdot\bld{a}_x + (M-1)\phi) ),
\end{eqnarray*}
and 
\begin{eqnarray*}
B = \mqty(0 & \dots & \dots & 0 \\
\vdots & \ddots & & \vdots\\
0 & & \ddots & \vdots\\
-t_y &  0 & \dots & 0 ).
\end{eqnarray*}
From now on, let us use $N$ to denote the number of unit cells in the $y$-direction. We can visualize this system in Fig.~\ref{fig:boundedSystem} for $M=3$ and $N=2$ unit cells.

\begin{figure}
    \centering
    \begin{subfigure}[t]{0.5\textwidth}
        \centering
        \includegraphics[height=1.7in]{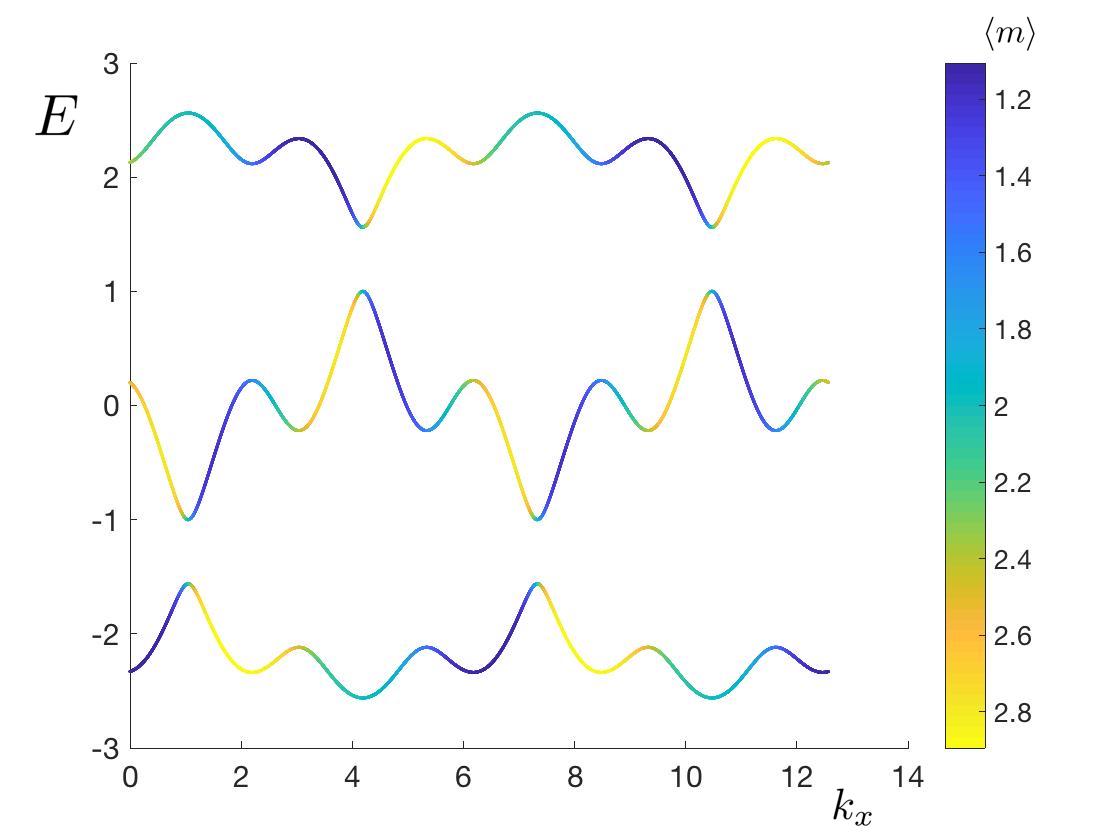}
        \caption{$M=3$, $N=1$ and $t_y=1$.}
        \label{fig:M31UCty1}
    \end{subfigure}%
    ~ 
    \begin{subfigure}[t]{0.5\textwidth}
        \centering
        \includegraphics[height=1.7in]{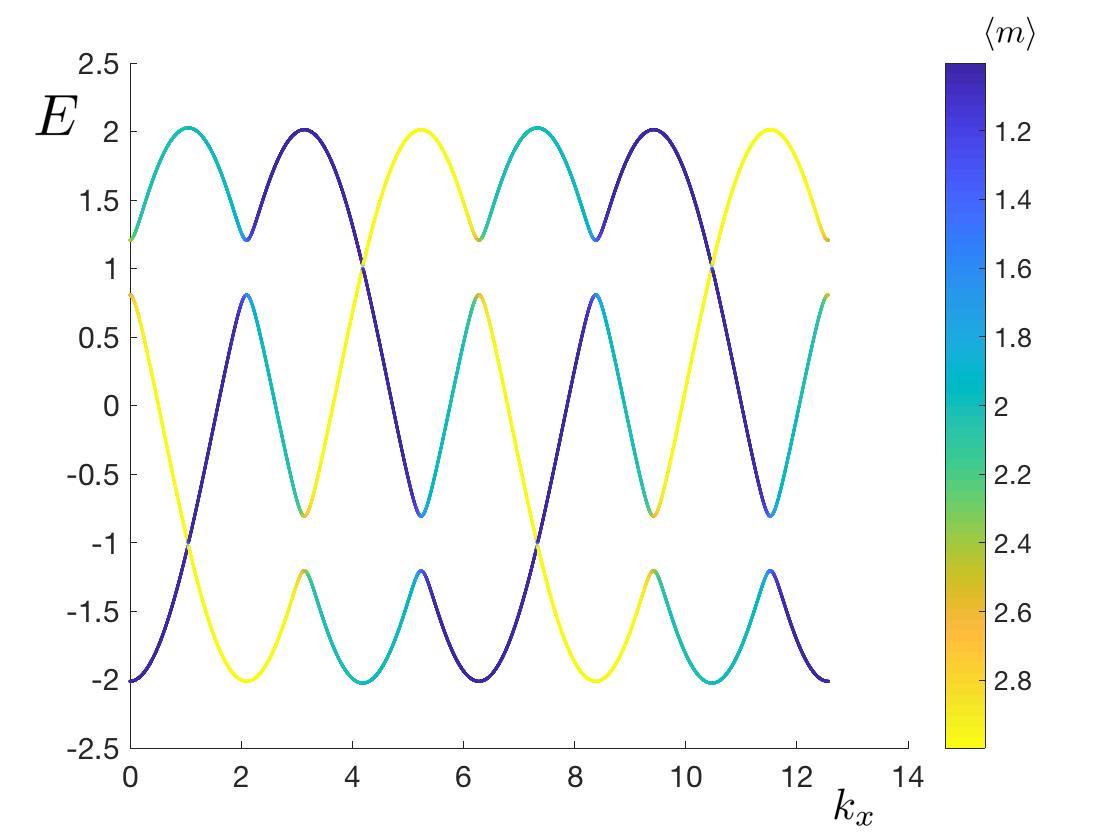}
        \caption{$M=3$, $N=1$ and $t_y=\frac{1}{5}$.}
        \label{fig:M31UCty15}
    \end{subfigure}
    
    \begin{subfigure}[t]{0.5\textwidth}
        \centering
        \includegraphics[height=1.7in]{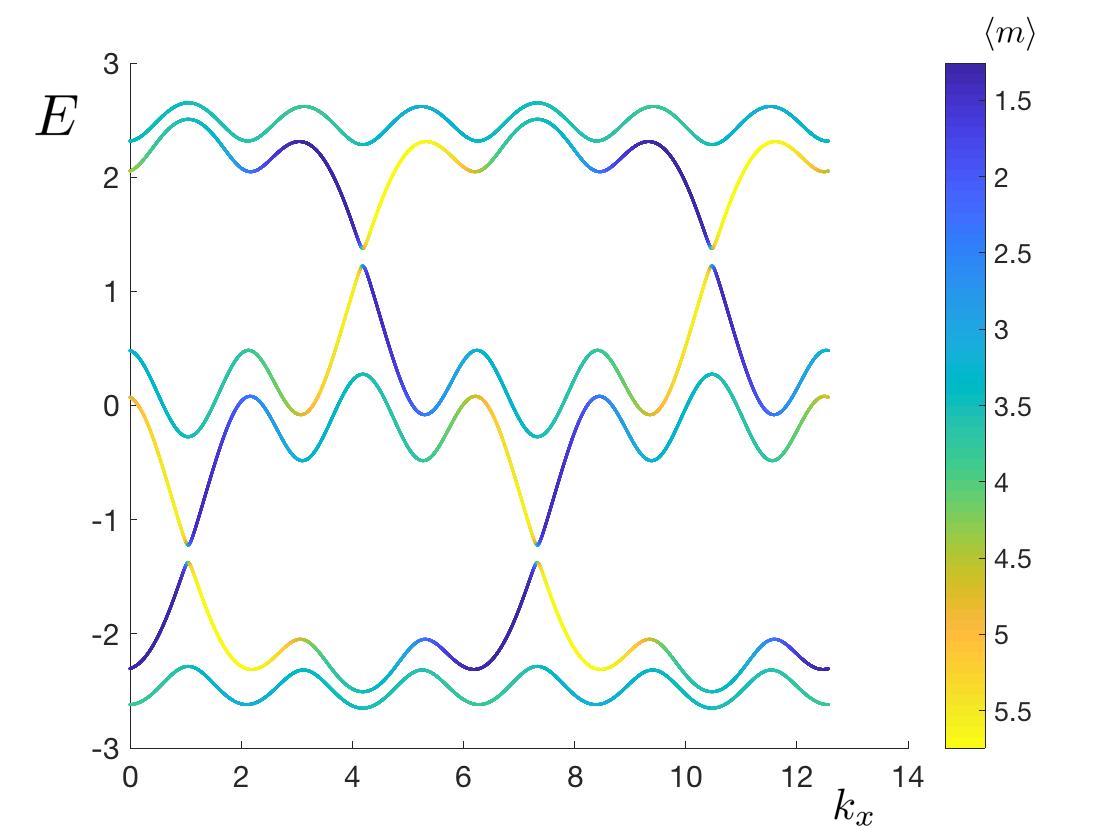}
        \caption{$M=3$, $N=2$ and $t_y=1$.}
        \label{fig:M32UCty1}
    \end{subfigure}%
    ~
    \begin{subfigure}[t]{0.5\textwidth}
        \centering
        \includegraphics[height=1.7in]{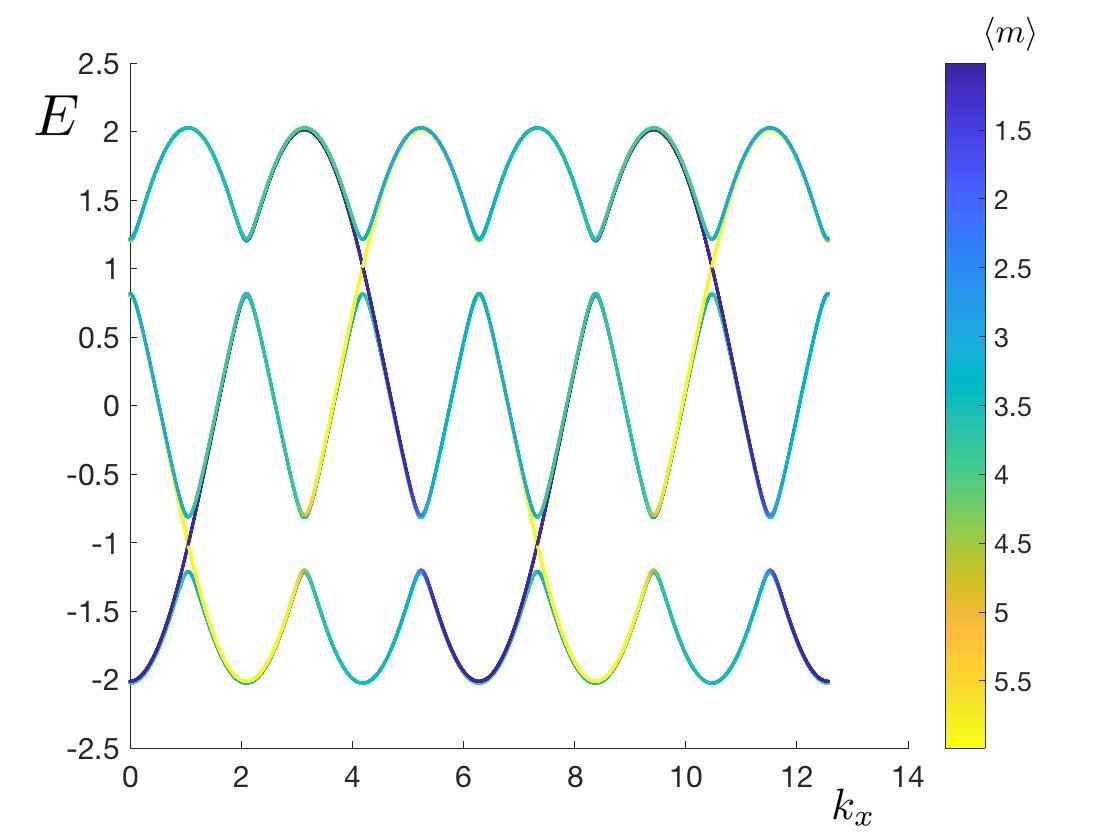}
        \caption{$M=3$, $N=2$ and $t_y=\frac{1}{5}$.}
        \label{fig:M32UCty15}
    \end{subfigure}
    
    \begin{subfigure}[t]{0.5\textwidth}
        \centering
        \includegraphics[height=1.7in]{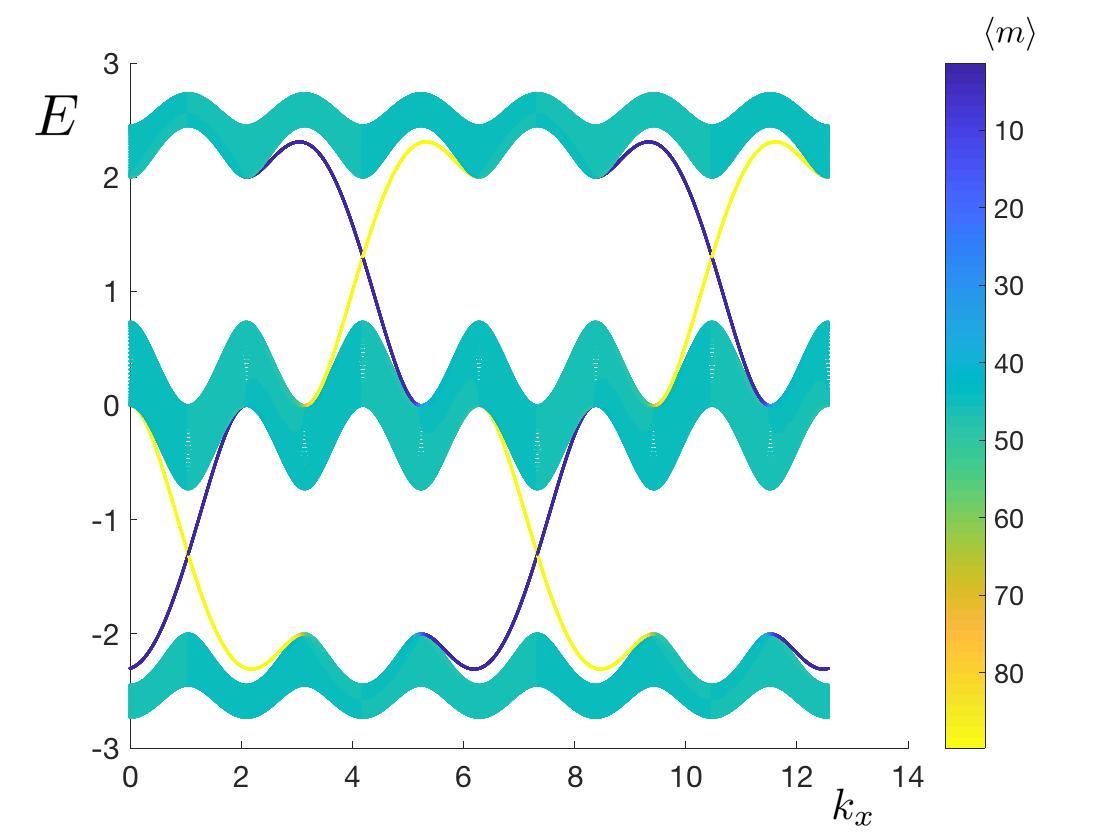}
        \caption{$M=3$, $N=30$ and $t_y=1$.}
        \label{fig:M330UCty1}
    \end{subfigure}%
    ~
    \begin{subfigure}[t]{0.5\textwidth}
        \centering
        \includegraphics[height=1.7in]{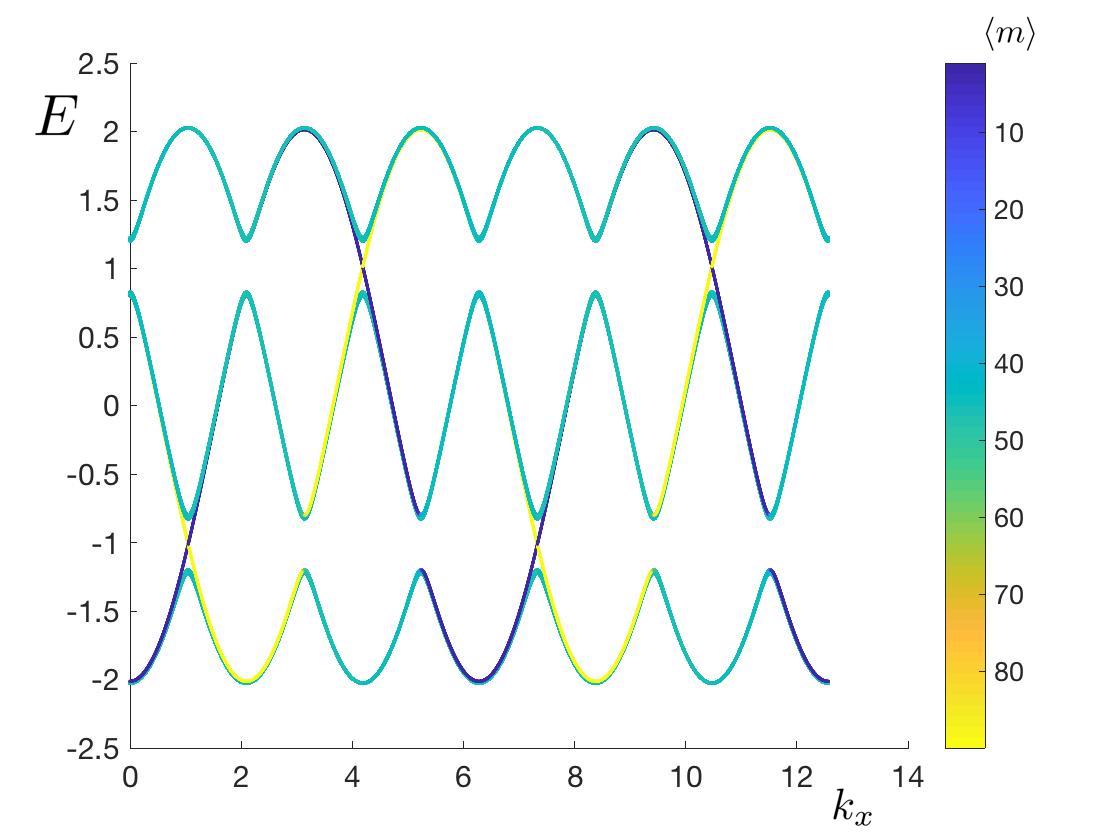}
        \caption{$M=3$, $N=30$ and $t_y=\frac{1}{5}$.}
        \label{fig:M330UCty15}
    \end{subfigure}
    
    \begin{subfigure}[t]{0.55\textwidth}
        \centering
        \includegraphics[height=1.7in]{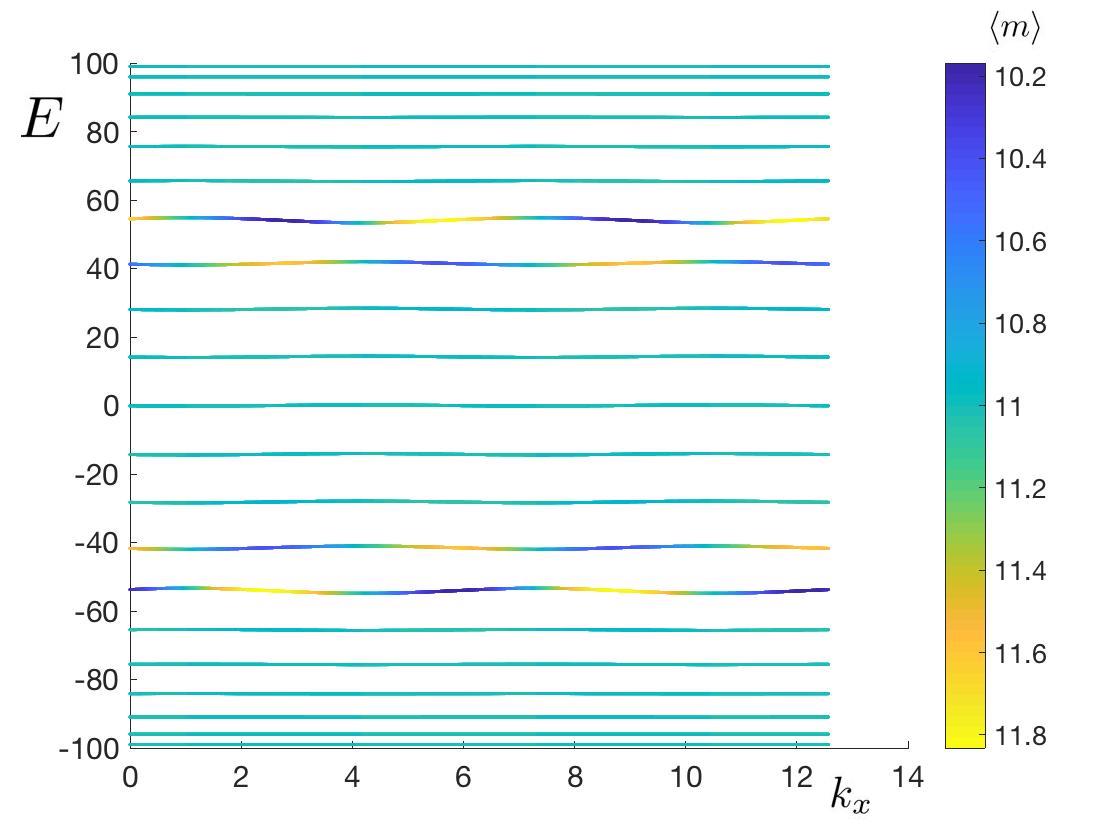}
        \caption{$M=3$, $N=7$ and $t_y=50$. Note that even for big systems the hopping value $t_y$ influences the gap opening. In this case, the eigenstates' mean position is mainly centred in the bulk (observe that the bar code just cover the sites between $10$ and $12$).}
        \label{fig:M37UCty50}
    \end{subfigure}\hfill
    \begin{minipage}{.40\linewidth}
        \caption{Dispersion relation of a bounded system. The gap in the dispersion relation of edge states decreases as the number of unit cells ($N$) increases (compare (a), (c) and (e)) or as the hopping value in the $y$-direction ($t_y$) decreases (compare (a) and (b)).}
        \label{fig:withBoundary}
    \end{minipage}
\end{figure}

Let us explore numerically the dispersion relations when the Hamiltonian is given by (\ref{eq:hamBoundary}). Choosing $M=3$, $t_x=1$ and a system without impurities, we obtain the results shown in Fig.~\ref{fig:withBoundary}. Note that the colours in the dispersion relation represent the localization of the corresponding eigenstate's mean position in the stripe, $\expval{m}$. For example, yellow means that the state is located in the bottom of the system. These results can be compared with similar outcomes in several articles \cite{DM_Stuhl_2015,DM_Mugel_2017,DM_Hatsugai_2016,DM_Celi_2014,DM_Hugel_2014,YH_1993}.

Comparing Fig.~\ref{fig:M31UCty1}, Fig.~\ref{fig:M32UCty1} and Fig.~\ref{fig:M330UCty1} we observe that as the number of unit cells increases the gap between energy bands decreases. There is a point where these gaps disappear and two edge modes are created. This is directly related to what we discussed in section \ref{Edge_States}. In fact, when the gap is closed if the Fermi Energy is set between two energy bands it will only intersect the energy levels of the edge states, leading to the appearance of edge currents \cite{Halperin_1982,YH_1993,Lacki_2016}. This suggests that small systems have a band structure similar to insulators and big systems are actually insulators in the bulk with conducting edges (topological insulators).

\begin{figure}
    \centering
    \begin{subfigure}[t]{0.46\linewidth}
        \centering
        \includegraphics[height=0.75in]{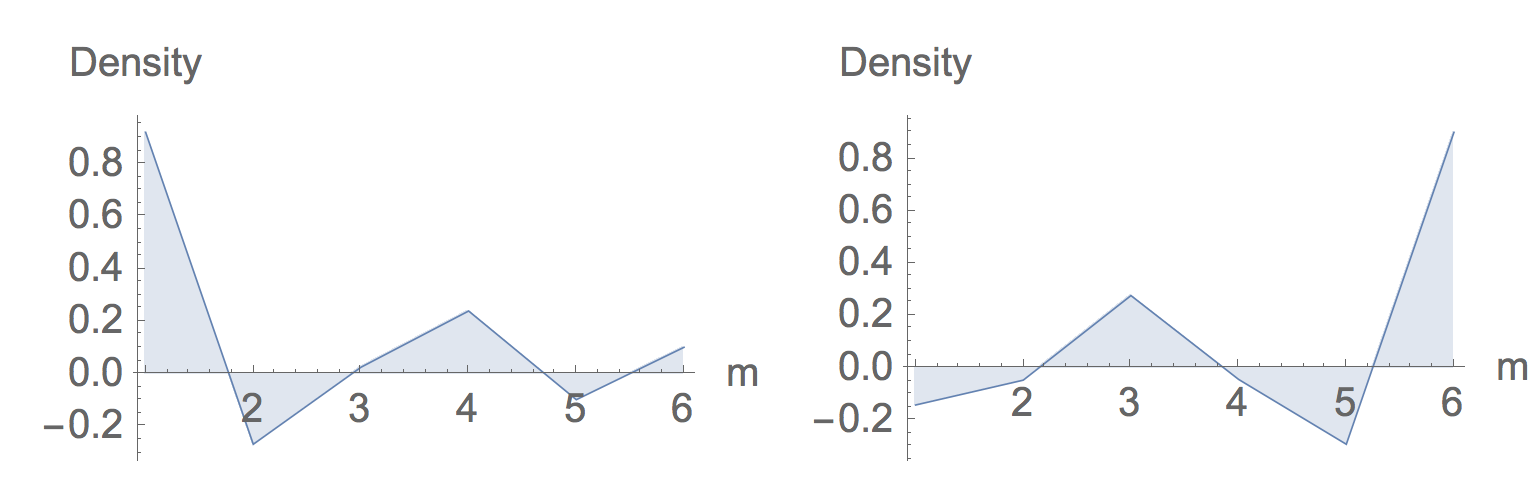}
        \caption{Top and bottom edge state, $t_y=1$.}
        \label{fig:edgeStatesTy1}
    \end{subfigure} \begin{subfigure}[t]{0.46\linewidth}
        \centering
        \includegraphics[height=0.75in]{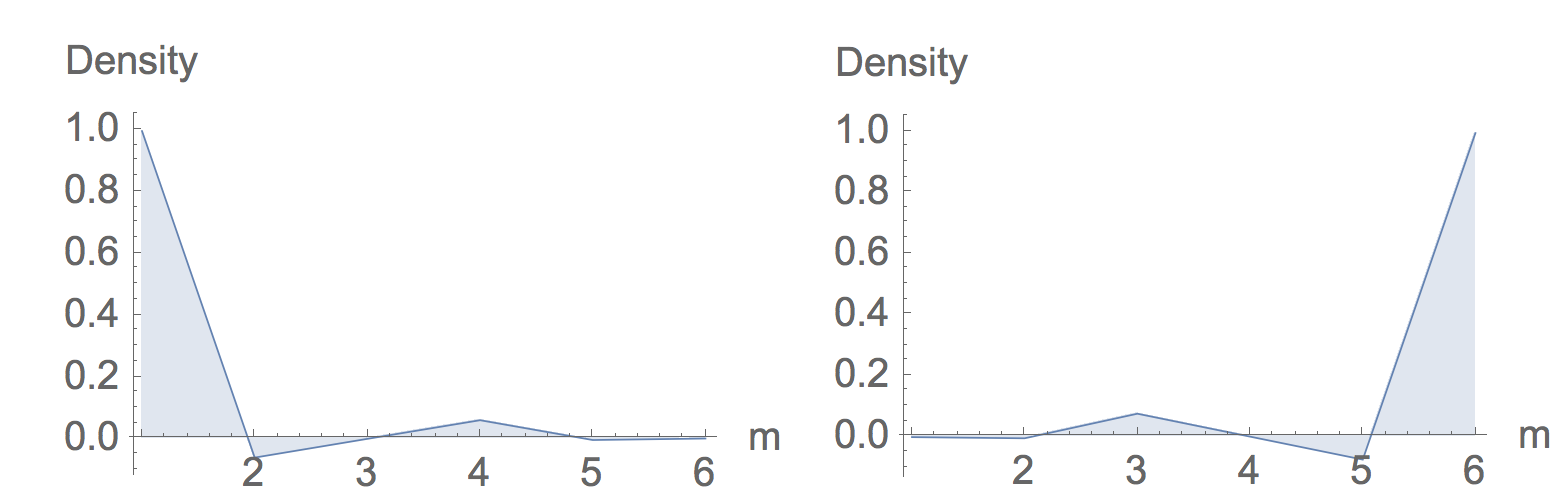}
        \caption{Top and bottom edge state, $t_y=\frac{1}{5}$.}
        \label{fig:edgeStatesTy15}
    \end{subfigure} 
    \caption{The density of edge states in each site for $k_x=4$ with different values of $t_y$. Comparing (a) and (b), we see that the extended edge states are more concentrated in the edge sites in the case of $t_y=\frac{1}{5}$. In fact, for big values of $t_y$ (e.g. $t_y=50$ in Fig.~\ref{fig:M37UCty50}) the states spread all over the sites and edge modes are not created.}
    \label{fig:edgeStatestys}
\end{figure}

A plausible explanation for the gap opening has to do with the interference of the chiral edge currents with the opposite boundary. When an electron with a cyclotron motion is reflected from the edge, it may hit the other side for small systems \cite{DM_Stuhl_2015}. Moreover, if the likeliness of a particle to hop in the $y$-direction decreases ($t_y\searrow$), the number of atoms that hit the other wall decreases as well. In fact, for $t_y=\frac{1}{5}$ (Fig.~\ref{fig:edgeStatesTy15}) the edge wavefunctions are more concentrated in the edges than for $t_y=1$ (Fig.~\ref{fig:edgeStatesTy1}). This explanation is in accordance with the numeric results presented by Fig.~\ref{fig:M31UCty1} and \ref{fig:M31UCty15}, where we see that the gap is almost closed in the latter whereas the gap for $t_x=1$ is of the order of the bulk energy gap for $t_y=\frac{1}{5}$.

\

These numerical results support the conjecture that there is a relation between the length of the system and the gap opening in the dispersion relation. In section \ref{problemSection} we explore this relation aiming to find a formula that describes the gap.

\newpage

\subsection{In-site impurities}

In this section, we explore numerically the robustness of the edge states when disorder is introduced. Following Halperin's work \cite{Halperin_1982}, we test that the edge states are protected from impurities. Proceeding as in \cite{YH_1993} the Hamiltonian can be written as

\begin{eqnarray}
\label{eq:HamBoundImp}
\bld{H} = H + H_{Imp},
\end{eqnarray}
where $H$ is given by (\ref{eq:HamArt2ndQuant}) and $H_{Imp} = \sum_{\bld{R}} \sum_{j=1}^{M} V(j) \bld{c}_{\bld{R},j}^\dagger \bld{c}_{\bld{R},j}$. The potential represents the impurities. It is constant in the $x$-direction and follows an uniform random distribution in the $y$-direction with values in the interval $[-V_{b},V_{b}]$. Following our previous work, we write the Hamiltonian (\ref{eq:HamBoundImp}) in the momentum space and obtain:

\begin{eqnarray*}
\bld{H} = \sum_{\bld{k}} \mqty(\bld{c}_{\bld{k},1}^\dagger & ... & \bld{c}_{\bld{k},M}^\dagger)\ \mathcal{H}'\  \mqty(\bld{c}_{\bld{k},1} \\ ... \\ \bld{c}_{\bld{k},M}),\\
\end{eqnarray*}
with $\mathcal{H}'$ given by

\begin{eqnarray*}
\mathcal{H}' = \mqty(A' & B & 0 & ...  & 0 \\
B^\dagger & A' & B &  & \vdots\\
0 & B^\dagger & \ddots &  & 0\\
\vdots & & & & B\\
0 & \dots & 0 & B^\dagger & A' ),
\end{eqnarray*}
where $B$ is given as in the previous sections and
\begin{eqnarray*}
A' = \left(\begin{smallmatrix}
-2 t_x \cos(\bld{k}\cdot\bld{a}_x) + V(1) & -t_y & 0 & ... & 0 \\
-t_y & \ddots &  &  & \vdots\\
0 & &  & & 0\\
\vdots & & & \ddots & -t_y\\
0 &  \dots & 0 & -t_y & -2 t_x \cos(\bld{k}\cdot\bld{a}_x + (M-1)\phi) + V(M) 
\end{smallmatrix}\right)
\end{eqnarray*}

\

\begin{figure}
    \centering
    \begin{subfigure}[t]{0.49\linewidth}
        \centering
        \includegraphics[height=1.9in]{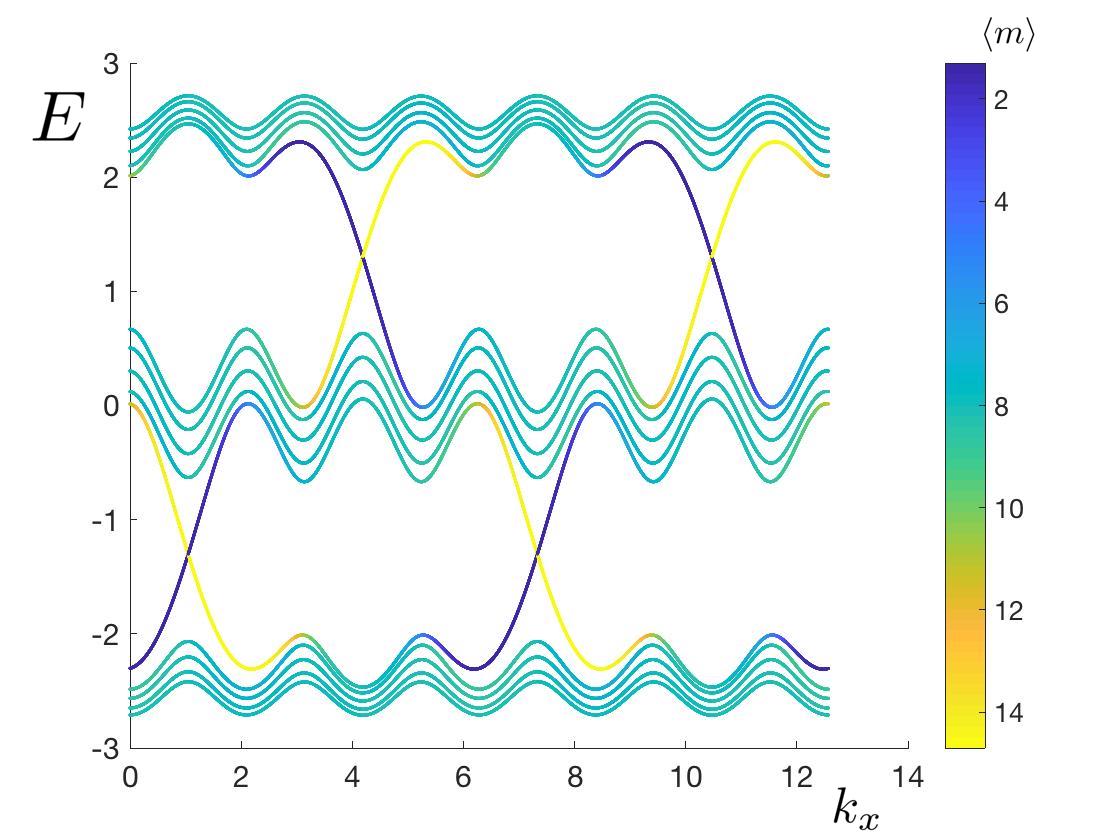}
        \caption{$V_b = 0$ and $t_y=1$.}
        \label{fig:ty0Vb0}
    \end{subfigure} \begin{subfigure}[t]{0.49\linewidth}
        \centering
        \includegraphics[height=1.9in]{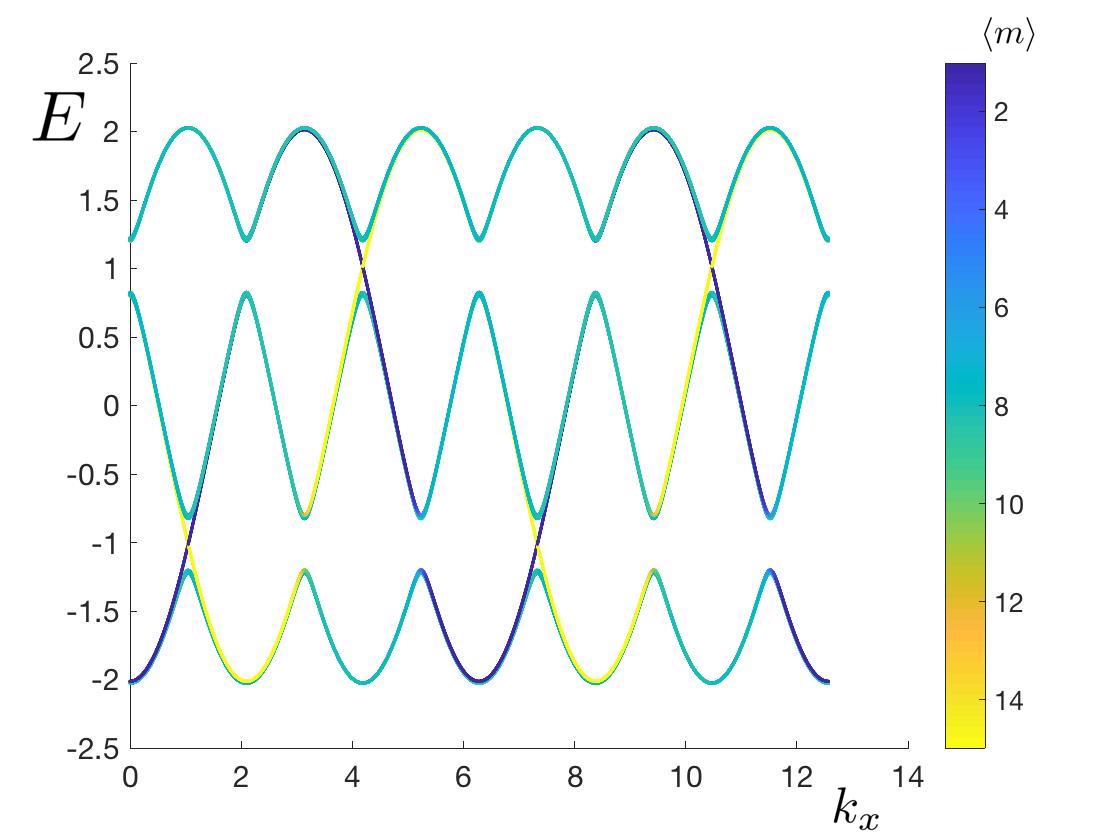}
        \caption{$V_b = 0$ and $t_y=\frac{1}{5}$.}
        \label{fig:ty15Vb0}
    \end{subfigure}
    ~
    \begin{subfigure}[t]{0.49\linewidth}
        \centering
        \includegraphics[height=1.9in]{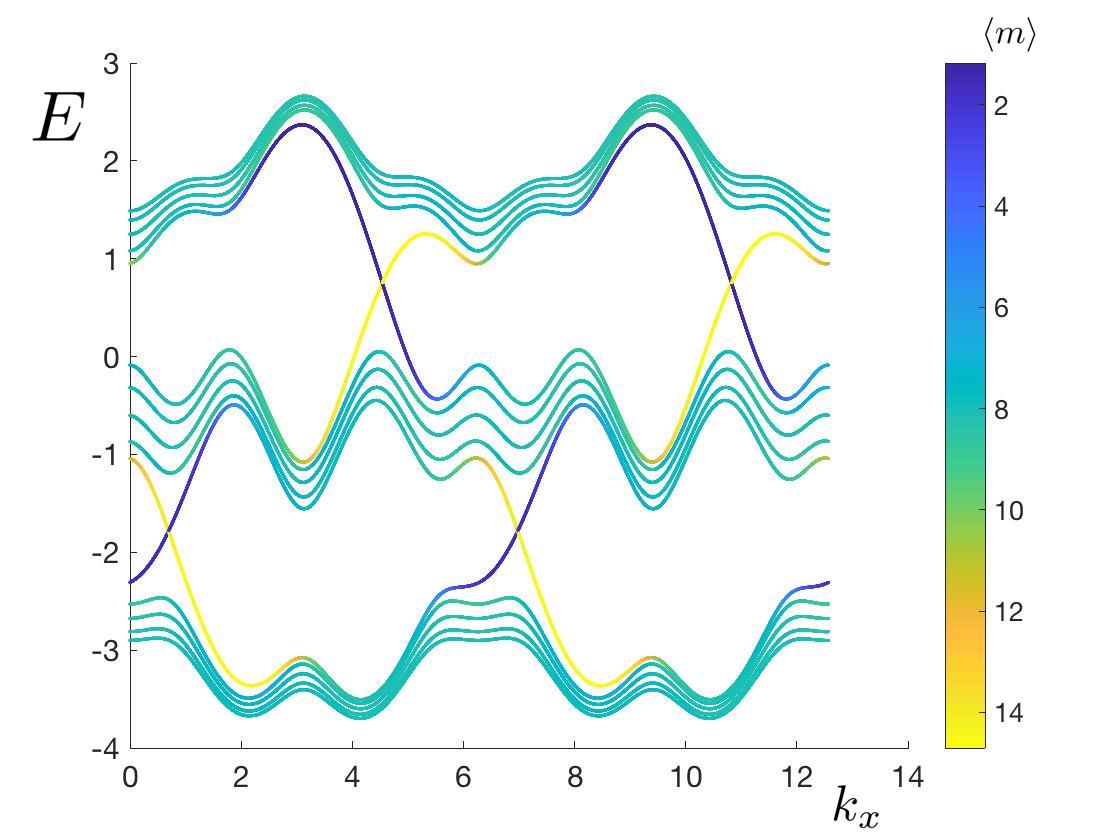}
        \caption{$V_b = 1.5$ and $t_y=1$.}
        \label{fig:ty0Vb15}
    \end{subfigure} \begin{subfigure}[t]{0.49\linewidth}
        \centering
        \includegraphics[height=1.9in]{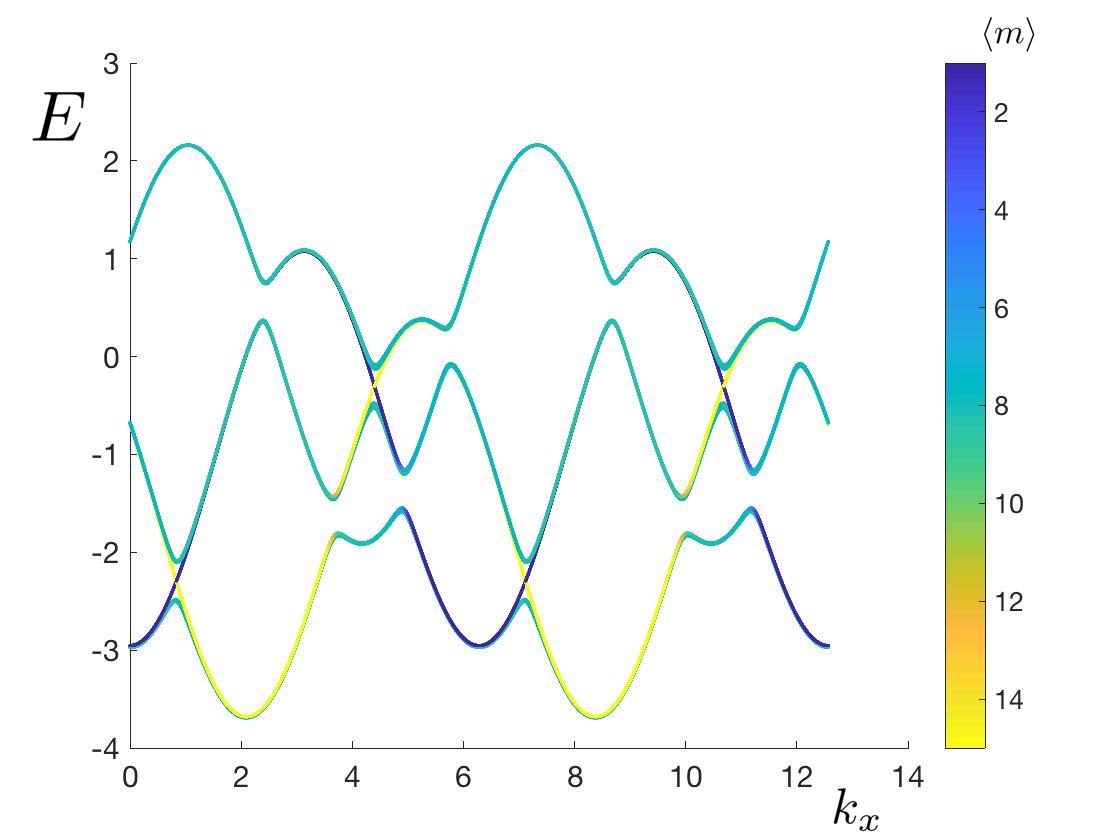}
        \caption{$V_b = 2.5$ and $t_y=\frac{1}{5}$.}
        \label{fig:ty15Vb25}
    \end{subfigure}
    ~
    \begin{subfigure}[t]{0.49\linewidth}
        \centering
        \includegraphics[height=1.9in]{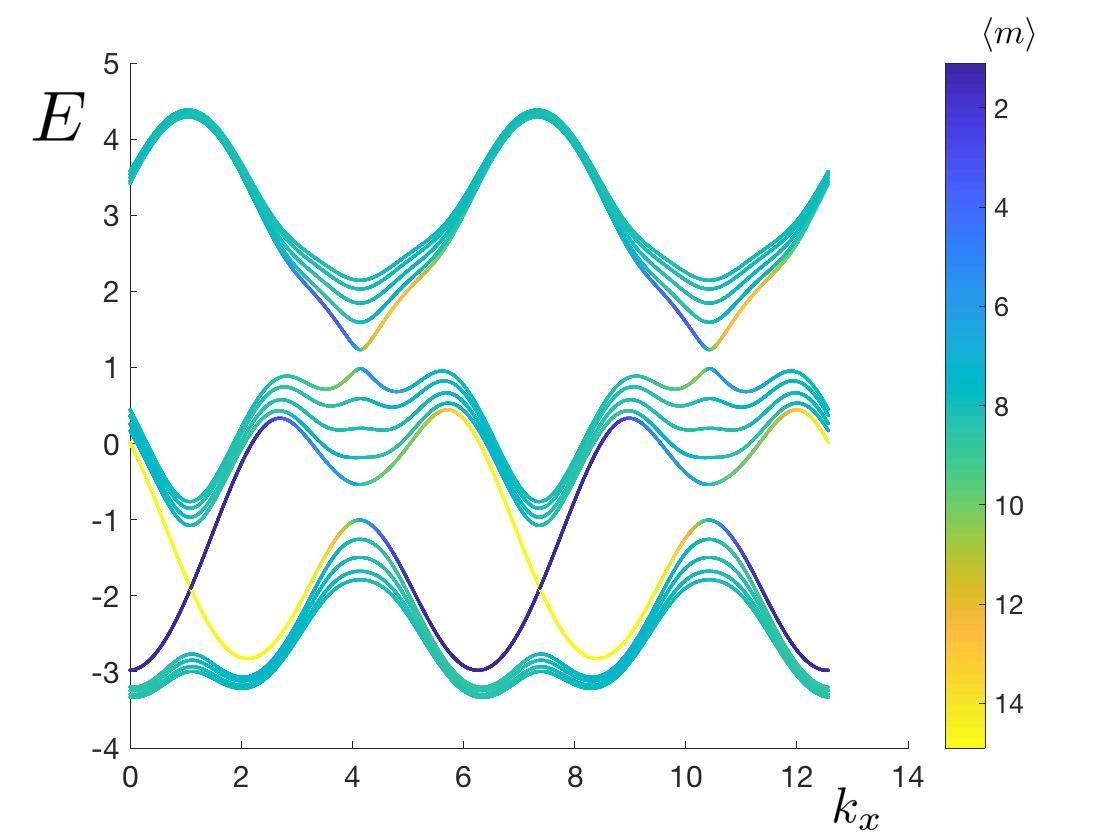}
        \caption{$V_b = 2.5$ and $t_y=1$.}
        \label{fig:ty0Vb25}
    \end{subfigure} \begin{subfigure}[t]{0.49\linewidth}
        \centering
        \includegraphics[height=1.9in]{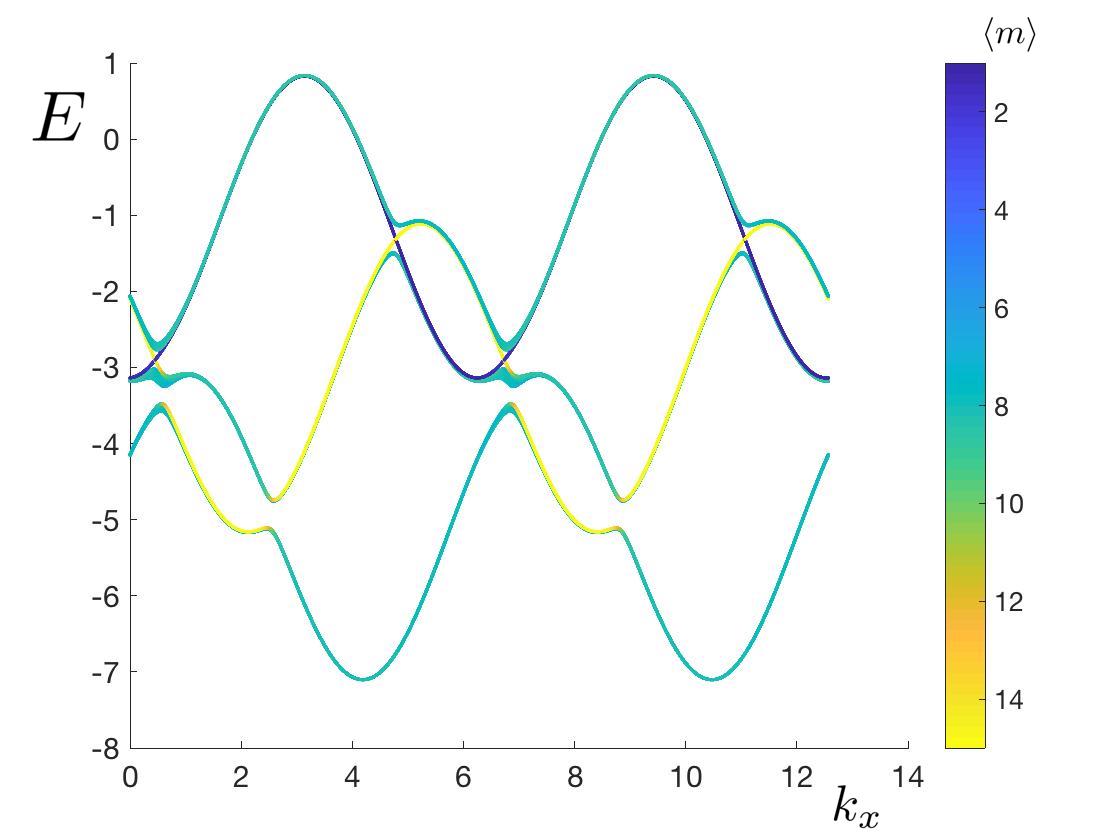}
        \caption{$V_b = 6$ and $t_y=\frac{1}{5}$.}
        \label{fig:ty15Vb6}
    \end{subfigure}
    \caption{Dispersion relation with impurities of a system with $M=3$ and $N=5$ unit cells. In (c) and (d) we observe that edge states are robust when disorder is introduced into the system. However, for a certain amount of impurities, a gap is opened.}
    \label{fig:dispRelImp}
\end{figure}

In Fig.~\ref{fig:dispRelImp} we see that the edges states keep the gap closed until a certain amount of disorder. For $V_b=1.5$ and $t_y=1$ (Fig.~\ref{fig:ty0Vb15}) the edge states cross exactly in the same values of $k_x$ as for $V_b=0$ (Fig.~\ref{fig:ty0Vb0}) and with a similar shape. Until this $V_b$ the chiral edge currents can still overcome the impurities without interfering with the other edge. When $V_b$ is increased to $2.5$ (Fig.~\ref{fig:ty0Vb25}) some gaps are opened, meaning that the impurities were too strong for the edge particles to overcome them without interaction. Furthermore, we see that when $t_y$ is decreased (Fig.~\ref{fig:ty15Vb25} and \ref{fig:ty15Vb6}) the amount of impurities necessary to break the intersection of edge states is bigger when compared to $t_y=1$. Again, this is in accordance with the supposition that the gap opening comes from the interaction of the edges states with the other boundary. These results can be compared with similar simulations \cite{YH_1993,Imura_2017}. Note that as we are considering a random potential, there are different outputs of numerical simulations with equal inputs. 

\

As we discussed in the chapter \ref{theory}, the non-zero Chern number predicts robust edge states for large systems. Indeed, that is what we get from our numerical simulations. However, the topological properties are mainly present for large system sizes. As we saw, when the size decreases a gap opens and the system is no longer a conductor. In the next chapter, we discuss the experimental procedure based on cold atoms in optical lattices and how they represent the Hofstadter model.

\newpage

\section{Experimental procedure} \label{expProc}

To better understand the Hofstadter model let us give a general overview of the experimental procedure used to obtain it. 

As mentioned above, the observation of IQH effects is subjected to the existence of powerful magnetic fields and extremely low temperatures. Consequently, the experimental effort to produce such conditions is very high, making it attractive to search for new ways to represent and mimic these systems. One of these ways is the development of cold atoms in optical lattices \cite{DM_Stuhl_2015,DM_Celi_2014,Mancini_2015,Lin_2009,Jim_2012,Aidelsburger_2013}.

The major advantage of this method has to do with the possibility to generate strong magnetic fields in a square lattice with cold atoms by an artificial gauge field \cite{DM_Celi_2014}. In this lattice, it is possible to control precisely the hopping terms enabling the cold atoms to behave like electrons \cite{Aidelsburger_2013}. This is a key feature that allows relating some effects in these systems with Quantum Hall effects. Moreover, the wave-behaviour and quantum effects are more visible in temperatures near $0$ Kelvin. As a consequence, the realization of cold atoms is useful due to the possibility to control some of its features with the help of optical devices.

In \cite{Oliver_2006}, Oliver Morsch describes a way to cool down atoms which leads to a Bose-Einstein condensate (BEC). Following the common practice in some experiments\cite{Oliver_2006,DM_Stuhl_2015,Jim_2012,DM_Celi_2014}, we will consider a cloud of BECs of an isotope of Rubidium, $^{87}$Rb.

\begin{figure}[h]
    \centering
    \begin{subfigure}[t]{0.5\textwidth}
        \centering
        \includegraphics[height=1.2in]{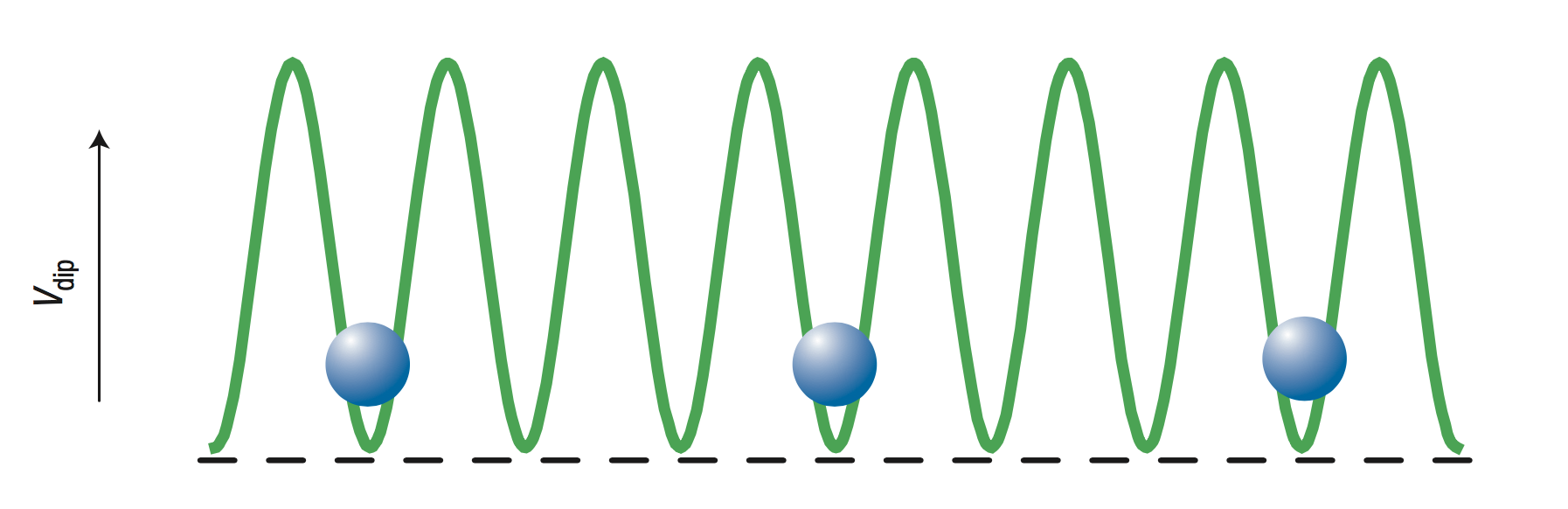}
        \caption{Potential trap, image take from \cite{Bloch_2005}.}
        \label{fig:potentialTrap}
    \end{subfigure}%
    ~ 
    \begin{subfigure}[t]{0.5\textwidth}
        \centering
        \includegraphics[height=1.2in]{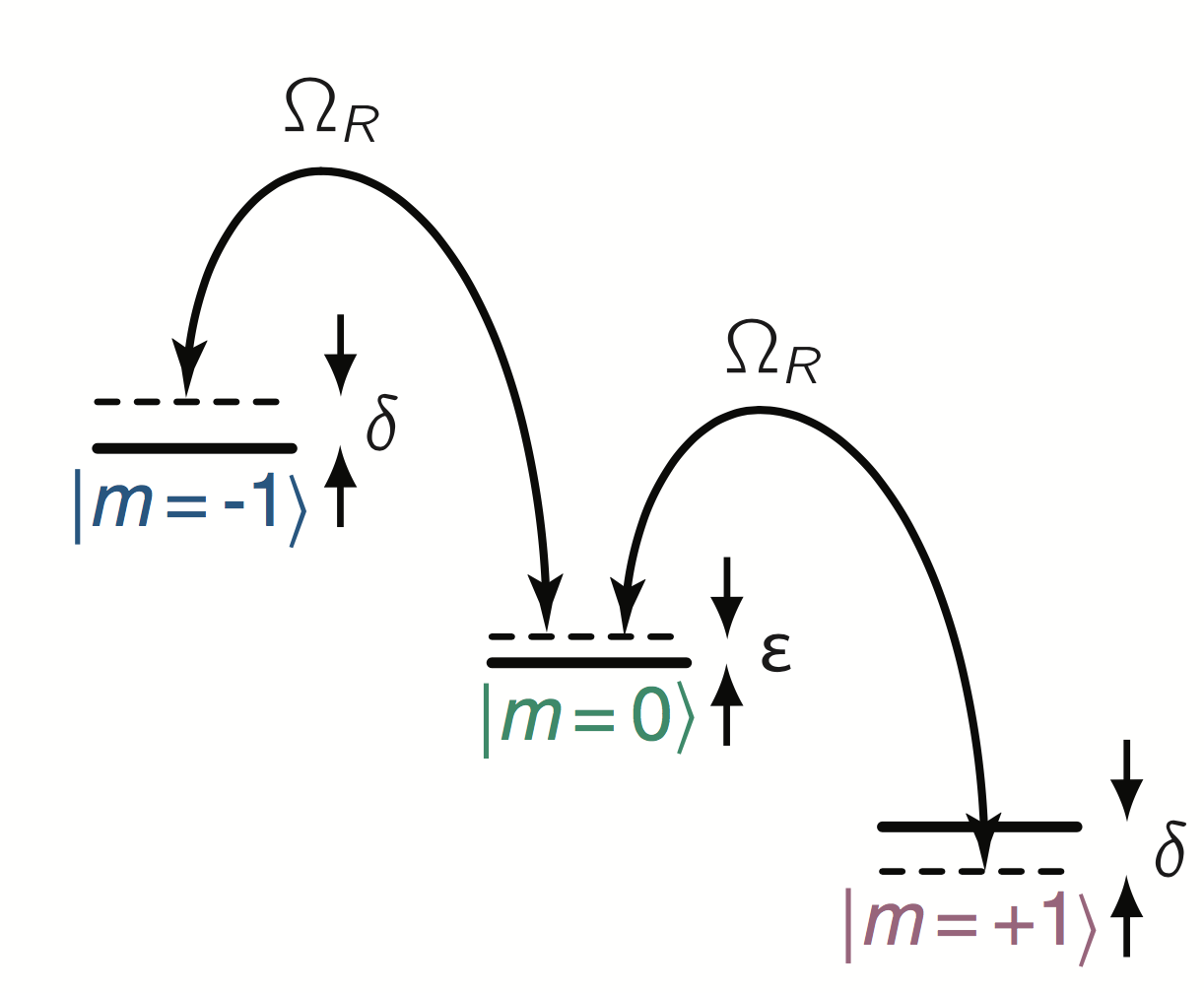}
        \caption{Internal transitions, image take from \cite{DM_Stuhl_2015}.}
        \label{fig:internalDegree}
    \end{subfigure}
\caption{Optical lattice and synthetic dimension.}
\end{figure}

As it is defined in \cite{Jim_2012} an \textit{optical lattice} results from the trapping of an atom in the electric field of an optical standing wave (Fig.~\ref{fig:potentialTrap}). We are interested in describing a one-dimensional chain of atoms with three internal spin states loaded in an optical lattice. Thus we have what is called a ``real" dimension and a ``synthetic" dimension. The first describes the actual position of the atoms and the hopping, $t$, 'along the $x$-direction. The latter describes an extra dimension along the $y$-direction where the hopping terms are determined by the frequency at which the atoms change from one internal state to another (Fig.~\ref{fig:internalDegree}). This frequency is called  \textit{Rabi frequency} and it is commonly denoted by $\Omega_R$.

\

\

\begin{figure}[h]
    \centering
    \begin{subfigure}[t]{0.5\textwidth}
    \centering
    \includegraphics[height=1.2in]{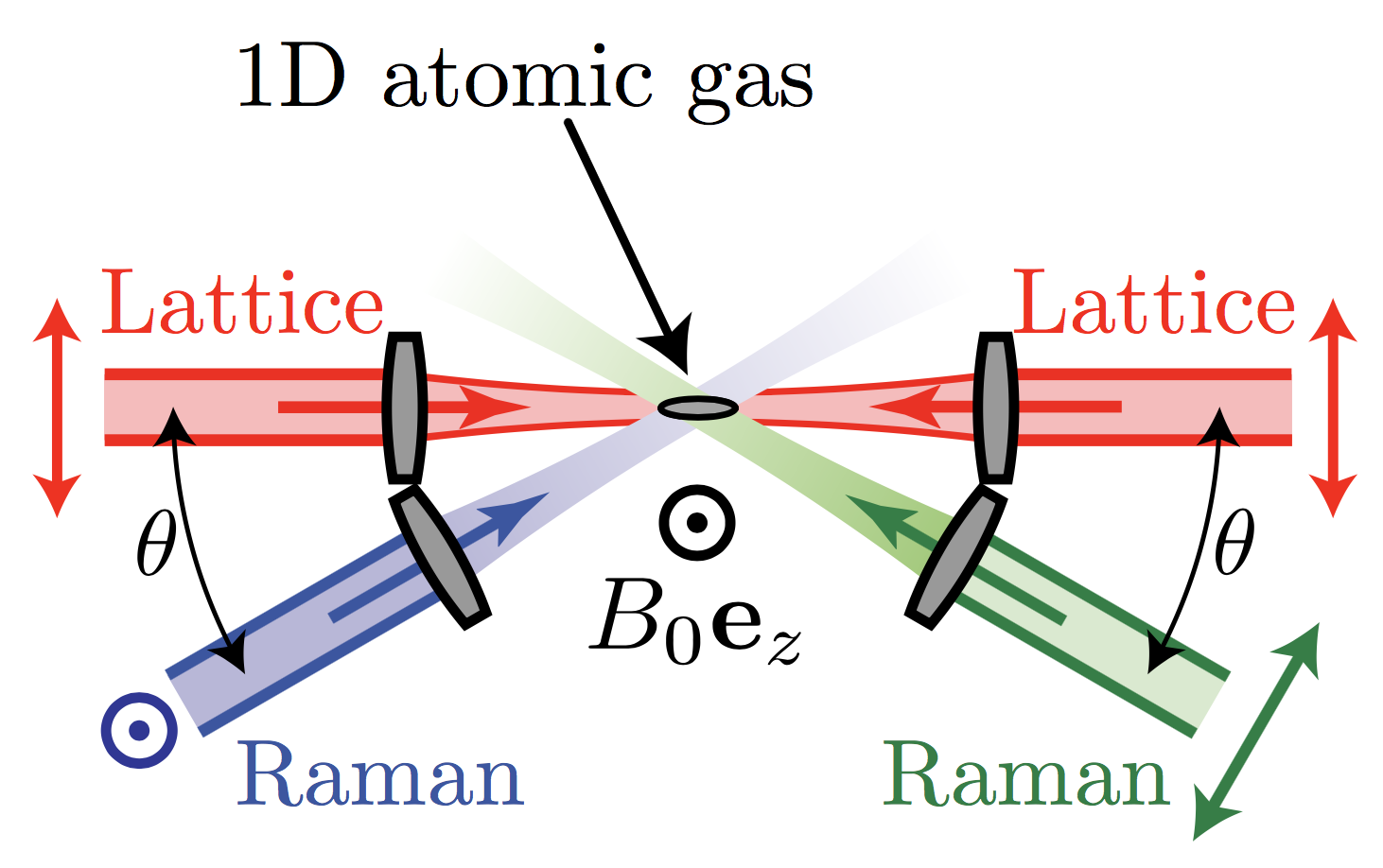}
    \caption{The apparatus, image take from \cite{DM_Celi_2014}.}
    \label{fig:apparatus}
    \end{subfigure}%
    ~ 
    \begin{subfigure}[t]{0.5\textwidth}
        \centering
        \includegraphics[width=7cm, height=2.5cm]{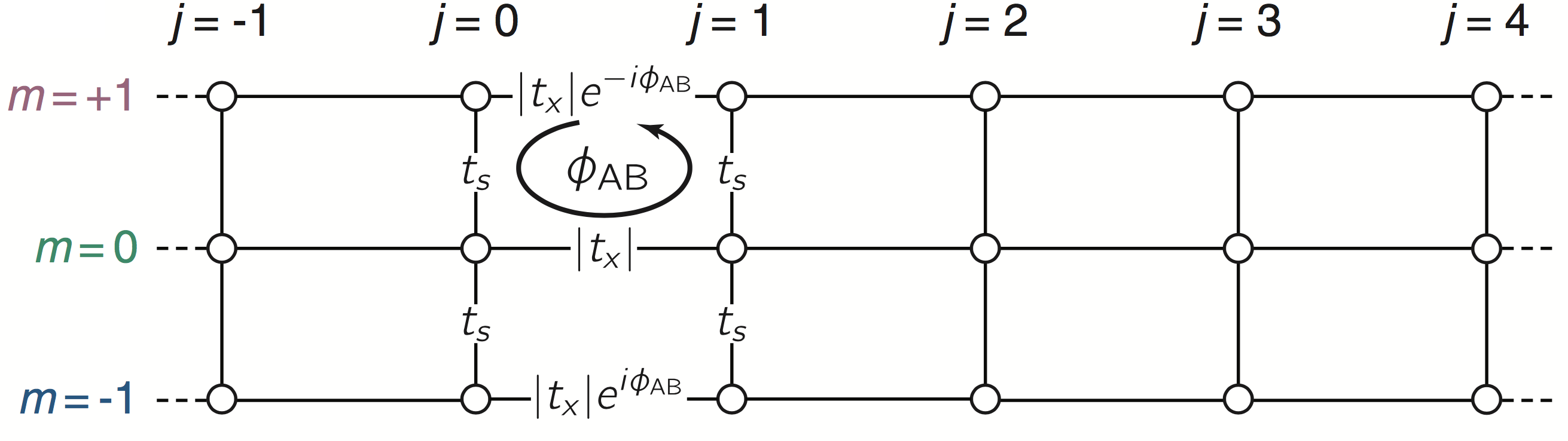}
        \caption{2D lattice, image take from \cite{DM_Stuhl_2015}.}
        \label{fig:2DlatticeExp}
    \end{subfigure}
\caption{Experimental lattice scheme.}
\end{figure}

As it is shown in Fig.~\ref{fig:apparatus}, two laser beams (red strips) with wavelength $\lambda = 1064$ nm and in opposite directions create a 1D optical lattice strong enough to use the tight-binding approximation \cite{DM_Stuhl_2015,DM_Celi_2014}.  In this conditions the optical lattice has a period of $a=\lambda/2$. The internal degrees of freedom of the atom are controled by the angle $\theta$ at which a special type of laser, \textit{Raman lasers}, shoot the 1D atomic gas (Fig.~\ref{fig:apparatus}). So, for a Raman laser with wavelength $\lambda_R=790$ nm \cite{DM_Stuhl_2015,DM_Celi_2014} and angle $\theta$ from $\bld{e}_x$, the hopping in the synthetic direction is given by: 
$$|t_x| e^{-im\phi_{AB}}$$
where $\phi_{AB} = 2k_R a$, $k_R=2\pi\cos(\theta)/\lambda_R$ is called the Raman recoil momentum, $m=-1,0,1$ and $t_x$ is given by the transitions between internal atomic states.
Note that in Fig.~\ref{fig:2DlatticeExp} the hopping terms represented in the $x$-direction (chain direction) are the internal hopping terms instead of the inter-atomic hopping term. This is obtained by a gauge transformation of the real system. With this procedure, we arrive at the system described by the Hamiltonian defined in section \ref{bounded_systems} with $M=3$ and just one unit cell, i.e.

\begin{eqnarray*}
H = \sum_{\bld{k}} \mqty(\bld{c}_{\bld{k},1}^\dagger & \bld{c}_{\bld{k},0}^\dagger & \bld{c}_{\bld{k},-1}^\dagger)\ \mathcal{H}\  \mqty(\bld{c}_{\bld{k},1} \\ \bld{c}_{\bld{k},0} \\ \bld{c}_{\bld{k},-1}),\\
\end{eqnarray*}
where 
\begin{eqnarray*}
\mathcal{H} = \mqty(-2 t_x \cos(\bld{k}\cdot\bld{a}_x - \phi_{AB}) & -t_y & 0\\
-t_y &-2 t_x \cos(\bld{k}\cdot\bld{a}_x) & -t_y\\
0 & -t_y & -2 t_x \cos(\bld{k}\cdot\bld{a}_x + \phi_{AB}) ).
\end{eqnarray*}

Apart from this method, there are other ways to control the parameters $t_x$ and $\phi_{AB}$. In \cite{Jim_2012} it is given a very brief overview of other alternatives. In resume, these parameters can also be controlled in driven optical lattices \cite{Ciampini_2011}, by using rotating optical lattices \cite{Williams_2010} or by Raman-assisted tunnelling in an optical superlattice \cite{Aidelsburger_2011}.

To conclude this small overview of the experimental procedure we just comment on the way the dynamic of the atoms is visualized. The most common technique used is known as time-of-flight (TOF) \cite{DM_Stuhl_2015,Jim_2012}. In \cite{Oliver_2006} it is briefly explained that this ``method consists in simply switching off the trapping field (magnetic or optical) at time t=0 and taking an image of the BEC a few (typically 5 to 25) milliseconds later". As an example, in the experiment performed in \cite{DM_Stuhl_2015} the time-of-flight was around 18 ms whether in \cite{Jim_2012} it was 28.2 ms.

\newpage

\section{Dispersion gap}\label{problemSection}

We saw in previous sections that the gap opening of the eigenenergies is related to the size of the system and to the value of the hopping terms, $t_y$ and $t_x$. As this opening effect changes the properties of the edges of the system, a description of it would help to engineer experiments that accurately describe topological insulators. In this chapter, we give an original exposition of the gap opening by starting with a deduction of a formula for the top edge state (TES) and the bottom edge state (BES) in a semi-infinite system. Then, we rely on perturbation theory to reveal the relationship between the size of a finite system and the energies of the TES and BES.

\subsection{Semi-infinite system} \label{semiInfinite}

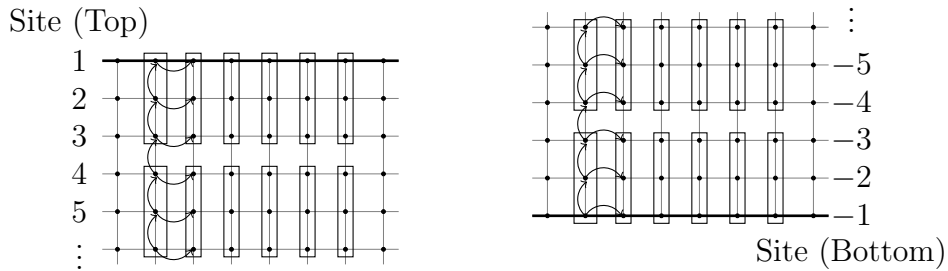
\begin{figure}[h]
    \centering
    \begin{subfigure}[t]{0.5\textwidth}
        \begin{center}
\begin{tikzpicture}

\draw[step=0.5cm,gray,very thin] (-0.2,-0.2) grid (3.7,2.5);

\draw[fill] (0,0) circle [radius=0.025];
\draw[fill] (0.5,0) circle [radius=0.025];
\draw[fill] (1,0) circle [radius=0.025];
\draw[fill] (1.5,0) circle [radius=0.025];
\draw[fill] (2,0) circle [radius=0.025];
\draw[fill] (2.5,0) circle [radius=0.025];
\draw[fill] (3,0) circle [radius=0.025];
\draw[fill] (3.5,0) circle [radius=0.025];
\draw[fill] (0,0.5) circle [radius=0.025];
\draw[fill] (0.5,0.5) circle [radius=0.025];
\draw[fill] (1,0.5) circle [radius=0.025];
\draw[fill] (1.5,0.5) circle [radius=0.025];
\draw[fill] (2,0.5) circle [radius=0.025];
\draw[fill] (2.5,0.5) circle [radius=0.025];
\draw[fill] (3,0.5) circle [radius=0.025];
\draw[fill] (3.5,0.5) circle [radius=0.025];
\draw[fill] (0,1) circle [radius=0.025];
\draw[fill] (0.5,1) circle [radius=0.025];
\draw[fill] (1,1) circle [radius=0.025];
\draw[fill] (1.5,1) circle [radius=0.025];
\draw[fill] (2,1) circle [radius=0.025];
\draw[fill] (2.5,1) circle [radius=0.025];
\draw[fill] (3,1) circle [radius=0.025];
\draw[fill] (3.5,1) circle [radius=0.025];
\draw[fill] (0,1.5) circle [radius=0.025];
\draw[fill] (0.5,1.5) circle [radius=0.025];
\draw[fill] (1,1.5) circle [radius=0.025];
\draw[fill] (1.5,1.5) circle [radius=0.025];
\draw[fill] (2,1.5) circle [radius=0.025];
\draw[fill] (2.5,1.5) circle [radius=0.025];
\draw[fill] (3,1.5) circle [radius=0.025];
\draw[fill] (3.5,1.5) circle [radius=0.025];
\draw[fill] (0,2) circle [radius=0.025];
\draw[fill] (0.5,2) circle [radius=0.025];
\draw[fill] (1,2) circle [radius=0.025];
\draw[fill] (1.5,2) circle [radius=0.025];
\draw[fill] (2,2) circle [radius=0.025];
\draw[fill] (2.5,2) circle [radius=0.025];
\draw[fill] (3,2) circle [radius=0.025];
\draw[fill] (3.5,2) circle [radius=0.025];
\draw[fill] (0,2.5) circle [radius=0.025];
\draw[fill] (0.5,2.5) circle [radius=0.025];
\draw[fill] (1,2.5) circle [radius=0.025];
\draw[fill] (1.5,2.5) circle [radius=0.025];
\draw[fill] (2,2.5) circle [radius=0.025];
\draw[fill] (2.5,2.5) circle [radius=0.025];
\draw[fill] (3,2.5) circle [radius=0.025];
\draw[fill] (3.5,2.5) circle [radius=0.025];

\node at (-0.5, 3) {Site (Top)};
\node at (-0.5, 2.5) {$1$};
\node at (-0.5, 2) {$2$};
\node at (-0.5, 1.5) {$3$};
\node at (-0.5, 1) {$4$};
\node at (-0.5, 0.5) {$5$};
\node at (-0.5, 0) {$\vdots$};


\draw [->] (0.5,0) arc (-150:-30:7.8pt);
\draw [->] (0.5,0.5) arc (-150:-30:7.8pt);
\draw [->] (0.5,1) arc (-150:-30:7.8pt);
\draw [->] (0.5,1.5) arc (-150:-30:7.8pt);
\draw [->] (0.5,2) arc (-150:-30:7.8pt);
\draw [->] (0.5,2.5) arc (-150:-30:7.8pt);

\draw [->] (0.5,0) arc (230:125:8.1pt);
\draw [->] (0.5,0.5) arc (230:125:8.1pt);
\draw [->] (0.5,1) arc (230:125:8.1pt);
\draw [->] (0.5,1.5) arc (230:125:8.1pt);
\draw [->] (0.5,2) arc (230:125:8.1pt);

\draw (0.35,-0.1) rectangle (0.65,1.1);
\draw (0.35,1.4) rectangle (0.65,2.6);

\draw (0.9,-0.1) rectangle (1.1,1.1);
\draw (0.9,1.4) rectangle (1.1,2.6);

\draw (1.4,-0.1) rectangle (1.6,1.1);
\draw (1.4,1.4) rectangle (1.6,2.6);

\draw (1.9,-0.1) rectangle (2.1,1.1);
\draw (1.9,1.4) rectangle (2.1,2.6);

\draw (2.4,-0.1) rectangle (2.6,1.1);
\draw (2.4,1.4) rectangle (2.6,2.6);

\draw (2.9,-0.1) rectangle (3.1,1.1);
\draw (2.9,1.4) rectangle (3.1,2.6);

\draw [line width=1pt] (-0.2,2.5)-- (3.7,2.5);

\end{tikzpicture}
\end{center}
\caption{Semi-infinite system with an edge in the top.}
\label{fig:topBoundedSystem}

    \end{subfigure}%
    ~ 
    \begin{subfigure}[t]{0.5\textwidth}
        \begin{center}
\begin{tikzpicture}

\draw[step=0.5cm,gray,very thin] (-0.2,0) grid (3.7,2.7);

\draw[fill] (0,0) circle [radius=0.025];
\draw[fill] (0.5,0) circle [radius=0.025];
\draw[fill] (1,0) circle [radius=0.025];
\draw[fill] (1.5,0) circle [radius=0.025];
\draw[fill] (2,0) circle [radius=0.025];
\draw[fill] (2.5,0) circle [radius=0.025];
\draw[fill] (3,0) circle [radius=0.025];
\draw[fill] (3.5,0) circle [radius=0.025];
\draw[fill] (0,0.5) circle [radius=0.025];
\draw[fill] (0.5,0.5) circle [radius=0.025];
\draw[fill] (1,0.5) circle [radius=0.025];
\draw[fill] (1.5,0.5) circle [radius=0.025];
\draw[fill] (2,0.5) circle [radius=0.025];
\draw[fill] (2.5,0.5) circle [radius=0.025];
\draw[fill] (3,0.5) circle [radius=0.025];
\draw[fill] (3.5,0.5) circle [radius=0.025];
\draw[fill] (0,1) circle [radius=0.025];
\draw[fill] (0.5,1) circle [radius=0.025];
\draw[fill] (1,1) circle [radius=0.025];
\draw[fill] (1.5,1) circle [radius=0.025];
\draw[fill] (2,1) circle [radius=0.025];
\draw[fill] (2.5,1) circle [radius=0.025];
\draw[fill] (3,1) circle [radius=0.025];
\draw[fill] (3.5,1) circle [radius=0.025];
\draw[fill] (0,1.5) circle [radius=0.025];
\draw[fill] (0.5,1.5) circle [radius=0.025];
\draw[fill] (1,1.5) circle [radius=0.025];
\draw[fill] (1.5,1.5) circle [radius=0.025];
\draw[fill] (2,1.5) circle [radius=0.025];
\draw[fill] (2.5,1.5) circle [radius=0.025];
\draw[fill] (3,1.5) circle [radius=0.025];
\draw[fill] (3.5,1.5) circle [radius=0.025];
\draw[fill] (0,2) circle [radius=0.025];
\draw[fill] (0.5,2) circle [radius=0.025];
\draw[fill] (1,2) circle [radius=0.025];
\draw[fill] (1.5,2) circle [radius=0.025];
\draw[fill] (2,2) circle [radius=0.025];
\draw[fill] (2.5,2) circle [radius=0.025];
\draw[fill] (3,2) circle [radius=0.025];
\draw[fill] (3.5,2) circle [radius=0.025];
\draw[fill] (0,2.5) circle [radius=0.025];
\draw[fill] (0.5,2.5) circle [radius=0.025];
\draw[fill] (1,2.5) circle [radius=0.025];
\draw[fill] (1.5,2.5) circle [radius=0.025];
\draw[fill] (2,2.5) circle [radius=0.025];
\draw[fill] (2.5,2.5) circle [radius=0.025];
\draw[fill] (3,2.5) circle [radius=0.025];
\draw[fill] (3.5,2.5) circle [radius=0.025];


\node at (4, 2.7) {$\vdots$};
\node at (4, 2) {$-5$};
\node at (4, 1.5) {$-4$};
\node at (4, 1) {$-3$};
\node at (4, 0.5) {$-2$};
\node at (4, 0) {$-1$};
\node at (4, -0.5) {Site (Bottom)};


\draw [->] (0.5,0) arc (150:30:7.8pt);
\draw [->] (0.5,0.5) arc (150:30:7.8pt);
\draw [->] (0.5,1) arc (150:30:7.8pt);
\draw [->] (0.5,1.5) arc (150:30:7.8pt);
\draw [->] (0.5,2) arc (150:30:7.8pt);
\draw [->] (0.5,2.5) arc (150:30:7.8pt);

\draw [->] (0.5,0) arc (230:125:8.1pt);
\draw [->] (0.5,0.5) arc (230:125:8.1pt);
\draw [->] (0.5,1) arc (230:125:8.1pt);
\draw [->] (0.5,1.5) arc (230:125:8.1pt);
\draw [->] (0.5,2) arc (230:125:8.1pt);

\draw (0.35,-0.1) rectangle (0.65,1.1);
\draw (0.35,1.4) rectangle (0.65,2.6);

\draw (0.9,-0.1) rectangle (1.1,1.1);
\draw (0.9,1.4) rectangle (1.1,2.6);

\draw (1.4,-0.1) rectangle (1.6,1.1);
\draw (1.4,1.4) rectangle (1.6,2.6);

\draw (1.9,-0.1) rectangle (2.1,1.1);
\draw (1.9,1.4) rectangle (2.1,2.6);

\draw (2.4,-0.1) rectangle (2.6,1.1);
\draw (2.4,1.4) rectangle (2.6,2.6);

\draw (2.9,-0.1) rectangle (3.1,1.1);
\draw (2.9,1.4) rectangle (3.1,2.6);

\draw [line width=1pt] (-0.2,0)-- (3.7,0);

\end{tikzpicture}
\end{center}
\caption{Semi-infinite system with an edge in the bottom.}
\label{fig:bottomBoundedSystem}
    \end{subfigure}
\caption{Semi-infinite systems. The sites for the bottom case are renumbered to make the deduction more natural.}
\label{fig:semiInfiniteSys}
\end{figure}

Let us first consider a semi-infinite system with a unit cell of length $M=3$ and just one edge. The two systems we will consider are represented in Fig.~\ref{fig:semiInfiniteSys}.

To analyze these two case we will have to use a semi-infinite matrix $\mathcal{H}$ similar to matrix (\ref{eq:hamBoundary}). Let us see the two cases separately.

\subsubsection{Top edge state}

In order to obtain a formula for the eigenenergy of the TES (Fig.~\ref{fig:topBoundedSystem}) we have to solve the following SE:

\begin{equation}
\label{eq:topSE}
\mathcal{H}_T\Psi = E_T\Psi,
\end{equation}
where

\begin{eqnarray*}
\mathcal{H}_T = \mqty(A & B & 0 & ...  \\
B^\dagger & A & B & \\
0 & B^\dagger &  & \ddots \\
\vdots &  & \ddots & \ddots) \textrm{,} \quad \Psi = \mqty( \psi_1 \\ \psi_2 \\ \vdots ) \textrm{,} \quad B = \mqty(0&0&0\\0&0&0\\-t_y&0&0) \quad \textrm{and}
\end{eqnarray*}

\begin{eqnarray*}
A = \mqty(-2 t_x \cos(\bld{k}\cdot\bld{a}_x) & -t_y & 0 \\
-t_y & -2 t_x \cos(\bld{k}\cdot\bld{a}_x + \frac{2\pi}{3}) & -t_y\\
0 & -t_y & -2 t_x \cos(\bld{k}\cdot\bld{a}_x + \frac{4\pi}{3}) ).
\end{eqnarray*}
From equation (\ref{eq:topSE}) we have that the bulk (i.e. for $n>1$) satisfies

\begin{equation}
\label{eq:ProbBulk}
E\psi_n =  B^\dagger\psi_{n-1} + A\psi_n + B\psi_{n+1},
\end{equation}
where $\psi_n$ are three dimensional vectors. Furthermore, in the edge, i.e. $n=1$, the wavefunction must satisfy

\begin{equation}
\label{eq:ProbEdgeRight}
E\psi_n = A\psi_n + B\psi_{n+1}.
\end{equation}
The difference between equations (\ref{eq:ProbBulk}) and (\ref{eq:ProbEdgeRight}) is the term $B^\dagger\psi_{n-1}$. Note that Eq.~(\ref{eq:ProbBulk}) is not translation invariant just because of the boundary condition. This means that if we had an infinite system we could use Bloch theorem \cite{Kittel_1986}. Having this in mind, we aim to use one plane wave to find the solution of Eq.~(\ref{eq:ProbEdgeRight}). Let us use the ansatz for $n>0$:

\begin{equation}
\label{eq:rightAnsats}
\psi_n = \kappa^{-n}\phi,
\end{equation}
under the following constraint:

\begin{equation}
\label{eq:rightConstraint}
B^\dagger\phi = 0.
\end{equation}

Note that for $|\kappa| < 1$ the ansatz (\ref{eq:rightAnsats}) does not accurately describe the TES because in that case the wavefunction would be exponentially increasing. So, we have that (\ref{eq:rightAnsats}) only describes this edge state for $|\kappa| > 1$. Thus, let us check for which values of $\kappa$ are Eq.~(\ref{eq:ProbBulk}) and Eq.~(\ref{eq:ProbEdgeRight}) satisfied. From (\ref{eq:rightConstraint}) we have that 

\begin{equation}
\phi = \mqty(\phi_1 \\ \phi_2 \\ 0).
\end{equation}
So, it follows from Eq.~(\ref{eq:ProbBulk}) that

\begin{eqnarray*}
E\phi &=& \kappa B^{\dagger}\phi + A\phi + \kappa^{-1}B\phi\\
E \mqty(\phi_1 \\ \phi_2 \\ 0) &=& \left(\begin{smallmatrix}
-2 t_x \cos(\bld{k}\cdot\bld{a}_x) & -t_y & -t_y \kappa \\
-t_y & -2 t_x \cos(\bld{k}\cdot\bld{a}_x+\frac{2\pi}{3}) & -t_y\\
-t_y\kappa^{-1} & -t_y & -2 t_x \cos(\bld{k}\cdot\bld{a}_x + \frac{4\pi}{3}) \end{smallmatrix}\right)  \mqty(\phi_1 \\ \phi_2 \\ 0).
\end{eqnarray*}
Solving this with respect to $\kappa$ we have

\begin{eqnarray*}
&&\begin{cases} 
    E \phi_1 &= -2 t_x \cos(\bld{k}\cdot\bld{a}_x)\phi_1 - t_y \phi_2 \\
    E \phi_2 &= -t_y \phi_1 - 2 t_x \cos(\bld{k}\cdot\bld{a}_x + \frac{2\pi}{3})\phi_2 \\
    0 &= -t_y\kappa^{-1} \phi_1 - t_y\phi_2
\end{cases}\\
&\iff& \begin{cases} 
      E \phi_1 &= -2 t_x \cos(\bld{k}\cdot\bld{a}_x)\phi_1 + t_y\kappa^{-1} \phi_1 \\
      -E \kappa^{-1} \phi_1 &= -t_y \phi_1 + 2 t_x \kappa^{-1} \cos(\bld{k}\cdot\bld{a}_x + \frac{2\pi}{3})\phi_1 \\
      \phi_2 &= -\kappa^{-1} \phi_1
\end{cases}\\
&\iff& \begin{cases} 
      \kappa^{-1} &= \frac{E + 2 t_x \cos(\bld{k}\cdot\bld{a}_x)}{t_y} \\
      \kappa &= \frac{E + 2 t_x \cos(\bld{k}\cdot\bld{a}_x + \frac{2\pi}{3}) }{t_y} \\
      \phi_2 &= -\kappa^{-1} \phi_1
\end{cases}.
\end{eqnarray*}
Subtracting the expressions of $\kappa$ and $\kappa^{-1}$ we get,

\begin{eqnarray*}
&&\kappa - \kappa^{-1} = 2t\Bigg(\cos(\bld{k}\cdot\bld{a}_x + \frac{2\pi}{3})-\cos(\bld{k}\cdot\bld{a}_x)\Bigg)\\
&\iff&\kappa^2-\kappa\ 2t\Bigg(\cos(\bld{k}\cdot\bld{a}_x + \frac{2\pi}{3})-\cos(\bld{k}\cdot\bld{a}_x) -1 = 0,
\end{eqnarray*}
where $t=\frac{t_x}{t_y}$. Thus we have

\begin{equation}
\label{eq:kappaValueUp}
\kappa_T^{\pm}(\bld{k}) = \frac{-b(\bld{k}) \pm \sqrt{b(\bld{k})^2 + 4}}{2},
\end{equation}
where $b(\bld{k})=-2t\big(\cos(\bld{k}\cdot\bld{a}_x + \frac{2\pi}{3})-\cos(\bld{k}\cdot\bld{a}_x)\big)$. The dispersion relation of the TES is then given by

\begin{equation}
\label{eq:dispersionRelProbUp}
E_T^{\pm}(\bld{k}) = t_y\kappa_T^{\pm}(\bld{k}) - 2 t_x \cos(\bld{k}\cdot\bld{a}_x + \frac{2\pi}{3}) .
\end{equation}

\subsubsection{Bottom edge state}

Let us deduce similarly the dispersion relation for the BES (Fig.~\ref{fig:bottomBoundedSystem}). In this case, we consider

\begin{equation}
\label{eq:bottomSE}
\mathcal{H}_B\Psi = E_B\Psi,
\end{equation}
where

\begin{eqnarray*}
\mathcal{H}_B = \mqty(\ddots & \ddots &  & \vdots  \\
\ddots &  & B & 0 \\
 & B^\dagger & A & B \\
\dots & 0 & B^\dagger & A) \textrm{,} \quad \Psi = \mqty( \vdots \\ \psi_{-2} \\ \psi_{-1} ) \textrm{,} \quad B = \mqty(0&0&0\\0&0&0\\-t_y&0&0)
\end{eqnarray*}
and $A$ and $B$ are given as in the previous section. In order to make the deduction more natural, we will consider $n\leq-1$. In the bulk, i.e. for $n<-1$, we also have

\begin{equation}
\label{eq:DProbBulk}
E\psi_n = B^\dagger\psi_{n-1} + A\psi_n + B\psi_{n+1},
\end{equation}
and in the edge, i.e. $n=-1$, the wavefunction must satisfy

\begin{equation}
\label{eq:ProbEdgeDown}
E\psi_n = B^\dagger\psi_{n-1} + A\psi_n.
\end{equation}
For this case we have instead

\begin{equation}
\label{eq:downAnsats}
\psi_n = \kappa^n\phi,
\end{equation}
with the following constraint:

\begin{equation}
\label{eq:downConstraint}
B\phi = 0.
\end{equation}
From that expression, we conclude that

\begin{equation}
\phi = \mqty(0 \\\phi_1 \\ \phi_2).
\end{equation}
As above, we have that the ansatz (\ref{eq:downAnsats}) only describes correctly the BES for $|\kappa|>1$. Again, aiming to find the allowed values of $\kappa$ we consider Eq.~(\ref{eq:DProbBulk}):

\begin{eqnarray*}
E\phi &=& \kappa^{-1} B^{\dagger}\phi + A\phi + \kappa B\phi\\
E \mqty(0 \\ \phi_1 \\ \phi_2) &=& \left(\begin{smallmatrix}-2 t_x \cos(\bld{k}\cdot\bld{a}_x) & -t_y & -t_y \kappa^{-1} \\
-t_y & -2 t_x \cos(\bld{k}\cdot\bld{a}_x+\frac{2\pi}{3}) & -t_y\\
-t_y\kappa & -t_y & -2 t_x \cos(\bld{k}\cdot\bld{a}_x + \frac{4\pi}{3})\end{smallmatrix}\right) \mqty(0 \\ \phi_1 \\ \phi_2).
\end{eqnarray*}
Solving this with respect to $\kappa$ we have

\begin{eqnarray*}
&&\begin{cases} 
    0 &= -t_y\phi_1 - t_y\kappa^{-1}\phi_2\\
    E \phi_1 &= -2 t_x \cos(\bld{k}\cdot\bld{a}_x+ \frac{2\pi}{3})\phi_1 - t_y \phi_2 \\
    E \phi_2 &= -t_y \phi_1 - 2 t_x \cos(\bld{k}\cdot\bld{a}_x + \frac{4\pi}{3})\phi_2 
\end{cases}\\
&\iff& \begin{cases}
      \phi_1 &= -\kappa^{-1} \phi_2\\
      -E\kappa^{-1} \phi_2 &= 2 t_x \cos(\bld{k}\cdot\bld{a}_x + \frac{2\pi}{3})\kappa^{-1}\phi_2 - t_y\phi_2 \\
      E \phi_2 &= t_y\kappa^{-1} \phi_2 - 2 t_x \cos(\bld{k}\cdot\bld{a}_x + \frac{4\pi}{3})\phi_2
\end{cases}\\
&\iff& \begin{cases} 
      \phi_1 &= -\kappa^{-1} \phi_2\\
      \kappa &= \frac{E + 2 t_x \cos(\bld{k}\cdot\bld{a}_x+ \frac{2\pi}{3})}{t_y} \\
      \kappa^{-1} &= \frac{E + 2 t_x \cos(\bld{k}\cdot\bld{a}_x + \frac{4\pi}{3}) }{t_y}
\end{cases}.
\end{eqnarray*}
Subtracting the expressions of $\kappa$ and $\kappa^{-1}$ we get,

\begin{eqnarray*}
&&\kappa - \kappa^{-1} = -2t\Bigg(\cos(\bld{k}\cdot\bld{a}_x+ \frac{2\pi}{3})-\cos(\bld{k}\cdot\bld{a}_x + \frac{4\pi}{3})\Bigg)\\
&\iff&\kappa^2-\kappa\ 2t\Bigg(\cos(\bld{k}\cdot\bld{a}_x + \frac{2\pi}{3})-\cos(\bld{k}\cdot\bld{a}_x + \frac{4\pi}{3})\Bigg) -1 = 0.
\end{eqnarray*}
where $t=\frac{t_x}{t_y}$. Thus we have

\begin{equation}
\label{eq:kappaValueDown}
\kappa_B^{\pm}(\bld{k}) = \frac{-b(\bld{k}) \pm \sqrt{b(\bld{k})^2 + 4}}{2},
\end{equation}
where $b(\bld{k})=-2t\big(\cos(\bld{k}\cdot\bld{a}_x + \frac{2\pi}{3})-\cos(\bld{k}\cdot\bld{a}_x + \frac{4\pi}{3})\big)$. The dispersion relation of the BES is then given by

\begin{equation}
\label{eq:dispersionRelProbDown}
E_B^{\pm}(\bld{k}) = t_y\kappa_B^{\pm}(\bld{k}) - 2 t_x \cos(\bld{k}\cdot\bld{a}_x + \frac{2\pi}{3}).
\end{equation}
\begin{figure}[!h]
    \begin{subfigure}[t]{\textwidth}
        \centering
        \includegraphics[width=7.5cm, height=6cm]{Img/ty-1_Nm-3_Nsys-30_imp-0.jpg}
        \caption{Numerical solution of the dispersion relations for $t_x=t_y=1$, $M=3$ and $N=30$.}
        \label{fig:numericalDispersion}
    \end{subfigure}

    \begin{subfigure}[t]{0.49\textwidth}
        \centering
        \includegraphics[width=6.5cm,height=9cm]{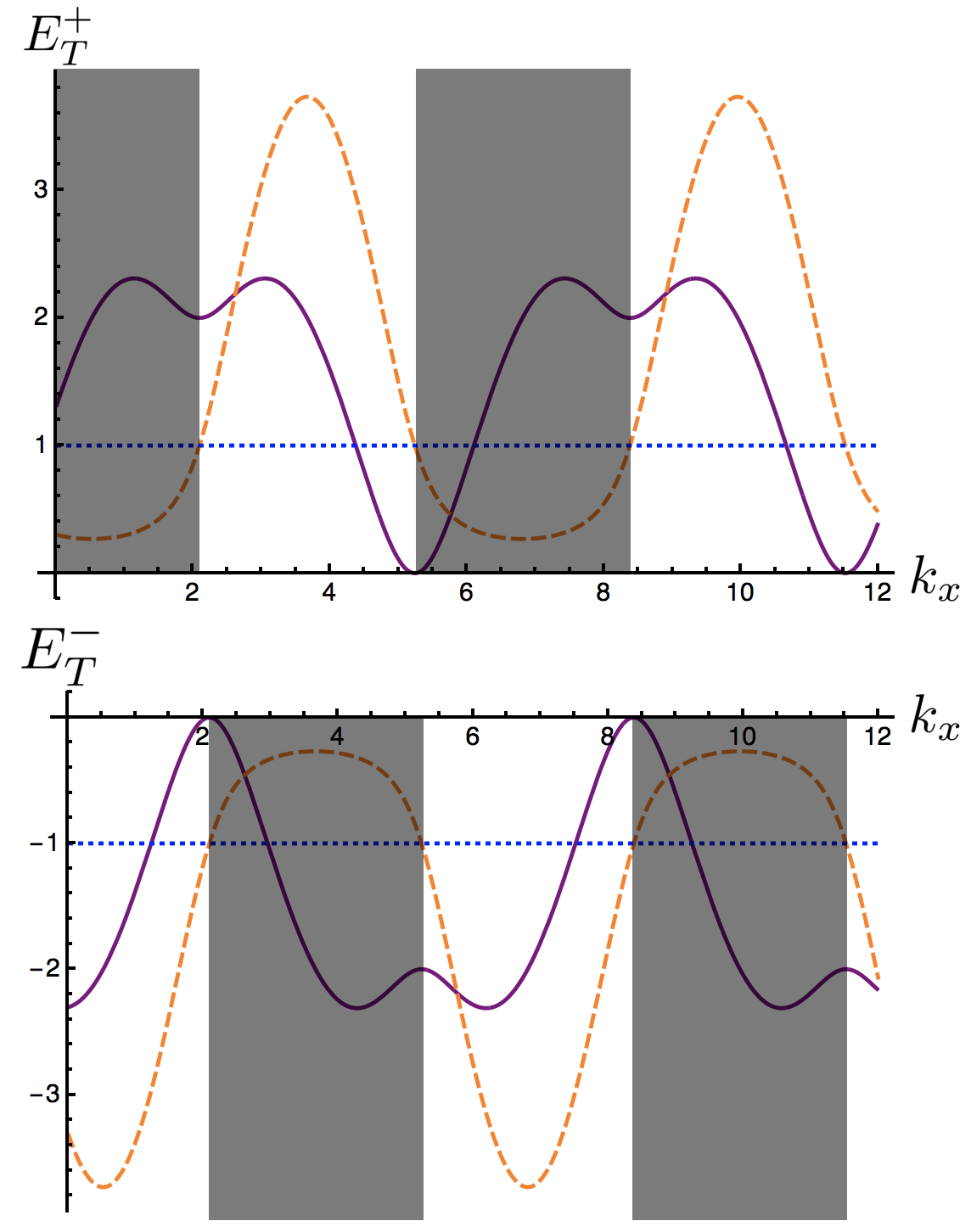}
        \caption{Purple: Top edge dispersion relation, (\ref{eq:dispersionRelProbUp}); Orange: $\kappa^{\pm}_B(\bld{k})$.}
        \label{fig:upEdgeDispersion}
    \end{subfigure}
    ~
    \begin{subfigure}[t]{0.49\textwidth}
        \centering
        \includegraphics[width=6.5cm,height=9cm]{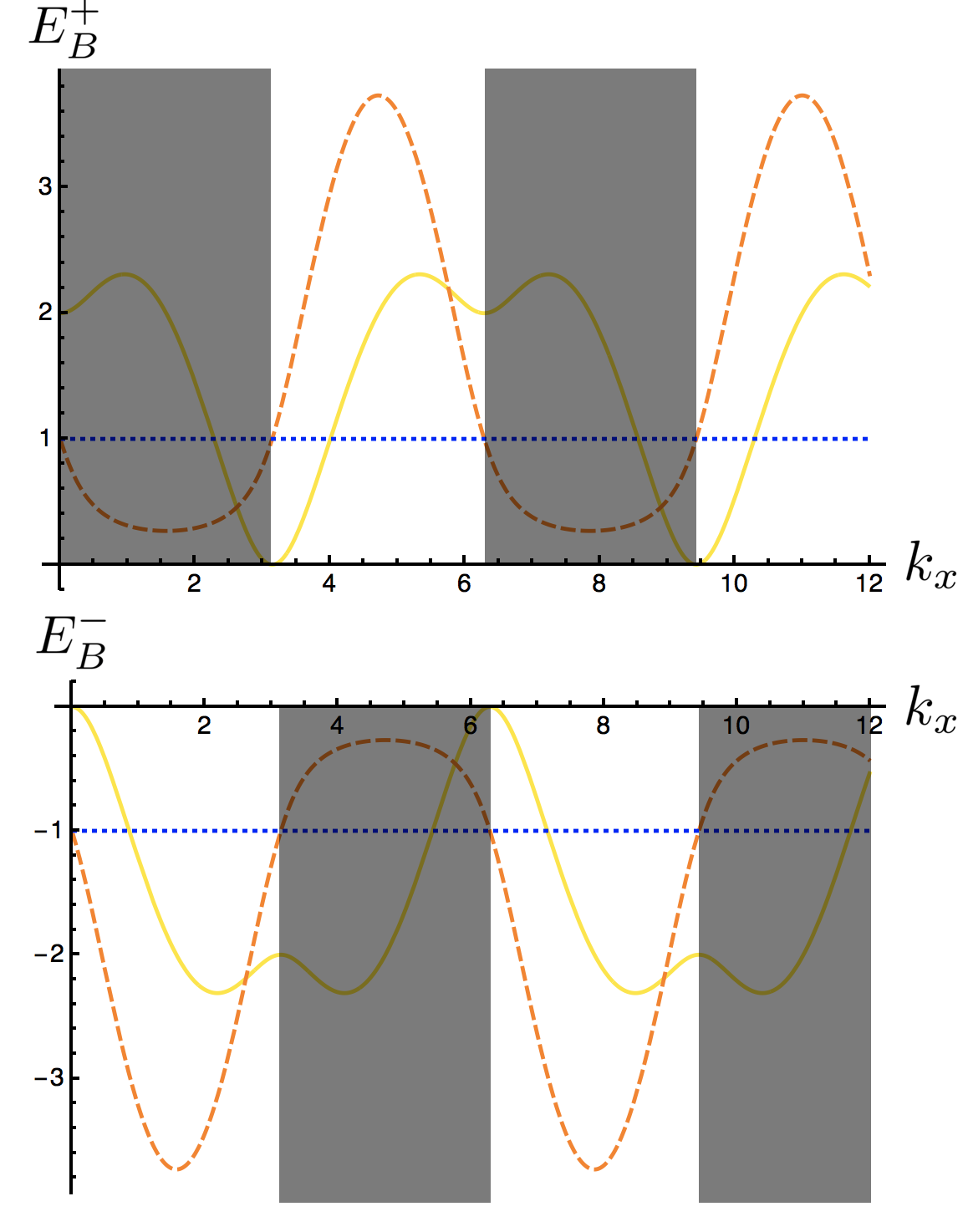}
        \caption{Yellow: Bottom edge dispersion relation, (\ref{eq:dispersionRelProbDown}); Orange: $\kappa^{\pm}_B(\bld{k})$.}
        \label{fig:downEdgeDispersion}
    \end{subfigure}
    \caption{Dispersion relation of the edge states for $t_x=t_y=1$, $M=3$. Compare the purple and yellow graphs between the numerical solution and the analytic expressions obtained for $E^\pm_T$~(\ref{eq:dispersionRelProbUp}) and $E^\pm_B$~(\ref{eq:dispersionRelProbDown}). Observe that exists a perfect match between the areas not shaded and the numerical eigenenergies.}
    \label{fig:dispRelTB}
\end{figure}

Let us compare the dispersion relations obtained for both edge states (\ref{eq:dispersionRelProbUp}, \ref{eq:dispersionRelProbDown}) and the numerical solution. In Fig.~\ref{fig:dispRelTB} it is represented the dispersion relations obtained previously. As we noted before, the ansatzs (\ref{eq:rightAnsats}) and (\ref{eq:downAnsats}) only describe their respective edge state for $|\kappa|>1$. In Fig.~\ref{fig:upEdgeDispersion} and \ref{fig:downEdgeDispersion} the shaded areas cover the areas where $|\kappa|<1$ and it is clear that the dispersion relations of both edge states match the numerical predictions when $|\kappa|>1$.

\subsection{Finite system}

\begin{figure}[h]
\begin{center}
\begin{tikzpicture}

\draw[step=0.5cm,gray,very thin] (-0.2,0) grid (3.7,2.5);

\draw[fill] (0,0) circle [radius=0.025];
\draw[fill] (0.5,0) circle [radius=0.025];
\draw[fill] (1,0) circle [radius=0.025];
\draw[fill] (1.5,0) circle [radius=0.025];
\draw[fill] (2,0) circle [radius=0.025];
\draw[fill] (2.5,0) circle [radius=0.025];
\draw[fill] (3,0) circle [radius=0.025];
\draw[fill] (3.5,0) circle [radius=0.025];
\draw[fill] (0,0.5) circle [radius=0.025];
\draw[fill] (0.5,0.5) circle [radius=0.025];
\draw[fill] (1,0.5) circle [radius=0.025];
\draw[fill] (1.5,0.5) circle [radius=0.025];
\draw[fill] (2,0.5) circle [radius=0.025];
\draw[fill] (2.5,0.5) circle [radius=0.025];
\draw[fill] (3,0.5) circle [radius=0.025];
\draw[fill] (3.5,0.5) circle [radius=0.025];
\draw[fill] (0,1) circle [radius=0.025];
\draw[fill] (0.5,1) circle [radius=0.025];
\draw[fill] (1,1) circle [radius=0.025];
\draw[fill] (1.5,1) circle [radius=0.025];
\draw[fill] (2,1) circle [radius=0.025];
\draw[fill] (2.5,1) circle [radius=0.025];
\draw[fill] (3,1) circle [radius=0.025];
\draw[fill] (3.5,1) circle [radius=0.025];
\draw[fill] (0,1.5) circle [radius=0.025];
\draw[fill] (0.5,1.5) circle [radius=0.025];
\draw[fill] (1,1.5) circle [radius=0.025];
\draw[fill] (1.5,1.5) circle [radius=0.025];
\draw[fill] (2,1.5) circle [radius=0.025];
\draw[fill] (2.5,1.5) circle [radius=0.025];
\draw[fill] (3,1.5) circle [radius=0.025];
\draw[fill] (3.5,1.5) circle [radius=0.025];
\draw[fill] (0,2) circle [radius=0.025];
\draw[fill] (0.5,2) circle [radius=0.025];
\draw[fill] (1,2) circle [radius=0.025];
\draw[fill] (1.5,2) circle [radius=0.025];
\draw[fill] (2,2) circle [radius=0.025];
\draw[fill] (2.5,2) circle [radius=0.025];
\draw[fill] (3,2) circle [radius=0.025];
\draw[fill] (3.5,2) circle [radius=0.025];
\draw[fill] (0,2.5) circle [radius=0.025];
\draw[fill] (0.5,2.5) circle [radius=0.025];
\draw[fill] (1,2.5) circle [radius=0.025];
\draw[fill] (1.5,2.5) circle [radius=0.025];
\draw[fill] (2,2.5) circle [radius=0.025];
\draw[fill] (2.5,2.5) circle [radius=0.025];
\draw[fill] (3,2.5) circle [radius=0.025];
\draw[fill] (3.5,2.5) circle [radius=0.025];

\node at (-0.5, 3) {Site ($m$)};
\node at (-0.5, 2.5) {$1$};
\node at (-0.5, 2) {$2$};
\node at (-0.5, 1.5) {$3$};
\node at (-0.5, 1) {$4$};
\node at (-0.5, 0.5) {$5$};
\node at (-0.5, 0) {$6$};


\draw [->] (0.5,0) arc (150:30:7.8pt);
\draw [->] (0.5,0.5) arc (150:30:7.8pt);
\draw [->] (0.5,1) arc (150:30:7.8pt);
\draw [->] (0.5,1.5) arc (-150:-30:7.8pt);
\draw [->] (0.5,2) arc (-150:-30:7.8pt);
\draw [->] (0.5,2.5) arc (-150:-30:7.8pt);

\draw [->] (0.5,0) arc (230:125:8.1pt);
\draw [->] (0.5,0.5) arc (230:125:8.1pt);
\draw [->] (0.5,1) arc (230:125:8.1pt);
\draw [->] (0.5,1.5) arc (230:125:8.1pt);
\draw [->] (0.5,2) arc (230:125:8.1pt);

\draw (0.35,-0.1) rectangle (0.65,1.1);
\draw (0.35,1.4) rectangle (0.65,2.6);

\draw (0.9,-0.1) rectangle (1.1,1.1);
\draw (0.9,1.4) rectangle (1.1,2.6);

\draw (1.4,-0.1) rectangle (1.6,1.1);
\draw (1.4,1.4) rectangle (1.6,2.6);

\draw (1.9,-0.1) rectangle (2.1,1.1);
\draw (1.9,1.4) rectangle (2.1,2.6);

\draw (2.4,-0.1) rectangle (2.6,1.1);
\draw (2.4,1.4) rectangle (2.6,2.6);

\draw (2.9,-0.1) rectangle (3.1,1.1);
\draw (2.9,1.4) rectangle (3.1,2.6);

\draw [line width=1pt] (-0.2,0)-- (3.7,0);
\draw [line width=1pt] (-0.2,2.5)-- (3.7,2.5);

\end{tikzpicture}
\end{center}
\caption{Finite system with $m=6$ sites and $N=2$ unit cells.}
\label{fig:finiteSys}
\end{figure}
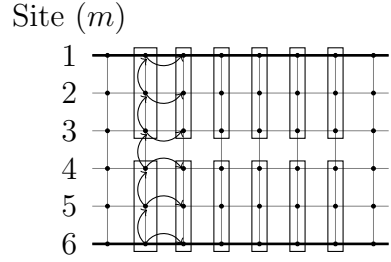

Until now we deduced the eigenenergies of the edge states in the semi-infinite system. Now, based on the numerical simulations previously performed, we aim to find an expression that depicts the gap opening. As it was discussed, the gap arises when the system is small enough. Consequently, we have to deal with a finite system instead. In Fig.~\ref{fig:finiteSys} it can be seen a finite system with $m=6$ sites in the $y$-direction.

In order to find the eigenenergies that describe the gap opening in the bounded system (\ref{eq:hamBoundary}) we will consider a linear combination of two degenerate states of the semi-infinite system. We can observe in Fig.~\ref{fig:numericalDispersion} that the intersections of the dispersion relation of the edge states occur at $k_x=\frac{\pi}{3}+\pi n$, $n\in\ZZ$. This can be confirmed by expressions (\ref{eq:dispersionRelProbUp}) and (\ref{eq:dispersionRelProbDown}):

\begin{eqnarray*}
b_T\big(\frac{\pi}{3}+\pi n,0\big) &=& -2 t\big(\cos\big(\frac{\pi}{3}+\pi n + \frac{2\pi}{3}\big) - \cos\big(\frac{\pi}{3}+\pi n + \frac{4\pi}{3}\big)\big)\\
&=& -2 t\big( \cos( \pi(n+1) ) - \cos\big( \pi(n+1) \frac{2\pi}{3} \big)  \big)\\
&=& -2 t((-1)^{n+1} - \frac{1}{2}(-1)^n)\\
&=& (-1)^n 3 t,
\end{eqnarray*}
where $t=\frac{t_x}{t_y}$. Using the same reasoning we conclude that $b_B\big(\frac{\pi}{3}+\pi n,0\big) = (-1)^n 3 t$. So, we conclude that

\begin{eqnarray*}
E_T^{+}\big(\frac{\pi}{3}+\pi n,0\big) &=& E_B^{+}\big(\frac{\pi}{3}+\pi n,0\big) \quad \textrm{$n$ odd,} \\
E_T^{-}\big(\frac{\pi}{3}+\pi n,0\big) &=& E_B^{-}\big(\frac{\pi}{3}+\pi n,0\big) \quad \textrm{$n$ even.}
\end{eqnarray*}

Without loss of generality let us restrict to the case where $n$ is odd. Consider again the equations:

\begin{eqnarray}
E\psi_1 &=& A\psi_1 + B\psi_{2} \label{eq:boundaryTop},\\
E\psi_n &=& B^\dagger\psi_{n-1} + A\psi_n + B\psi_{n+1} \quad \textrm{for $1<n<N$}, \label{eq:bulkFinite} \\
E\psi_N &=& B^\dagger\psi_{N-1} + A\psi_N. \label{eq:boundaryBottom}
\end{eqnarray}
where $A$ and $B$ are given as in the previews section \ref{semiInfinite}. Our aim is to find a state that satisfies these three expressions. Recall that the top and bottom edge states are given respectively by

\begin{eqnarray*}
\psi_n^T &=& \kappa^{-n}\phi^T\\
\psi_n^B &=& \kappa^{n}\phi^B,
\end{eqnarray*}
where $|\kappa|>1$. As we want to find two different states with the same eigenenergy, let us substitute $\psi_n^T$ and $\psi_n^B$ in equation (\ref{eq:bulkFinite}):

\begin{eqnarray}
E\phi^T &=& \kappa B^{\dagger}\phi^T + A\phi^T + \kappa^{-1}B\phi^T \label{eq:topRelation},\\
E\phi^B &=&\kappa^{-1} B^{\dagger}\phi^B + A\phi^B + \kappa B\phi^B. \label{eq:bottomRelation}
\end{eqnarray}
Again, we get the equations that describe the top and bottom relation. Note that when $n$ is odd, we can write (\ref{eq:topRelation}) and (\ref{eq:bottomRelation}) as follows:

\begin{eqnarray*}
E\phi^T &=& H(\kappa)\phi^T,\\
E\phi^B &=& H(1/\kappa)\phi^B,
\end{eqnarray*}
where $H$ is given by

\begin{eqnarray*}
H(\kappa) &=& \kappa B^\dagger + A + \kappa^{-1}B \\
&=& \mqty(-2 t_x \cos(k_x) & -t_y & -t_y \kappa \\
-t_y & -2 t_x \cos(k_x+\frac{2\pi}{3}) & -t_y\\
-t_y\kappa^{-1} & -t_y & -2 t_x \cos(k_x + \frac{4\pi}{3}) )\\
&=& \mqty( t_x & -t_y & -t_y \kappa \\
-t_y & -2 t_x & -t_y\\
-t_y\kappa^{-1} & -t_y & t_x ).
\end{eqnarray*}
For $n$ even we would get a similar matrix:

\begin{eqnarray*}
H'(\kappa) = \mqty( -t_x & -t_y & -t_y \kappa \\
-t_y & 2 t_x & -t_y\\
-t_y\kappa^{-1} & -t_y & -t_x ).
\end{eqnarray*}
It is clear that we have the following relation,

$$\tau H(\kappa) \tau = H(1/\kappa)$$
where $\tau = \mqty( 0 & 0 & 1 \\
0 & 1 & 0\\
1 & 0 & 0 )$. Since $\tau = \tau^{-1}$, the eigenvalues of $H(\kappa)$ are equal to the eigenvalues of $H(1/\kappa)$,

$$H(\kappa)\phi = E\phi \longrightarrow H(1/\kappa)\tau\phi = E\tau\phi.$$
Thus, we found that the states of the form $\kappa^{-n}\phi$ and $\kappa^{n}\tau\phi$ have the same eigenenergy for Eq.~(\ref{eq:bulkFinite}). So, let us take a liner combination of these two degenerate states and deduce the restrictions imposed on them by the boundary conditions (\ref{eq:boundaryTop}, \ref{eq:boundaryBottom}):

\begin{equation}
\label{eq:degenerateSolution}
\psi_n = (\alpha\kappa^{N+1-n} + \beta\kappa^n\tau)\phi =  (\alpha\kappa^{N+1-n} + \beta\kappa^{n}\tau)\mqty(x \\ y \\ z).
\end{equation}

\begin{figure}[h]
    \begin{subfigure}[t]{0.5\textwidth}
        \centering
        \includegraphics[width=5cm,height=3cm]{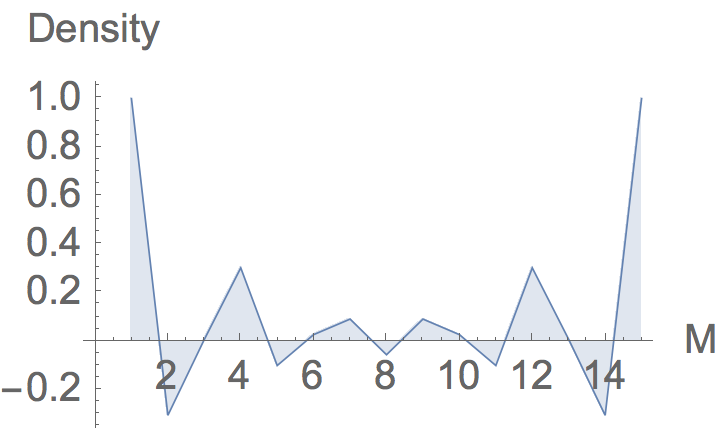}
        \caption{Symmetric result for $N=5$ unit cells.}
        \label{fig:symEdge}    
    \end{subfigure}
    ~
    \begin{subfigure}[t]{0.5\textwidth}
        \centering
        \includegraphics[width=5cm,height=3cm]{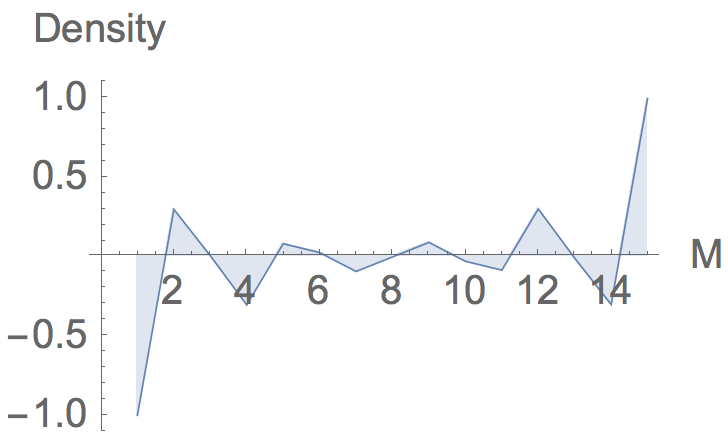}
        \caption{Anti-symmetric result for $N=5$ unit cells.}
        \label{fig:antiSymEdge}    
    \end{subfigure}
    \caption{Superposition of edge states for $k_x=\frac{4\pi}{3}$.}
    \label{fig:edgeStatesAntiAndSym}
\end{figure}

In fact, we can check numerically the eigenstates related to the eigenenergies for which we aim an expression. The densities of these states are represented in Fig.~\ref{fig:edgeStatesAntiAndSym} for $t_x=t_y=1$. We clearly observe that there is a symmetric (Fig.~\ref{fig:symEdge}) and an anti-symmetric (Fig.~\ref{fig:antiSymEdge}) superposition of edge states. 

Using the same reasoning as before and comparing the top boundary condition (\ref{eq:boundaryTop}) with the bulk condition (\ref{eq:bulkFinite}) for $n=1$, we conclude that we must have $B^\dagger\psi_0 = 0$:

\begin{eqnarray*}
&&\mqty(0 & 0 & -t_y \\
0 & 0 & 0 \\
0 & 0 & 0) \psi_0 = 0\\
&\iff&\mqty(0 & 0 & -t_y \\
0 & 0 & 0 \\
0 & 0 & 0)\Bigg[ \alpha \kappa^{N+1} \mqty(x \\ y \\ z) + \beta \mqty(z \\ y \\ x) \Bigg] = 0\\
&\iff& \alpha \kappa^{N+1} z + \beta x = 0.
\end{eqnarray*}
Also, comparing the bottom boundary condition (\ref{eq:boundaryBottom}) with the bulk condition (\ref{eq:bulkFinite}) for $n=N$ we get $B\psi_{N+1} = 0$:

\begin{eqnarray*}
&&\mqty(0 & 0 & 0 \\
0 & 0 & 0 \\
-t_y & 0 & 0) \psi_{N+1} = 0\\
&\iff&\mqty(0 & 0 & 0 \\
0 & 0 & 0 \\
-t_y & 0 & 0)\Bigg[ \alpha \mqty(x \\ y \\ z) + \beta\kappa^{N+1} \mqty(z \\ y \\ x) \Bigg] = 0\\
&\iff& \alpha x + \beta \kappa^{N+1} z = 0.
\end{eqnarray*}
From these we conclude that:

\begin{eqnarray} 
\label{eq:restrictionsTB}
\begin{cases} 
    \alpha \kappa^{N+1} z + \beta x &= 0\\
    \alpha x + \beta \kappa^{N+1} z &= 0
\end{cases}.
\end{eqnarray}
Let us set $x=1$ and $\alpha=1$. This is valid because one can always adjust the state to be normalized. So, the only unknown variables in (\ref{eq:restrictionsTB}) are $z$ and $\beta$. Thus,

\begin{eqnarray} 
\label{eq:restrictionsTB2}
\begin{cases} 
    \kappa^{N+1} z + \beta &= 0\\
    1 + \beta \kappa^{N+1} z &= 0
\end{cases} \iff \begin{cases} 
   \beta &= \pm 1\\
    \kappa^{N+1} z &= - \beta
\end{cases}.
\end{eqnarray}
Note that we are dealing with the case where $|\kappa|>1$. As we pointed out, if we require $\phi$ to be an eigenstate of Eq.~(\ref{eq:topRelation}) we also have that $\tau\phi$ is an eigenstate of Eq.~(\ref{eq:bottomRelation}). This ensures that $\psi_n$ (\ref{eq:degenerateSolution}) satisfies all the finite system relations (\ref{eq:boundaryTop}), (\ref{eq:bulkFinite}) and (\ref{eq:boundaryBottom}) So, let us require that

\begin{eqnarray}
&& H(\kappa)\phi  = E\phi \nonumber\\
&\iff& \mqty( t_x & -t_y & -t_y \kappa \\
-t_y & -2 t_x & -t_y\\
-t_y\kappa^{-1} & -t_y & t_x )\mqty(1 \\ y \\ z) = E \mqty(1 \\ y \\ z). \label{eq:restrictionsMatrix}
\end{eqnarray}
Joining (\ref{eq:restrictionsTB2}) and (\ref{eq:restrictionsMatrix}) we obtain the following set of expressions:

\begin{subnumcases}{\label{eq:restrictions}}
    \beta &$= \pm 1$ \label{eq:1rest}\\
    z &$= -\beta\kappa^{-(N+1)}$ \label{eq:2rest}\\
    t_x-y\ t_y - z\ t_y \kappa &$= E$ \label{eq:3rest}\\
    -t_y - y\ 2t_x - z\ t_y &$= y E $ \label{eq:4rest}\\
    -t_y\kappa^{-1} - y\ t_y + z\ t_x &$= z E$ \label{eq:5rest}
\end{subnumcases}
Our aim is to find a formula of the energy $E$ depending on the system size $N$. Let us attempt a perturbative solution of (\ref{eq:restrictions}). We expand the unknowns as follows:

\begin{eqnarray*} 
&z &= z_0 + z_1 r^N + z_2 r^{2 N} + z_3 r^{3 N} + ...\\
&y &= y_0 + y_1 r^N + y_2 r^{2 N} + y_3 r^{3 N} + ...\\
&\kappa &= \kappa_0 + \kappa_1 r^N + \kappa_2 r^{2 N} + \kappa_3 r^{3 N} + ...\\
&E &= E_0 + E_1 r^N + E_2 r^{2 N} + E_3 r^{3 N} + ... ,
\end{eqnarray*}
where $r$ is a small parameter to be determined.

\

\bld{Zeroth order}

In the semi-infinite case ($N\rightarrow \infty$) we have that $z = 0$. This can be confirmed through Eq.~(\ref{eq:2rest}). Thus, we conclude that $z_0 = 0$. Taking the zeroth order coefficients, we get the expressions for the energy $E$ obtained in the previous section:

\begin{eqnarray*} 
&&\begin{cases} 
     z_0 &= 0\\
     t_x-y_0\ t_y &= E_0\\
     -t_y - y_0\ 2t_x &= y_0\ E_0\\
     -t_y\kappa^{-1}_0 - y_0\ t_y&= 0
\end{cases} \\
&\iff& \begin{cases} 
     t_x + \kappa^{-1}_0\ t_y &= E_0\\
     -t_y +\kappa^{-1}_0\ 2t_x &= -\kappa^{-1}_0\ E_0\\
     y_0 &= -\kappa^{-1}_0
\end{cases} \iff \begin{cases} 
    \kappa^{-1}_0 &= \frac{E_0 - t_x}{t_y}\\
    \kappa_0 &= \frac{E_0 + 2t_x}{t_y}\\
     y_0 &= -\kappa^{-1}_0
\end{cases}.
\end{eqnarray*}
As before, we have

\begin{eqnarray*} 
&& \kappa^2_0 - \kappa_0 \frac{3 t_x}{t_y} - 1 = 0\\
&\Longrightarrow& \kappa_0 = \frac{3 t_x + \sqrt{9 t_x^2 + 4 t_y^2}}{2 t_y},
\end{eqnarray*}
where we take the root such that $|\kappa_0|>1$. Finally we have:

\begin{eqnarray} 
\label{eq:firstAproc}
\begin{cases} 
     z_0 &= 0\\
     y_0 &= \kappa_0\\
     \kappa_0 &= \frac{3 t_x + \sqrt{9 t_x^2 + 4 t_y^2}}{2 t_y}\\
     E_0 &= t_y\ \kappa_0 - 2t_x 
\end{cases}.
\end{eqnarray}
Thus, as a zeroth order approximation we have:

\begin{equation}
\label{eq:zerothOrderEnergy}
E(N) = E_0.
\end{equation}

\

\bld{First order}

Let us first insert the expansion of the unknowns into equations (\ref{eq:3rest}), (\ref{eq:4rest}) and (\ref{eq:5rest}) and then set the terms of $r^N$ equal to one another:

\begin{eqnarray*} 
&&\begin{cases} 
     \begin{split}t_x - (y_0 + y_1 r^N + ...) t_y - (z_0 + z_1 r^N + ...) t_y (k_0 + k_1 r^N + ...) \\ = E_0 + E_1 r^N + ...
     \end{split}\\
     \begin{split}-t_y - (y_0 + y_1 r^N + ...)\ 2t_x - (z_0 + z_1 r^N + ...)t_y\\ = (y_0 + y_1 r^N + ...)(E_0 + E_1 r^N + ...)
     \end{split}\\
     \begin{split}-t_y(\kappa_0 + \kappa_1 r^N + ...)^{-1} - (y_0 + y_1 r^N + ...) t_y + (z_0 + z_1 r^N + ...)t_x\\ = (z_ 0 + z_1 r^N + ...)(E_0 + E_1 r^N + ...)
     \end{split}
\end{cases} \\
&\Longrightarrow& \begin{cases} 
    -y_1 t_y - t_y z_1\kappa_0 &= E_1\\
    -2 y_1 t_x - t_y z_1 &= y_0 E_1 + y_1 E_0\\
    - t_y(y_0\kappa_1 + y_1\kappa_0) + t_x z_1\kappa_0 &= z_1 E_0 \kappa_0
\end{cases},
\end{eqnarray*}
since $z_0=0$. From Eq.~(\ref{eq:2rest}) we also have that:

\begin{eqnarray*}
&&\kappa^{N+1} z = -\beta\\
&\iff& (\kappa_0 + \kappa_1 r^N +...)^{N+1}(z_0 + z_1 r^N +...) = -\beta\\
&\Longrightarrow& \kappa_0^{N+1}z_1r^N = -\beta.
\end{eqnarray*}
Thus, we set $r=\frac{1}{\kappa_0}$ and $z_1=-\frac{\beta}{\kappa_0}$. Until now we have the following expressions:

\begin{eqnarray*}
&&\begin{cases} 
    -y_1 t_y - t_y z_1\kappa_0 &= E_1\\
    -2 y_1 t_x - t_y z_1 &= y_0 E_1 + y_1 E_0\\
    - t_y(y_0\kappa_1 + y_1\kappa_0) + t_x z_1\kappa_0 &= z_1 E_0 \kappa_0\\
    z_1 &=-\frac{\beta}{\kappa_0}
\end{cases}\\
&\iff& \begin{cases}
     -y_1 t_y + \beta t_y &= E_1\\
     -2 y_1 t_x - \frac{t_y \beta}{\kappa_0} &= y_0 E_1 + y_1 E_0\\
     t_y (y_0 \kappa_1 + y_1 \kappa_0) - \beta t_x &= -\beta E_0\\
     &-
\end{cases}\\
&\iff& \begin{cases}
    E_1 &= \beta t_y - y_1 t_y\\
    -2 y_1 t_x - \frac{t_y \beta}{\kappa_0} &= y_0(\beta t_y - y_1 t_y) + y_1 E_0\\
    &-\\
    &-
\end{cases}\\
&\iff& \begin{cases}
    E_1 &= \beta t_y - y_1 t_y\\
    y_1 &= \beta \frac{\frac{t_y}{\kappa_0} - y_0 t_y}{E_0 - y_0 t_y + 2 t_x}\\
    \kappa_1 &= \frac{\beta t_x - \beta E_0 - t_y \kappa_0 y_1}{y_0 t_y}\\
    z_1 &=-\frac{\beta}{\kappa_0}
\end{cases}.\\
\end{eqnarray*}
Hence, as a first order approximation we have:

\begin{equation}
\label{eq:firstOrderEnergy}
E^\beta(N) = E_0 + E_1^\beta r^N,
\end{equation}
where $r=\frac{1}{\kappa_0}$. 

Finally, taking into account that $E^\beta_1 = \beta \frac{3 t_x t_y}{\sqrt{4 t_y^2 + 9 t_x^2}}$, we get a first order approximation of the formula that describes the gap opening:

\begin{equation}
\label{eq:gap}
\boxed{\Delta_{gap}(N) = \frac{6 t_x t_y}{\sqrt{4 t_y^2 + 9 t_x^2}} \bigg(\frac{1}{\kappa_0}\bigg)^N},
\end{equation}
where $\kappa_0 = \frac{3 t_x + \sqrt{9 t_x^2 + 4 t_y^2}}{2 t_y}$. Note that an analogous computation with $n$ even would yield the same result for the gap.

\begin{figure}[!h]
    \begin{subfigure}[t]{0.5\textwidth}
\begin{center}
\definecolor{cqcqcq}{rgb}{0.7529411764705882,0.7529411764705882,0.7529411764705882}
\begin{tikzpicture}[line cap=round,line join=round,>=triangle 45,x=1.0cm,y=1.0cm]
\clip (8,-8) rectangle (12,-4);
\draw [line width=1pt,color=cqcqcq] (6.,-10.)-- (14.,-2.);
\draw [line width=1pt,color=cqcqcq] (6.,-2.)-- (14.,-10.);
\draw [samples=50,domain=-0.99:0.99,rotate around={90.:(10.,-6.)},xshift=10.cm,yshift=-6.cm,line width=1pt] plot ({0.3527293310910264*(1+(\x)^2)/(1-(\x)^2)},{0.35437553384521386*2*(\x)/(1-(\x)^2)});
\draw [samples=50,domain=-0.99:0.99,rotate around={90.:(10.,-6.)},xshift=10.cm,yshift=-6.cm,line width=1.2pt] plot ({0.3527293310910264*(-1-(\x)^2)/(1-(\x)^2)},{0.35437553384521386*(-2)*(\x)/(1-(\x)^2)});

\node at (11.35,-6) {$\Delta \sim r^N$};
\draw [line width=0.2pt,dash pattern=on 0.5pt off 1pt] (10.004380496149475,-5.647243721678609)-- (10.998103238136006,-5.65334781326973);
\draw [line width=0.2pt,dash pattern=on 0.5pt off 1pt] (10.004741478676287,-6.3527609023780265)-- (10.998103238136006,-6.354746150578567);
\draw [->,line width=0.4pt] (10.5,-5.65) -- (10.5,-6.35);
\draw [->,line width=0.4pt] (10.5,-6.35) -- (10.5,-5.65);

\draw [-,line width=0.7pt] (10,-5.6) -- (10,-5.7);
\draw [-,line width=0.7pt] (10,-6.30) -- (10,-6.40);

\node at (10,-5.1) {$E^{+}(N)$};
\node at (10,-6.9) {$E^{-}(N)$};

\end{tikzpicture}
\end{center}
        \caption{Energy gap opening for $k_x=\frac{4\pi}{3}$.}
        \label{fig:gapA}
    \end{subfigure}
    ~
    \begin{subfigure}[t]{0.5\textwidth}
        \centering
        \includegraphics[width=7.6cm,height=3.7cm]{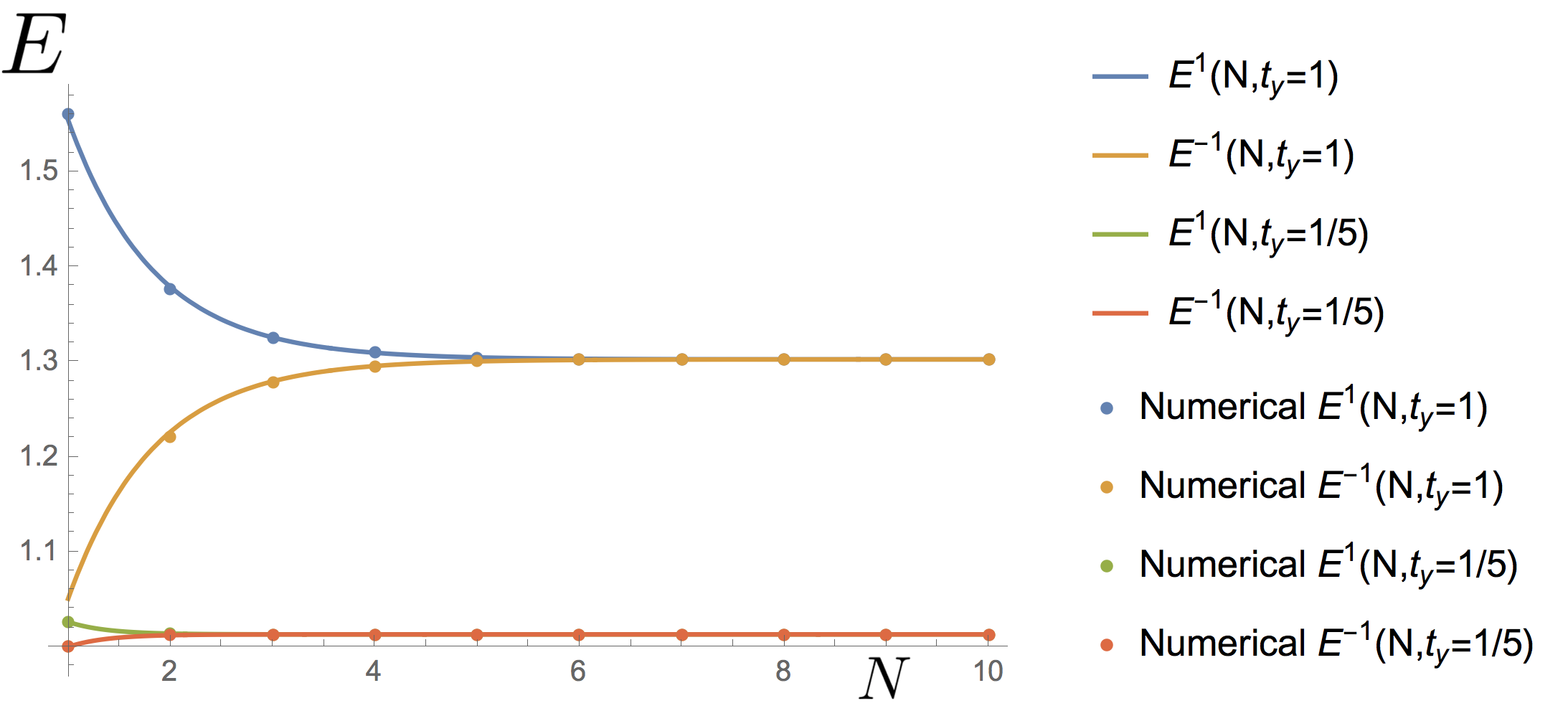}
        \caption{Numerical results and first order formula.}
        \label{fig:gapB}    
    \end{subfigure}
    
    \begin{subfigure}[t]{0.5\textwidth}
        \centering
        \includegraphics[width=6.6cm,height=3.6cm]{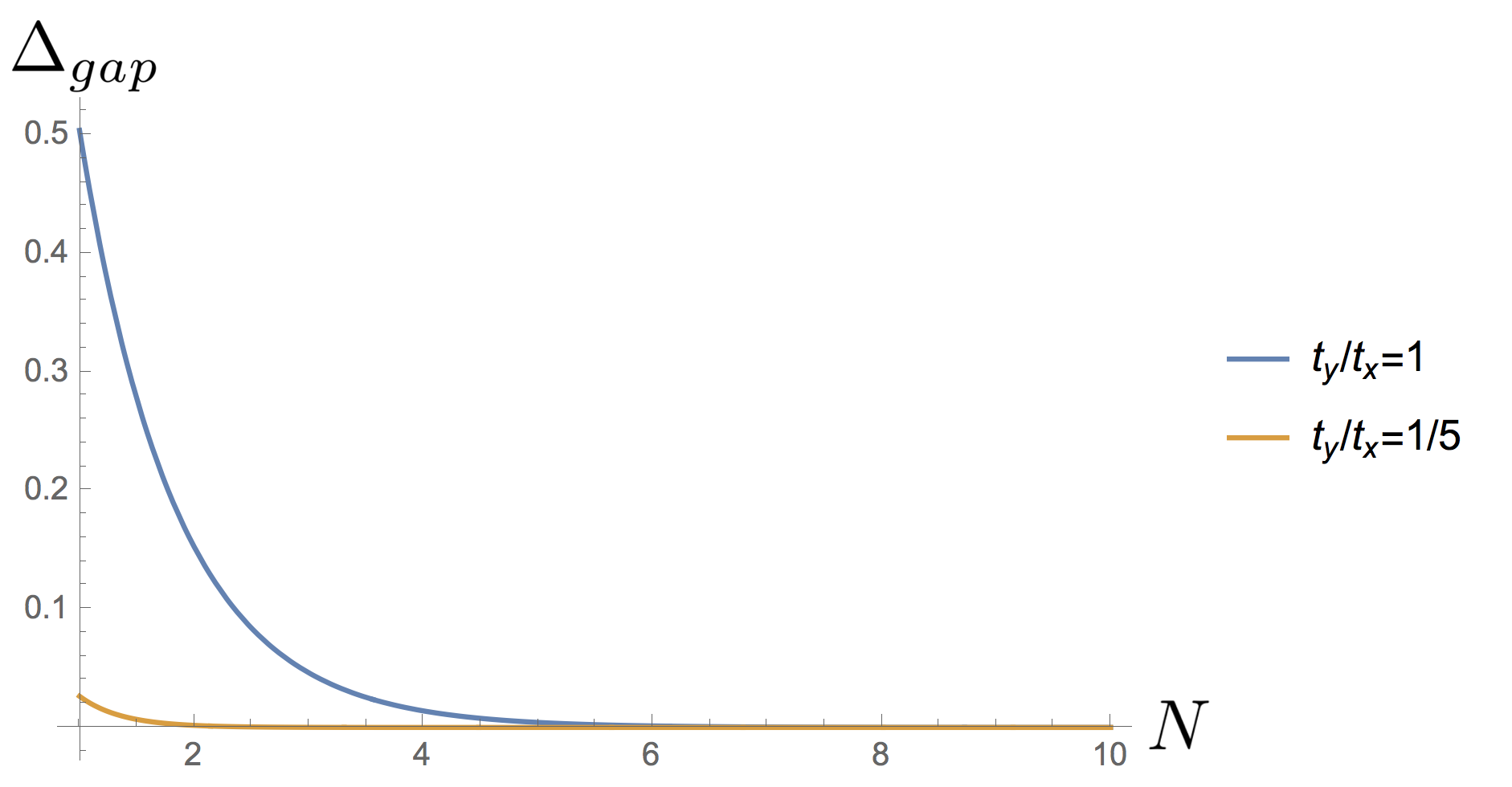}
        \caption{Gap opening for two different hopping ratios.}
        \label{fig:gapC}    
    \end{subfigure}
    ~
    \begin{subfigure}[t]{0.5\textwidth}
        \centering
        \includegraphics[width=5.7cm,height=3.6cm]{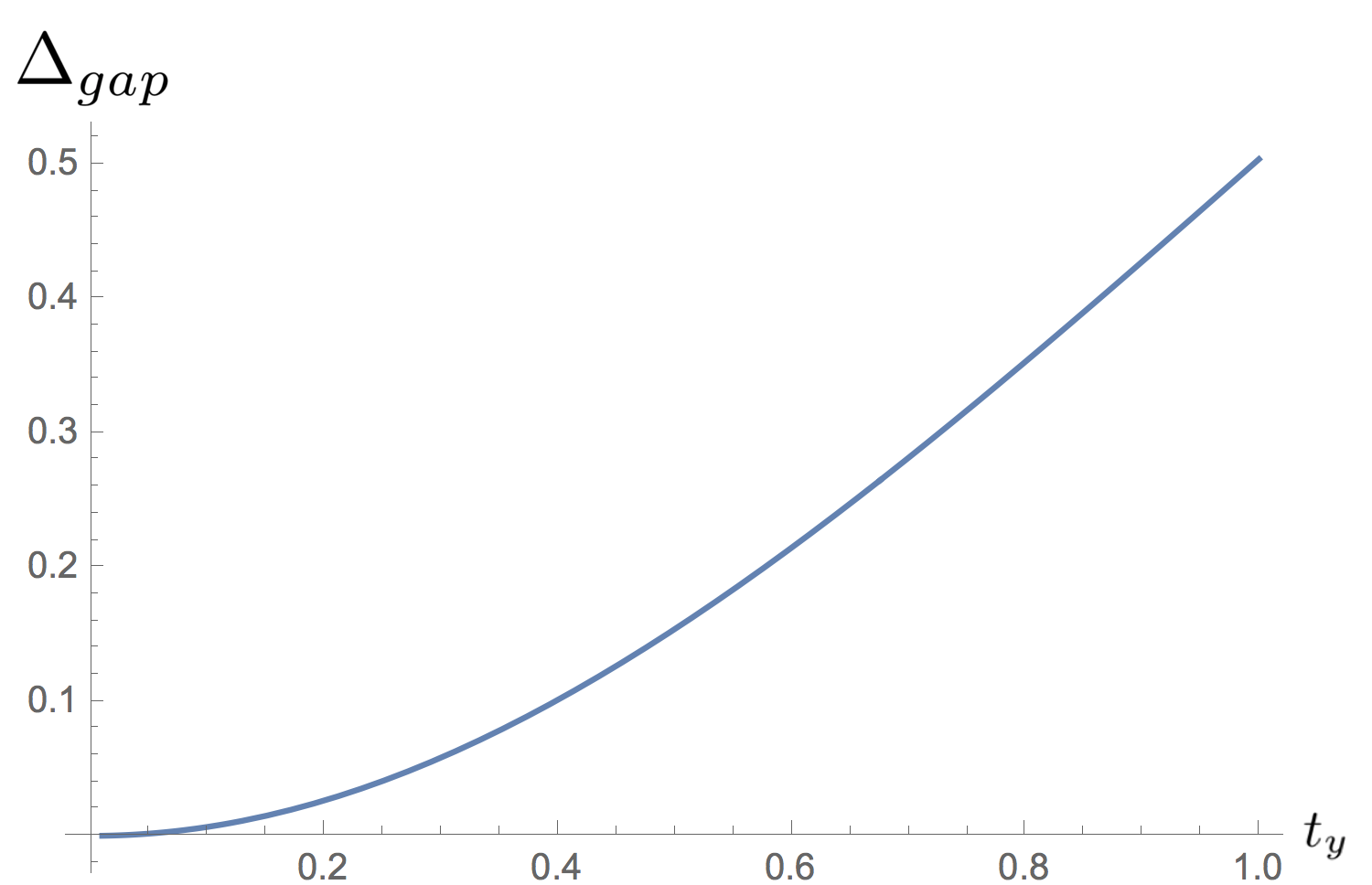}
        \caption{Gap opening with respect to $t_y$ for $N=1$.}
        \label{fig:gapD}    
    \end{subfigure}
    \caption{Energy gap opening for $k_x=\frac{4\pi}{3}$. We observe that the gap opens for small systems and the bigger the $y$-direction hopping value ($t_y$) the wider the gap.}
    \label{fig:gapOpening}
\end{figure}
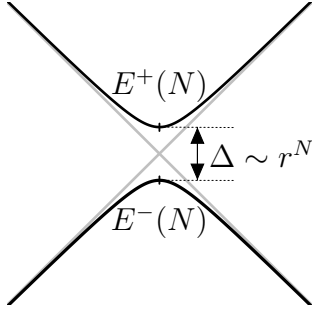
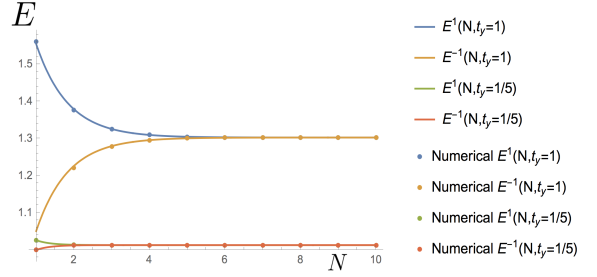
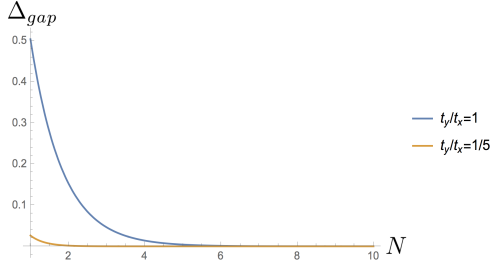
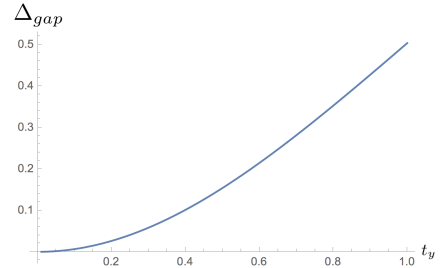

From this deduction, we can confirm that the gap opening is directly related to the size of the system and the hopping terms, $t_y$ and $t_x$. Let us set $t_x=1$. In Fig.~\ref{fig:gapB} we can see that the results obtained using formula (\ref{eq:firstOrderEnergy}) and the numerical results have a very good agreement. As expected, it is clear that the gap decreases as the system size increases.
Moreover, in Fig.~\ref{fig:gapB} we can compare the energy values for $t_y=1$ and $t_y=\frac{1}{5}$. In fact, the gap closes quicker for smaller values of $t_y$. This can also be confirmed by the plot of Eq.~(\ref{eq:gap}) for different $t_y$ values, Fig.~\ref{fig:gapC}. 

These results are in accordance with the numerical simulations presented in section \ref{bounded_systems}. We conjectured that the gap would increase due to the interaction of edge states with the other boundary. This would be caused mainly by small sized systems and by big values of the hopping term in the $y$-direction, $t_y$.

Concerned with this issue, Mugel's group analyzed to what extent it is possible to measure topological invariants of topological insulators in small systems \cite{DM_Mugel_2017}. They claim that the measurement of the Chern number gets less accurate when the ratio $t_y/t_x$ decreases, pointing the delocalization of the edge eigenstates as the main reason for that to happens. Now, with an analytic expression for the gap, we can understand for which values of $t_y$ and $N$ the system represents better a topological insulator and its topological invariants.

\newpage

\section{Conclusion}

In several articles \cite{Mancini_2015,Lin_2009,DM_Hugel_2014,Jim_2012,DM_Celi_2014,DM_Stuhl_2015} some features of topological insulators such as Chern number, edge currents and the robustness of the edge states are analyzed. They all base their measurements in small sized systems of ultra-cold atoms. Due to that, it is important to understand to what extent these systems are accurately describing the behaviour of topological insulators. In order to do it effectively, it is necessary to know how the gap changes with respect to different parameters. A formula describing this would give some clues to experimentalist about the type of Hofstadter model that better describes topological insulators.

In section \ref{Edge_States} we discussed the origin of edge currents in finite Quantum Hall systems and in section \ref{bounded_systems} we observed that the edge eigenenergies should close the gap between bulk bands. However, based on numerical simulations we saw that the gap opens depending on the size of the system and the value of the hopping terms, $t_y$ and $t_x$. This suggests that systems with a wide gap may not exactly mimic the behaviour of a topological insulator where edge currents must exist.

To allow a better understanding of the gap we aimed to find a formula that describes its opening. We first found the dispersion relation of the top and bottom edge states in the semi-infinite system:

\begin{eqnarray*}
E_T^{\pm}(\bld{k}) &=& t_y\kappa_T^{\pm}(\bld{k}) - 2 t_x \cos(\bld{k}\cdot\bld{a}_x + \frac{2\pi}{3})\\
E_B^{\pm}(\bld{k}) &=& t_y\kappa_B^{\pm}(\bld{k}) - 2 t_x \cos(\bld{k}\cdot\bld{a}_x + \frac{2\pi}{3}) 
\end{eqnarray*}
where we should have $|\kappa_i^{\pm}(\bld{k})|<1$ for $E_i^{\pm}(\bld{k})$ to describe the edge state properly, ($i=T,B$). 

Then, using a linear combination of opposite edge states of the semi-infinite system and perturbation theory, we concluded and confirmed that the size of the gap is directly related to the size of the system $N$ and the hopping terms, $t_y$ and $t_x$:

\begin{equation*}
\Delta_{gap}(N) = \frac{6 t_x t_y}{\sqrt{4 t_y^2 + 9 t_x^2}} \bigg(\frac{1}{\kappa_0}\bigg)^N
\end{equation*}
where $\kappa_0 = \frac{3 t_x + \sqrt{9 t_x^2 + 4 t_y^2}}{2 t_y}$. This result is valid for $k_x=\frac{\pi}{3}+\pi n$, $n\in\ZZ$.

The natural extension of this work would be to analyze and deduce an analytic formula that describes the gap opening with the introduction of in-site disorder. This would ease the understanding of the robustness of edge states to impurities in topological insulators.

\newpage

\nocite{*}
\bibliographystyle{abbrv}
\bibliography{biblio}

\end{document}